\documentclass[twocolumn,times]{aastex62}

\usepackage{CJKutf8}

\usepackage{amsmath}
\usepackage{amssymb}
\usepackage{amsfonts}
\usepackage{graphicx}
\usepackage{wasysym}
\usepackage{multirow}
\usepackage{mdframed}
\usepackage[T1]{fontenc}

\hypersetup{linkcolor=red,citecolor=blue,filecolor=cyan,urlcolor=magenta}

\newcommand{\methanol}{\mbox{CH$_3$OH}}

\newcommand{\water}{\mbox{H$_2$O}}
\newcommand{\amm}{\mbox{NH$_3$}}

\newcommand{\nthp}{\mbox{N$_2$H$^+$}}

\newcommand{\kms}{\mbox{km\,s$^{-1}$}}

\newcommand{\sqc}{\mbox{cm$^{-2}$}}
\newcommand{\cc}{\mbox{cm$^{-3}$}}
\newcommand{\lsol}{\mbox{$L_\odot$}}
\newcommand{\msol}{\mbox{$M_\odot$}}
\newcommand{\msolpyr}{\mbox{$M_\odot$\,yr$^{-1}$}}
\newcommand{\mjypbm}{\mbox{mJy\,beam$^{-1}$}}
\newcommand{\jypbm}{\mbox{Jy\,beam$^{-1}$}}
\newcommand{\hii}{\mbox{H\,{\sc ii}}}
\newcommand{\vlsr}{\mbox{$V_\text{lsr}$}}

\newcommand{\ctw}{the 20~\kms{} cloud}
\newcommand{\cfi}{the 50~\kms{} cloud}

\newcommand{\gzp}{G0.253+0.016}
\newcommand{\sgb}{Sgr~B1-off}

\received{- -, --}
\revised{- -, --}
\accepted{- -, --}
\submitjournal{ApJ}

\shorttitle{Star Formation Rates of Central Molecular Zone Clouds}
\shortauthors{Lu et al.}

\begin{document}
\begin{CJK}{UTF8}{gbsn}

\title{Star Formation Rates of Massive Molecular Clouds in the Central Molecular Zone}

\correspondingauthor{Xing Lu}
\email{xinglv.nju@gmail.com, xing.lu@nao.ac.jp}

\author[0000-0003-2619-9305]{Xing Lu (吕行)}
\affiliation{National Astronomical Observatory of Japan, 2-21-1 Osawa, Mitaka, Tokyo, 181-8588, Japan}

\author{Qizhou Zhang}
\affiliation{Center for Astrophysics | Harvard \& Smithsonian, 60 Garden Street, Cambridge, MA 02138, USA}

\author{Jens Kauffmann}
\affiliation{Haystack Observatory, Massachusetts Institute of Technology, 99 Millstone Road, Westford, MA 01886, USA}

\author{Thushara Pillai}
\affiliation{Boston University Astronomy Department, 725 Commonwealth Avenue, Boston, MA 02215, USA}

\author{Adam Ginsburg}
\affiliation{National Radio Astronomy Observatory, 1003 Lopezville Road, Socorro, NM 87801, USA}

\author{Elisabeth A.\ C.\ Mills}
\affiliation{Physics Department, Brandeis University, 415 South Street, Waltham, MA 02453, USA}

\author{J.\ M.\ Diederik Kruijssen}
\affiliation{Astronomisches Rechen-Institut, Zentrum f\"ur Astronomie der Universit\"at Heidelberg, M\"onchhofstra{\ss}e 12-14, D-69120 Heidelberg, Germany}

\author{Steven N.\ Longmore}
\affiliation{Astrophysics Research Institute, Liverpool John Moores University, 146 Brownlow Hill, Liverpool L3 5RF, UK}

\author{Cara Battersby}
\affiliation{University of Connecticut, Department of Physics, 2152 Hillside Road, Storrs, CT 06269, USA}

\author{Hauyu Baobab Liu}
\affiliation{European Southern Observatory, Karl-Schwarzschild-Stra{\ss}e 2, D-85748 Garching bei M\"unchen, Germany}

\author{Qiusheng Gu}
\affiliation{School of Astronomy and Space Science, Nanjing University, Nanjing, Jiangsu 210093, China}

\begin{abstract} 
We investigate star formation at very early evolutionary phases in five massive clouds in the inner 500~pc of the Galaxy, the Central Molecular Zone. Using interferometer observations of \water{} masers and ultra-compact \hii{} regions, we find evidence of ongoing star formation embedded in cores of 0.2~pc scales and $\gtrsim$10$^5$~\cc{} densities. Among the five clouds, Sgr~C possesses a high (9\%) fraction of gas mass in gravitationally bound and/or protostellar cores, and follows the dense ($\gtrsim$10$^4$~\cc{}) gas star formation relation that is extrapolated from nearby clouds. The other four clouds have less than 1\% of their cloud masses in gravitationally bound and/or protostellar cores, and star formation rates 10 times lower than predicted by the dense gas star formation relation.
At the spatial scale of these cores, the star formation efficiency is comparable to that in Galactic disk sources. We suggest that the overall inactive star formation in these Central Molecular Zone clouds could be because there is much less gas confined in gravitationally bound cores, which may be a result of the strong turbulence in this region and/or the very early evolutionary stage of the clouds when collapse has only recently started.
\end{abstract}

\keywords{Galatic: center --- stars: formation --- ISM: clouds}

\section{INTRODUCTION}\label{sec:intro}

The classic Kennicutt-Schmidt relation \citep{schmidt1959,kennicutt1998} describes a correlation between the star formation rate (SFR) per unit area and the total gas mass (including both molecular and atomic gases) in galaxies. One of its variations is a linear correlation between the SFR (traced by infrared luminosities or young stellar object counts) and the amount of dense ($\gtrsim$10$^4$~\cc{}) molecular gas found in both Galactic sources and external galaxies \citep{gao2004,wu2005,wu2010,lada2010,lada2012,zhangZY2014}, sometimes referred to as the \textit{dense gas star formation relation}. This linear correlation is suggested to be a result of constant star formation efficiency (SFE) in molecular gas of densities $\gtrsim$10$^4$~\cc{} \citep{lada2012}.

\begin{figure*}[!t]
\centering
\includegraphics[width=1\textwidth]{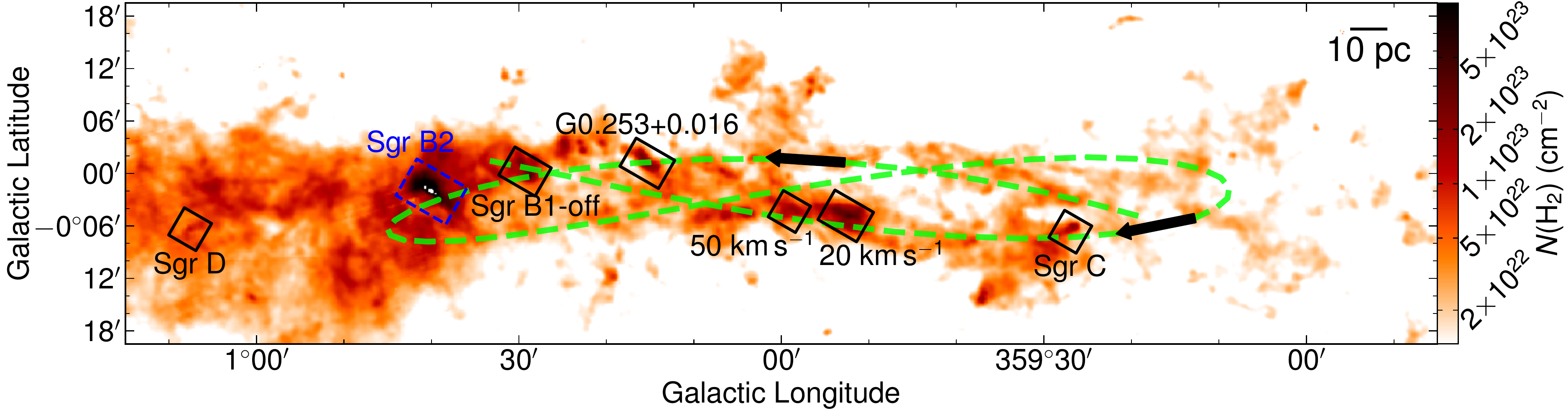} \\
\caption{An overview of the CMZ and the six clouds in our observations. The background image shows column densities derived from \textit{Herschel} data \citep{battersby2011}. The black boxes mark the same areas toward the six clouds in \autoref{fig:sample}. Sgr~B2 is not included in our observations but is discussed in Sections~\ref{subsec:disc_sfr} \& \ref{subsec:disc_sflaw}, and is marked by a blue box. The orbital model for the gas streams in the CMZ proposed by \citet{kruijssen2015} is shown by the green dashed curve, with two black arrows indicating the direction of the proposed orbital motion.}
\label{fig:overview}
\end{figure*}

The Central Molecular Zone (CMZ; \autoref{fig:overview}), the inner 500~pc of our Galaxy, does not fit into this correlation. It contains molecular gas of several times 10$^7$~\msol{} with mean densities of $\sim$10$^4$~\cc{} (\citealt{morris1996,ferriere2007}; potential multiple density components from 10$^3$ to 10$^7$~\cc{}, \citealt{walmsley1986,mills2018}), but the SFR is lower by at least an order of magnitude than expected from the dense gas star formation relation, both on the scale of the entire CMZ \citep[e.g.,][]{yusefzadeh2009,an2011,immer2012a,longmore2013a,barnes2017} and of individual clouds \citep[e.g.,][]{kauffmann2013a,kauffmann2017a,barnes2017}. \citet{kauffmann2017a,kauffmann2017b} studied star formation and dense gas content of several representative massive clouds in the CMZ, and concluded that star formation in a time scale of 1.1~Myr in some of the CMZ clouds is $\gtrsim$10 times lower than expected from the linear correlation of \citet{lada2010}.

A possible explanation for the low SFR in the CMZ clouds is that these clouds are at very early evolutionary phases and active star formation has not emerged yet \citep{kruijssen2014,KK2015,krumholz2017}, although this may not be able to account for individual clouds that already show signatures of late evolutionary phases \citep[e.g., \hii{} regions in several clouds;][]{kauffmann2017a}. Previous studies using infrared luminosities \citep[e.g.,][]{barnes2017} or free-free emission from \hii{} regions \citep[e.g.,][]{longmore2013a,kauffmann2017a} generally characterize star formation in a time scale of a few Myr. A very young generation of star formation deeply embedded in dense gas that is invisible in infrared or free-free emission could have been missed. This young generation of star formation can be traced by masers, ultra-compact (UC) \hii{} regions, and hot molecular cores that are usually associated with star formation that occurred in the last $<$1~Myr.

Another compelling reason to investigate star formation at very early evolutionary phases is the evolutionary cycling---that is, the potential time delay between the star formation indicated by observations and the observed gas \citep{KL2014}. The turbulent crossing time scale of the CMZ clouds is $\gtrsim$0.3~Myr \citep{kruijssen2014,federrath2016,kauffmann2013a,kauffmann2017a}.
Over a time scale of longer than this, the feedback from star formation may have altered the environment therefore the observed gas reservoir is not directly related to the observed star formation. Such disconnection between active star formation revealed by \hii{} regions and a lack of dense and massive clumps has been noted for \cfi{} \citep{kauffmann2017b}. If we only consider the protostellar population formed in the last 0.3~Myr (i.e., those formed within a time scale comparable to the crossing time), then the derived SFR should be more relevant to the observed gas, although the ratio between star formation and gas is still time-dependent and evolutionary cycling matters.

\begin{figure*}[!t]
\centering
\begin{tabular}{@{}p{0.333\textwidth}@{}@{}p{0.333\textwidth}@{}@{}p{0.333\textwidth}@{}}
\begin{tabular}[c]{@{}c@{}}
\includegraphics[width=0.333\textwidth]{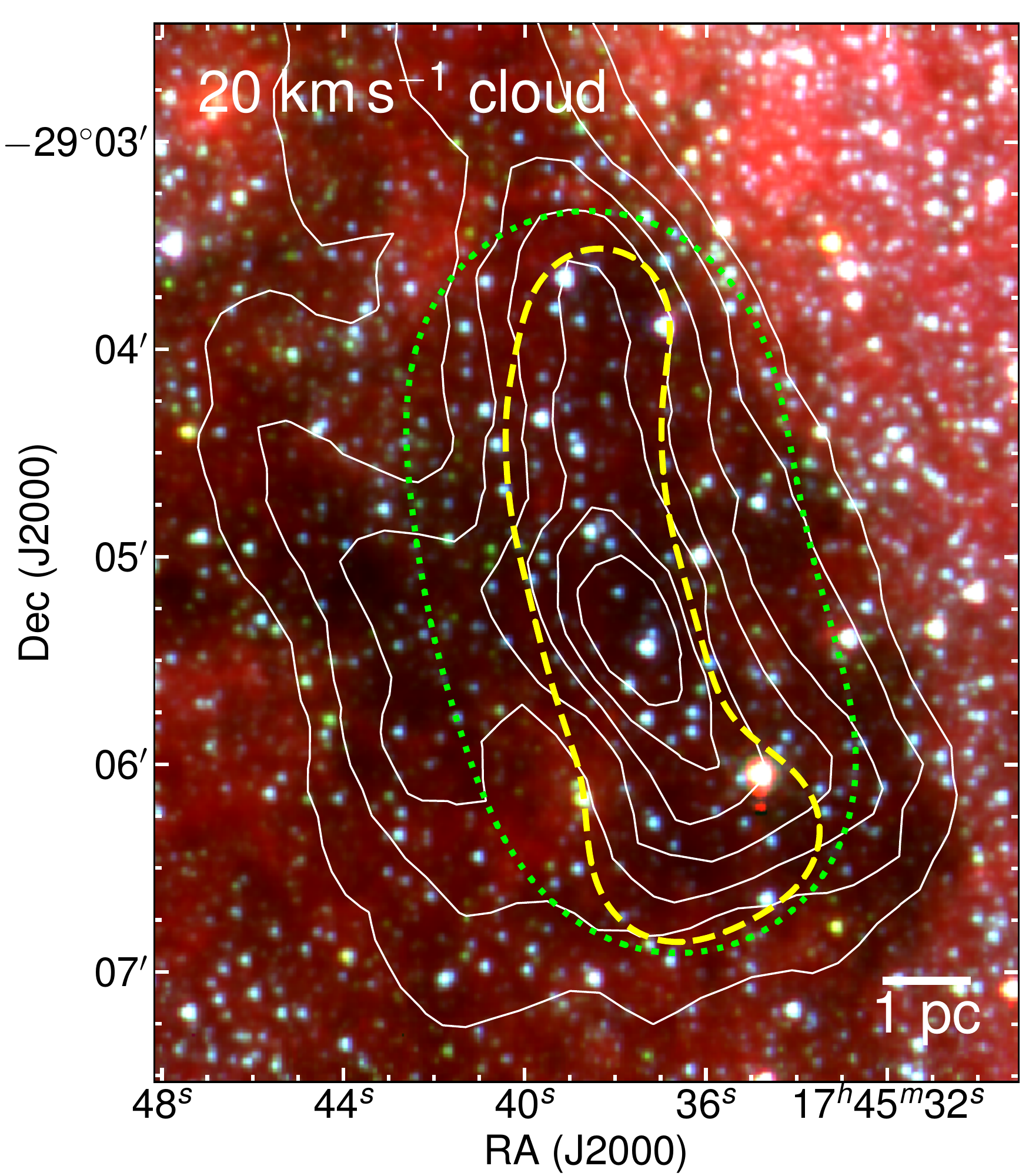} \\
\includegraphics[width=0.286\textwidth]{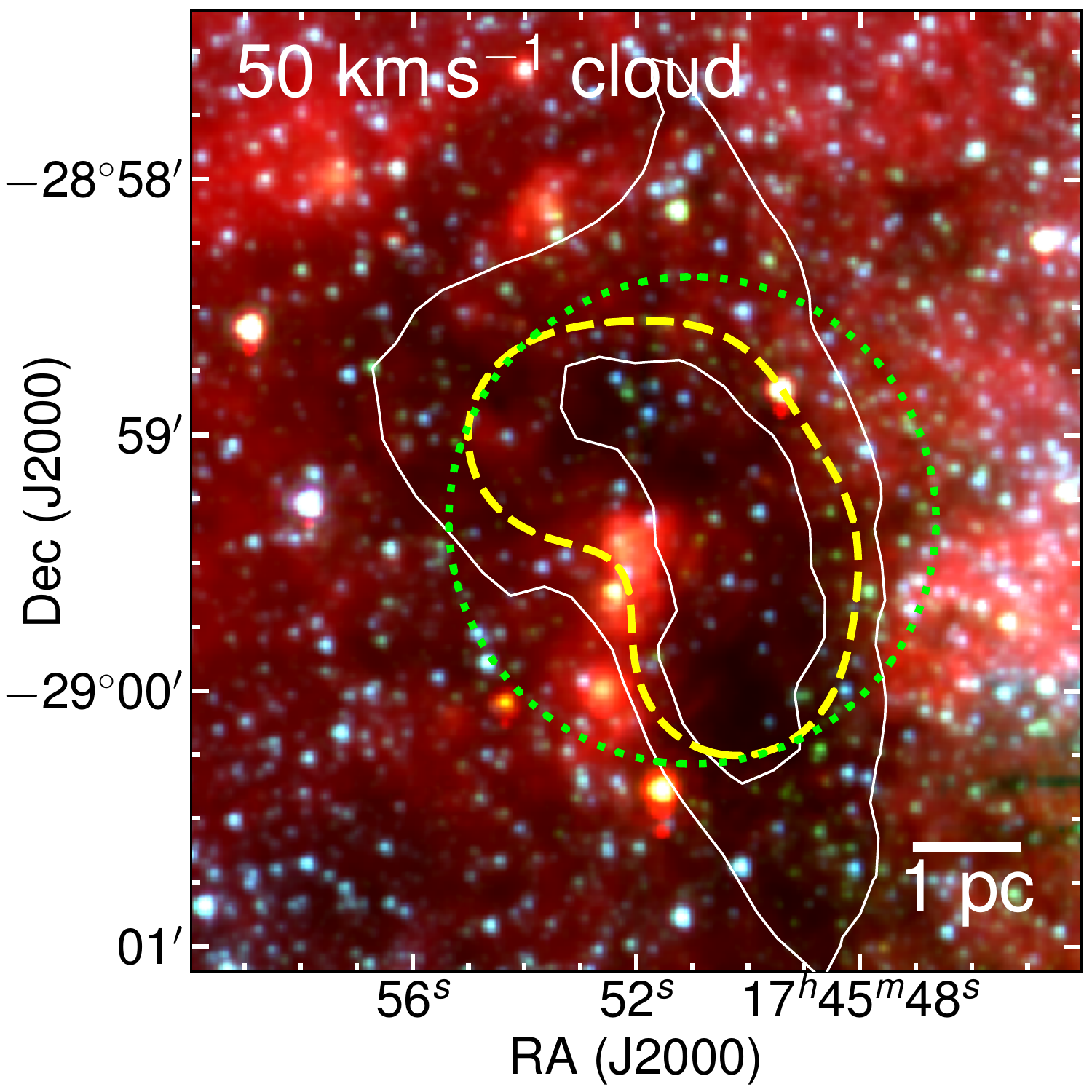}
\end{tabular}
&
\begin{tabular}[c]{@{}c@{}}
\includegraphics[width=0.314\textwidth]{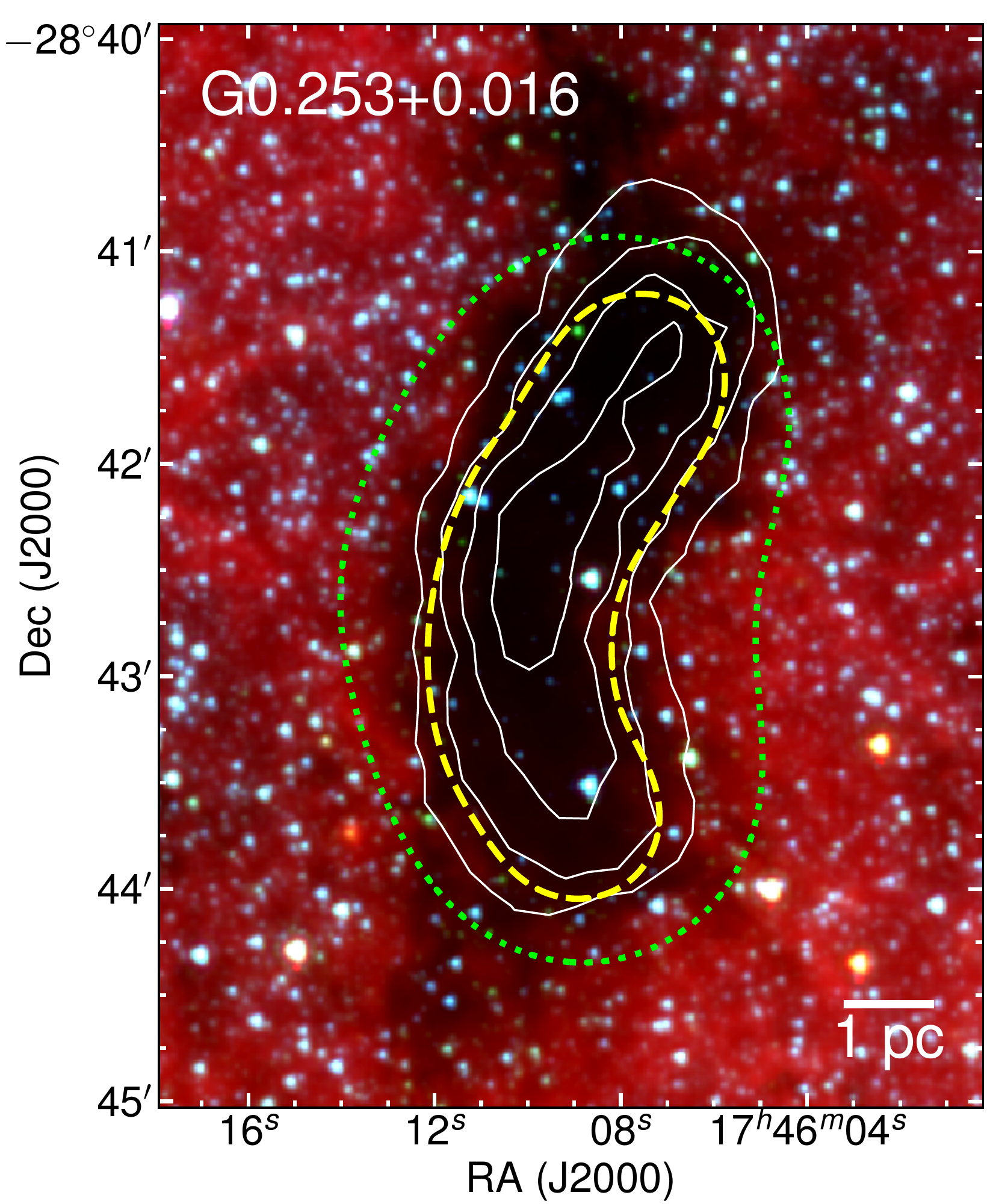} \\
\includegraphics[width=0.314\textwidth]{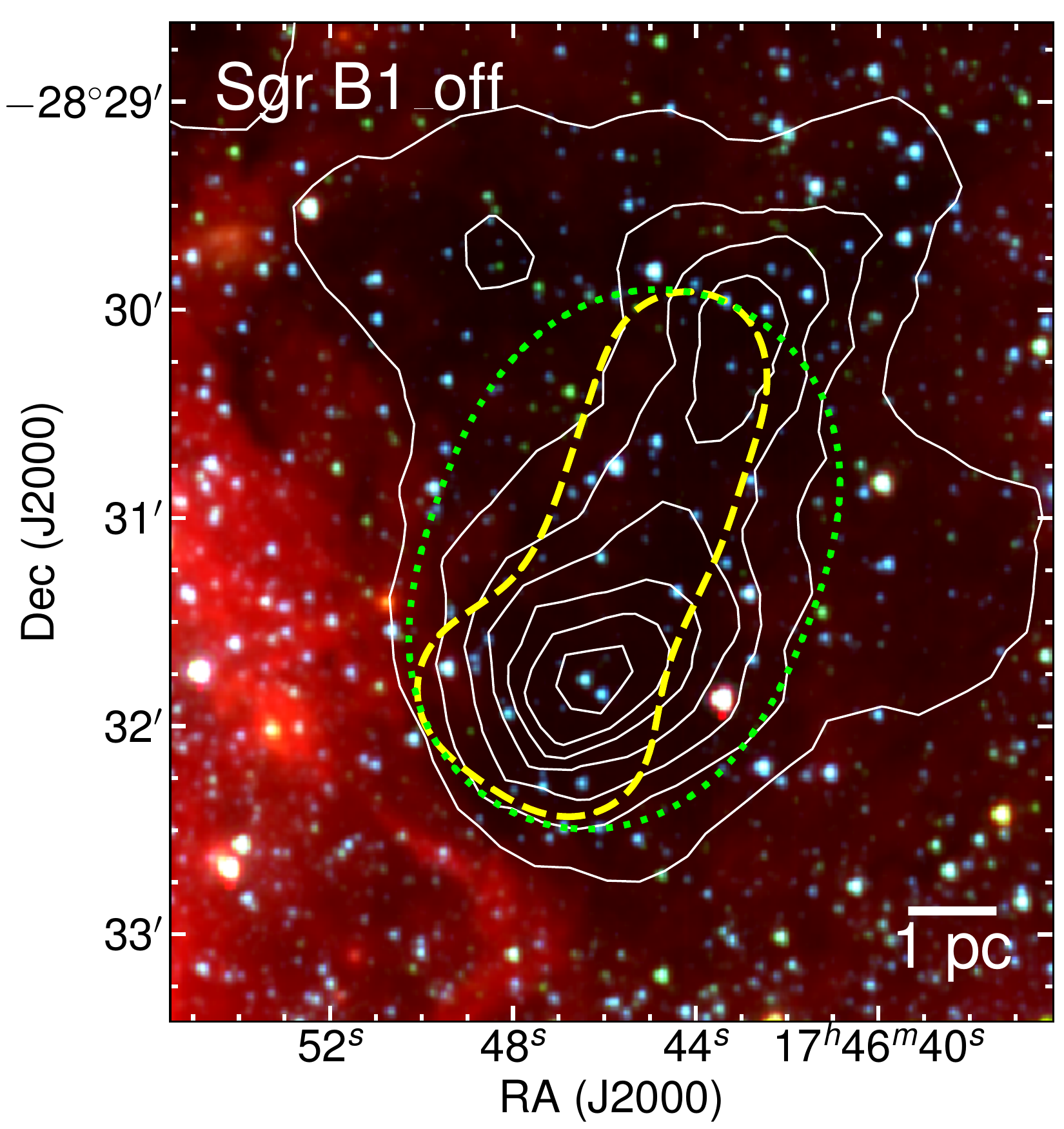}
\end{tabular}
&
\begin{tabular}[c]{@{}c@{}}
\includegraphics[width=0.286\textwidth]{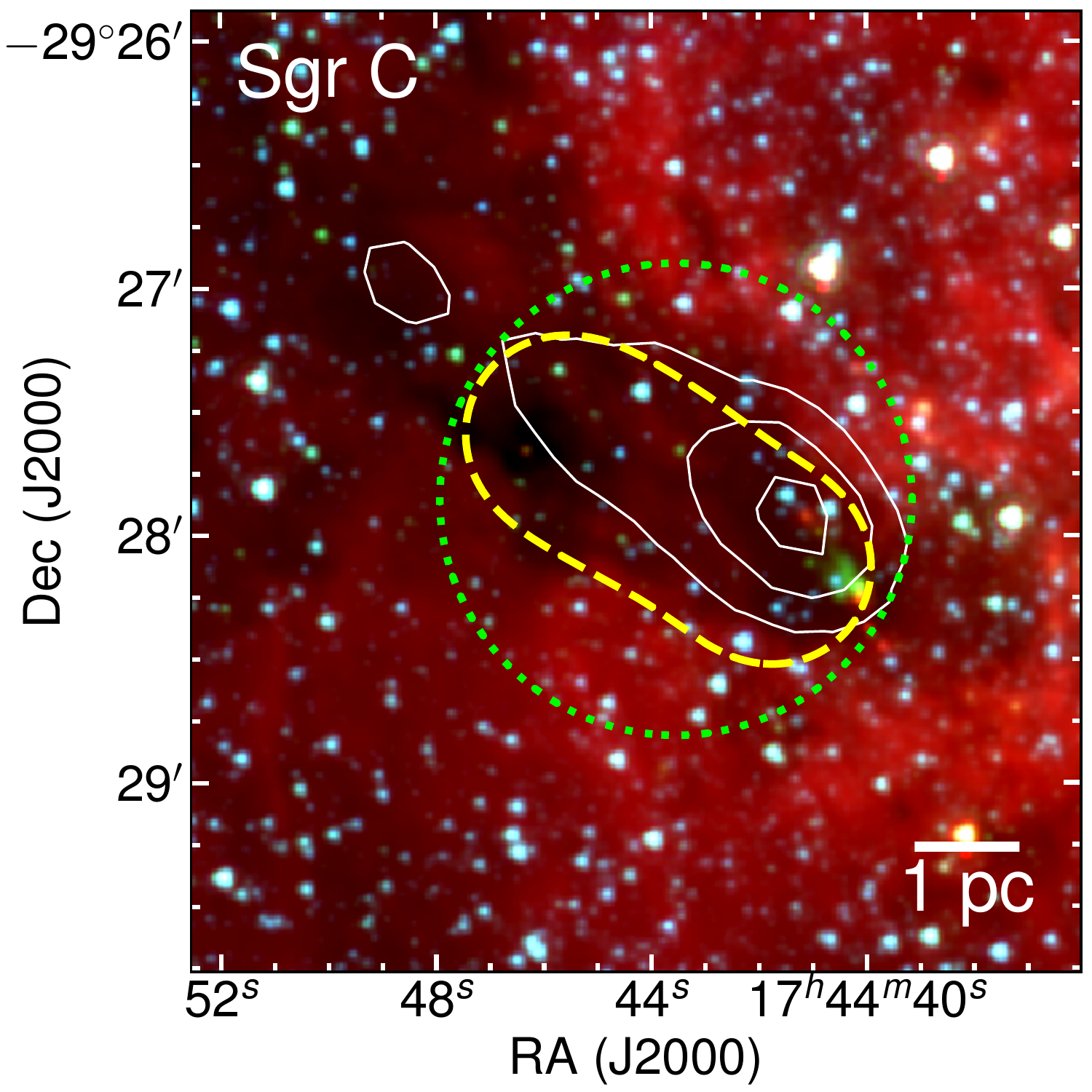} \\
\includegraphics[width=0.286\textwidth]{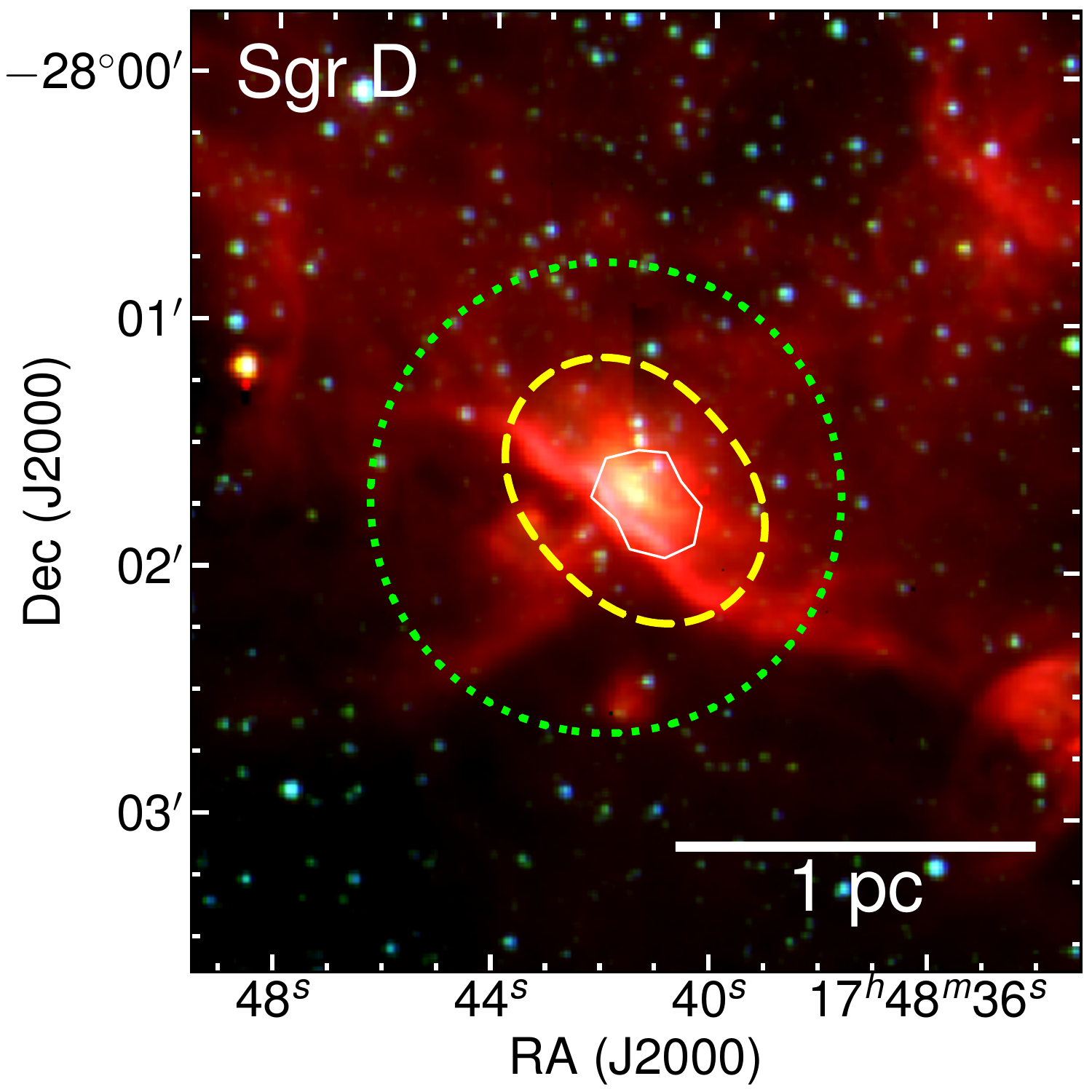}
\end{tabular}
\end{tabular}
\caption{The six clouds in the sample. Background three-color images show \textit{Spitzer} 3.6~\micron{} (blue), 4.5~\micron{} (green), and 8.0~\micron{} (red) emission. Contours show column densities derived from \textit{Herschel} data \citep{battersby2011}, starting from 10$^{23}$~\sqc{} in steps of 0.5$\times$10$^{23}$~\sqc{}. The yellow dashed and green dotted loops in each panel show the mosaic fields of the SMA and VLA, respectively.}
\label{fig:sample}
\end{figure*}

To investigate star formation at very early evolutionary phases in the CMZ clouds, we conducted observations using the Submillimeter Array (SMA) at 1.3~mm and the Karl J. Jansky Very Large Array (VLA) at the $K$-band toward a sample of six massive clouds (Figures~\ref{fig:overview} \& \ref{fig:sample}): \ctw{}, \cfi{}, \gzp{}, \sgb{} (also known as Dust Ridge clouds e/f), Sgr~C, and Sgr~D. This sample has been studied with the SMA at 280~GHz in \citet{kauffmann2017a,kauffmann2017b}. Five of them have high column densities ($>$10$^{23}$~\sqc{}; Figures~\ref{fig:overview} \& \ref{fig:sample}) and therefore are potential sites of star formation. One cloud, Sgr~D, which is associated with an \hii{} region in projection, has been suggested to reside outside of the CMZ \citep[e.g.,][]{sawada2009,sakai2017}. We include it here as a control object. In addition, Sgr~B2 is one of the most active star forming regions in the Galaxy and one that we cannot overlook in the CMZ, therefore we compile published data from the literature and include it in the discussion of star formation. Throughout this paper, we adopt a distance of 8.1~kpc \citep[the best-fit distance to Sgr~A* in][]{gravity2018}, except for Sgr~D, which we adopt 2.36~kpc \citep[the parallax distance from][]{sakai2017}.

In \autoref{sec:obs}, we introduce details of the SMA and VLA observations and data reduction. In \autoref{sec:results}, we present the SMA 1.3~mm continuum emission, based on which we identify cores at the 0.2~pc scale and estimate virial states of the cores. We also present VLA $K$-band radio continuum emission and \water{} masers. Then in \autoref{sec:disc}, we search for signatures of early phase star formation embedded in the cores using \water{} masers and UC~\hii{} regions. We then estimate SFRs of the clouds in \autoref{subsec:disc_sfr}, and compare with the dense gas star formation relation in \autoref{subsec:disc_sflaw}. The conclusions are in \autoref{sec:conclusions}. All the scripts used in the analyses in this paper are available at \url{https://github.com/xinglunju/CMZclouds}.

\begin{deluxetable*}{lcccccccc}
\tabletypesize{\scriptsize}
\tablecaption{Summary of the SMA and VLA observations.\label{tab:obs}}
\tablewidth{0pt}
\tablehead{
\colhead{\multirow{2}{*}{Project ID / PI}} & \multirow{2}{*}{Config.} & \# of unflagged  & \multirow{2}{*}{Date} & \multirow{2}{*}{Targets} & \# of & \multicolumn{3}{c}{Calibrators\tablenotemark{b}} \\
\cline{7-9} & & antennas & & & pointings\tablenotemark{a} & Bandpass & Flux & Gain
}
\startdata
\multicolumn{1}{l}{\underline{SMA 1.3~mm}} \\
2012B-S097 / Q.~Zhang & SUBCOM & 5 & 2013 May 21 & 20~\kms{}, Sgr~C & 8+3 & Q1 & Titan, Neptune & Q2, Q3 \\
                                        & SUBCOM & 5 & 2013 Aug 23 & 50~\kms{}, \sgb{} & 4+6 & Q1, Q4 & Neptune & Q2, Q3 \\
2013A-S049 / X.~Lu & COM   & 6 & 2013 Jul 24  & 20~\kms{}   & 8 & Q1, Q5 & Neptune & Q2, Q3 \\ 
                                 & COM   & 6 & 2013 Jul 25  & 50~\kms{}, Sgr~C & 4+3 & Q1 & Neptune & Q2, Q3 \\
                                 & COM   & 6 & 2013 Aug 01  & \sgb{}, Sgr~D & 6+2 & Q1 & Neptune & Q2, Q3 \\
                                 & COM   & 6 & 2013 Aug 02  & \sgb{}, Sgr~D & 6+2 & Q1 & MWC349A & Q2, Q3 \\
                                 & COM   & 5 & 2013 Aug 03  & 20~\kms{} & 8 & Q1, Q5 & Neptune & Q2, Q3 \\
                                 & COM   & 5 & 2013 Aug 09  & 20~\kms{} & 8 & Q1, Q6 & Neptune & Q2, Q3 \\
2013B-S083 / X.~Lu & SUBCOM & 7 & 2014 Mar 10 & \gzp{}, Sgr D & 6+2 & Q1 & Titan & Q2 \\
                                 & SUBCOM & 7 & 2014 Mar 21 & \gzp, Sgr D & 6+2 & Q1 & Titan & Q2 \\ \hline
\multicolumn{1}{l}{\underline{SMA 1.3~mm archival}} \\
2012A-S024 / K.~G.~Johnston & COM & 7 & 2012 Jun 09 & \gzp & 6 & Q1 & Titan & Q2 \\ \hline
\multicolumn{1}{l}{\underline{VLA K-band}}\\
AZ216 / Q.~Zhang   & DnC    & 23 & 2013 May 11  & \gzp{}, \sgb{} & 3+2 & Q1  & Q7  & Q3 \\
                                 & DnC    & 23 & 2013 May 12  & 20~\kms{}, 50~\kms{} & 3+1 & Q1  & Q7  & Q3 \\
                                 & DnC    & 22 & 2013 May 24  & Sgr~C, Sgr~D & 1+1 & Q1  & Q7  & Q3 \\
\enddata
\tablenotetext{a}{For shared-track observations, two numbers of pointings are shown for the two targets respectively.}
\tablenotetext{b}{Quasar calibrators: Q1: 3C279; Q2: 1733$-$130; Q3: 1744$-$312; Q4: 3C454.3; Q5: 3C84; Q6: 1924$-$292; Q7: 3C286.}
\end{deluxetable*}

\section{OBSERVATIONS AND DATA REDUCTION}\label{sec:obs}

\subsection{SMA Observations}\label{subsec:obs_sma}
The six clouds were observed with the SMA \citep{ho2004}\footnote{The SMA is a joint project between the Smithsonian Astrophysical Observatory and the Academia Sinica Institute of Astronomy and Astrophysics, and is funded by the Smithsonian Institution and the Academia Sinica.} in the compact and subcompact array configurations to obtain the 1.3~mm continuum and spectral lines (expect \gzp{} in the compact array configuration, for which we used the archival data). Each cloud was mosaiced with two to eight pointings to cover dense regions seen in the column density maps (see dashed loops in \autoref{fig:sample}). The ASIC correlator was configured to cover 217--221~GHz in the lower sideband and 229--233~GHz in the upper sideband, with a uniform channel width of 0.812~MHz (1.1~\kms{} at 1.3~mm). Part of the SMA observations toward the 20~\kms{} cloud has been published in \citet{lu2015b,lu2017}. Details of the observations are listed in \autoref{tab:obs}. 

In addition, we obtained the archival SMA 1.3~mm data in the compact array configuration toward \gzp{} (PI: K.~G.~Johnston), which have been published in \citet{johnston2014}.

The data from the two array configurations were calibrated using the IDL superset MIR\footnote{\url{https://www.cfa.harvard.edu/~cqi/mircook.html}}. Continuum visibility models were fit using line-free channels with MIRIAD \citep{sault1995}. Then the two datasets were combined and imaged to produce continuum maps with CASA~4.2.0 \citep{mcmullin2007}. Spectral lines were split from the continuum-subtracted visibility data and were imaged separately with a uniform channel width of 1.1~\kms{}. For all images, we used the Briggs weighting with a robustness of 0.5. We did not use multiscale CLEAN or combine with single-dish data, as in our previous work \citep{lu2017}, because in this paper we intended to study compact cores; therefore, we do not need information on extended structures.

The achieved rms and synthesized beam sizes are summarized in \autoref{tab:imaging}. The typical synthesized beam size (angular resolution) of continuum images is 5\arcsec$\times$3\arcsec{} (equivalent to 0.2~pc$\times$0.12~pc at the distance of 8.1~kpc), and the typical rms is 3~\mjypbm{}. The continuum images and selected spectral line images are publicly available at\dataset[https://doi.org/10.5281/zenodo.1436909]{https://doi.org/10.5281/zenodo.1436909}.

The images presented in figures throughout this paper are without primary beam corrections. These images have uniform rms levels across maps and are good for presentation, but the fluxes are attenuated toward the edge of the images. Therefore, when calculating densities and masses (e.g., in \autoref{subsec:results_cores}), we applied primary beam corrections to the images to have correct fluxes.

\begin{deluxetable*}{lccccccc}[!t]
\tabletypesize{\scriptsize}
\tablecaption{Properties of the SMA/VLA images. \label{tab:imaging}}
\tablewidth{0pt}
\tablehead{
\colhead{} & \multicolumn{3}{c}{Continuum} & & \multicolumn{3}{c}{Spectral lines}  \\
\cline{2-4} \cline{6-8} \colhead{Images} & Bandwidth & Beam size \& PA & RMS & & Channel width & Beam size \& PA & RMS \\
& (GHz) & (\arcsec$\times$\arcsec, \arcdeg) & (\mjypbm{}) & & (\kms{}) & (\arcsec$\times$\arcsec, \arcdeg) & (\mjypbm{})
}
\startdata
\multicolumn{1}{l}{\underline{SMA 1.3~mm}} \\
20~\kms{} & 8 & 4.9$\times$2.8, 5.2 & 3 & & 1.1 & 5.1$\times$2.8, 3.8 & 110 \\
50~\kms{} & 8 & 5.2$\times$2.9, 0.3 & 3 & & 1.1 & 5.5$\times$3.2, 1.5 & 110 \\
\gzp{}        & 8 & 4.8$\times$3.3, 10.9& 2 & & 1.1 & 5.6$\times$3.8, $-$13.6 & 60 \\
\sgb{}        & 8 & 5.2$\times$2.8, $-$9.6 & 3 & & 1.1 & 5.6$\times$2.9, $-$8.4 & 120 \\
Sgr~C       & 8 & 5.2$\times$2.9, 5.2 & 3 & & 1.1 & 5.4$\times$3.1, 5.0 & 120 \\
Sgr~D       & 8 & 6.8$\times$4.4, 26.5 & 4 & & 1.1 & 7.2$\times$4.6, 26.1 & 70 \\
\multicolumn{1}{l}{\underline{VLA K-band}}\\
20~\kms{} & 1 & 3.1$\times$2.1, 8.5 & 0.1 & & 0.2 & 3.5$\times$2.4, 5.5 & 5.5 \\
50~\kms{} & 1 & 2.8$\times$2.2, $-$0.3 & 0.07 & & 0.2 & 3.6$\times$2.4, $-$3.6 & 5.0 \\
\gzp{}        & 1 & 2.5$\times$2.0, 67.9 & 0.05 & & 0.2 & 2.8$\times$2.2, 67.5 & 4.8 \\
\sgb{}        & 1 & 2.4$\times$2.0, $-$74.9 & 0.035 & & 0.2 & 2.8$\times$2.2, $-$79.4 & 4.5 \\
Sgr~C       & 1 & 2.8$\times$2.1, $-$51.8 & 0.05 & & 0.2 & 3.1$\times$2.4, $-$54.8 & 4.5 \\
Sgr~D       & 1 & 2.5$\times$2.2, $-$61.2 & 0.2  & & 0.2 & 2.9$\times$2.3, $-$67.8 & 4.2
\enddata
\tablecomments{Beams and rms of the SMA spectral line images are measured for line-free channels of SiO 5--4 images not corrected for primary beam response, but they slightly vary between different lines. Beams and rms of the VLA spectral line images are measured for line-free channels of \water{} maser images. The rms of the SMA and VLA continuum images are measured in emission-free regions away from the emission peaks not corrected for primary beam response.}
\end{deluxetable*}

\subsection{VLA Observations}\label{subsec:obs_vla}
The sample was observed with the NRAO\footnote{The National Radio Astronomy Observatory is a facility of the National Science Foundation operated under cooperative agreement by Associated Universities, Inc.} Karl G. Jansky VLA in the DnC configuration, using a $K$-band setup that covers five metastable \amm{} lines from ($J$,~$K$)=(1,~1) to (5,~5), an \water{} maser line at 22.235~GHz, and 1~GHz wide continuum centered at $\sim$23~GHz. Part of the observations toward the 20~\kms{} cloud has been published in \citet{lu2015b,lu2017}, and details of the VLA observations can be found in \autoref{tab:obs}.

The data were calibrated using CASA~4.3.0. In \ctw{}, Sgr~C, and Sgr~D, bright ($>$1~Jy) \water{} masers are detected, so we performed self-calibration with the channel where the peak \water{} maser emission is found. We tried two or three rounds of phase-only self-calibration, until the image rms stopped to improve, and did a final round of phase and amplitude self-calibration. Then we applied the calibration tables to the data and produced images of the \water{} masers (see the next paragraph). We compared fluxes of the masers in the final image with those in the initial image to make sure the amplitude is consistent. The rms of channels with strong maser signals was significantly improved, and the achieved dynamic range is up to $\sim$3000. For Sgr~D where strong continuum emission is detected, we also applied the calibration tables from the self-calibration of masers to the continuum data to improve the dynamic range, and verified the amplitude consistency by comparing fluxes in images with and without applying the calibration tables.

The calibrated data were imaged using CASA~4.6.0. For the continuum, we used the multiscale CLEAN algorithm \citep{cornwell2008} to improve the imaging of spatially extended structures (e.g., filaments of $>$1~pc). The resulting continuum maps are still dynamic range limited, especially for that of Sgr~D, even after applying calibration tables from the self-calibration of \water{} masers. The typical achieved rms is 5~\mjypbm{} in 0.2~\kms{} for the \water{} maser, and 35--200~$\mu$Jy\,beam$^{-1}$ for the continuum depending on the target, with a beam size of 3\arcsec$\times$2\arcsec{}, as summarized in \autoref{tab:imaging}. The continuum and maser images are publicly available at\dataset[https://doi.org/10.5281/zenodo.1436909]{https://doi.org/10.5281/zenodo.1436909}.

\section{RESULTS}\label{sec:results}

\begin{figure*}[!t]
\begin{tabular}{@{}p{1.0\textwidth}@{}}
\centering
\includegraphics[width=0.9\textwidth]{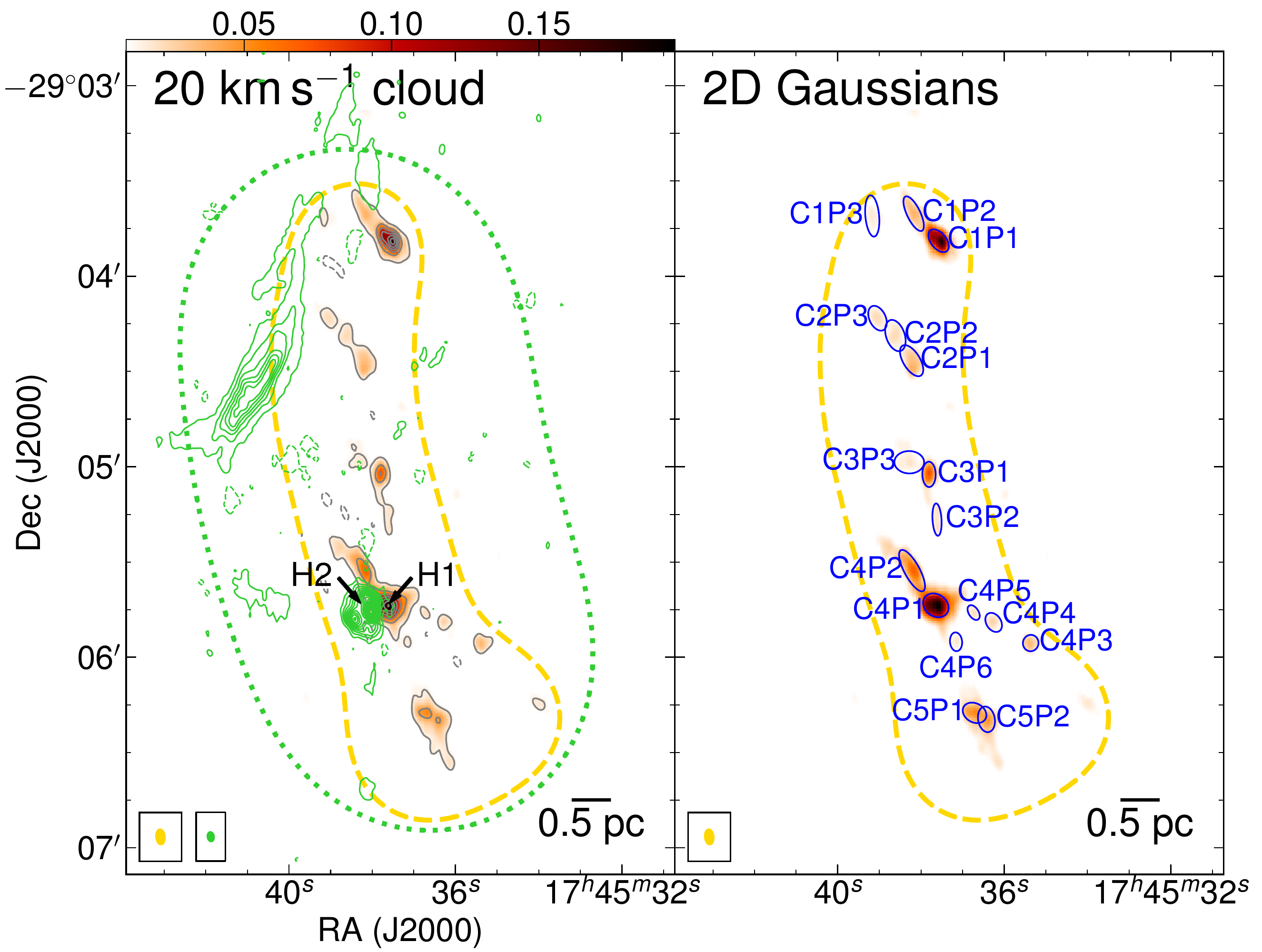} \\
\includegraphics[width=0.8\textwidth]{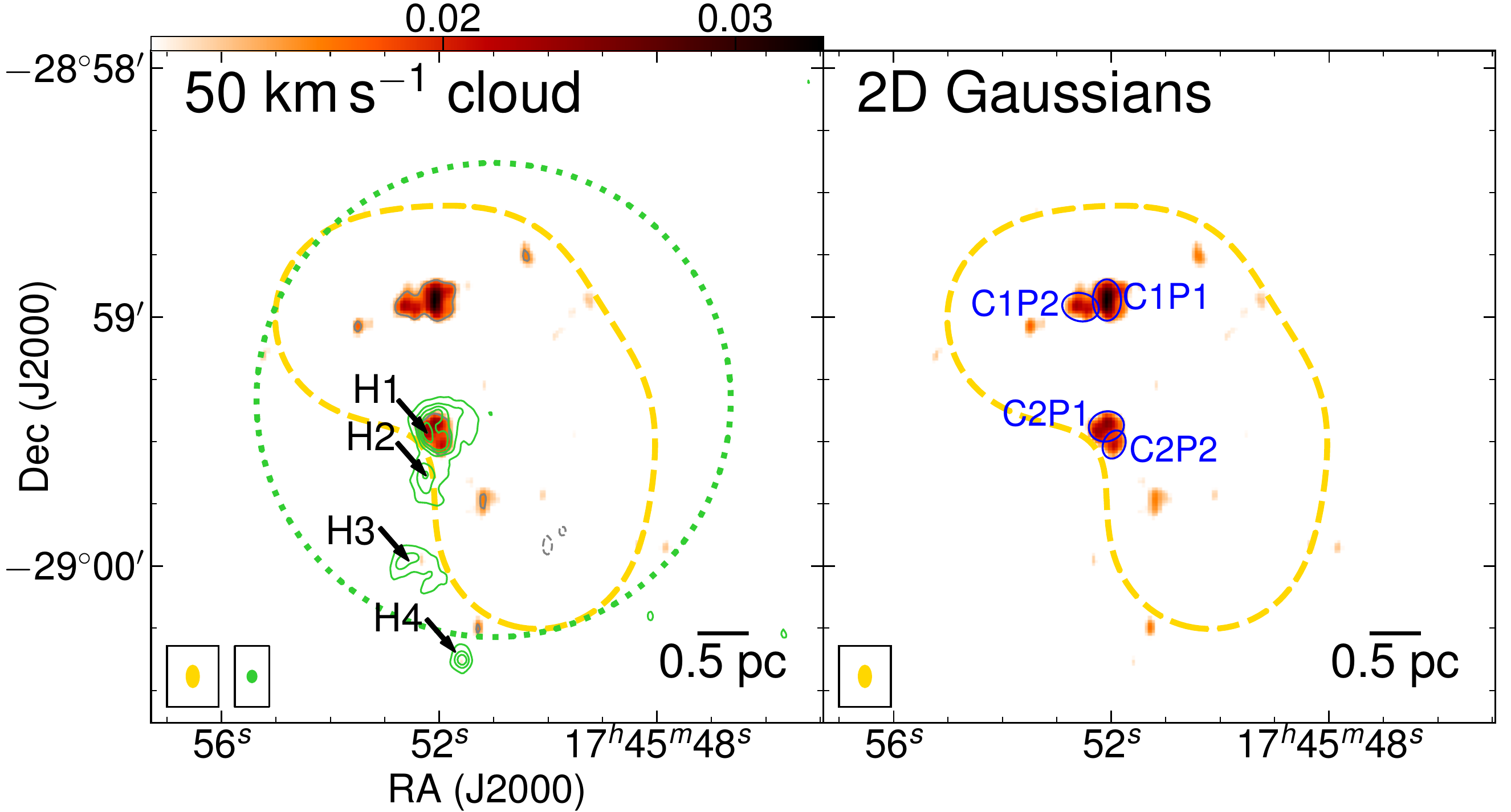}
\end{tabular}
\caption{The SMA 1.3~mm continuum and VLA radio continuum emission in the six clouds. On the left side of each panel, both the background image and the grey contours show the SMA 1.3~mm continuum emission, with the scale bar attached to the top in the unit of \jypbm{}. The contours start at the 5$\sigma$ level and increase in the step of 10$\sigma$, where 1$\sigma$ is 4~\mjypbm{} for Sgr~D, 2~\mjypbm{} for \gzp{}, and 3~\mjypbm{} for the other four clouds. The green contours show the VLA radio continuum emission, starting at the 5$\sigma$ level in the step of 10$\sigma$ (aside from Sgr~D where the step is 20$\sigma$), where 1$\sigma$ values for each map can be found in \autoref{tab:imaging}. For both the SMA and VLA continuum emissions, dashed contours at the $-$5$\sigma$ level are plotted to show the level of imaging artifacts. (UC) \hii{} regions identified in \autoref{subsec:results_vlacont} are marked by arrows and labeled. The dashed and dotted loops show the mosaic fields of the SMA and VLA, respectively. The synthesized beams of the SMA and VLA are shown in the bottom left corner. On the right side of each panel, the background image is identical with that on the left side, while the ellipses show the FWHM of 2D Gaussians fit to the cores.}
\label{fig:smacont}
\end{figure*}

\addtocounter{figure}{-1}
\begin{figure*}[!t]
\begin{tabular}{@{}p{1.0\textwidth}@{}}
\centering
\includegraphics[width=0.9\textwidth]{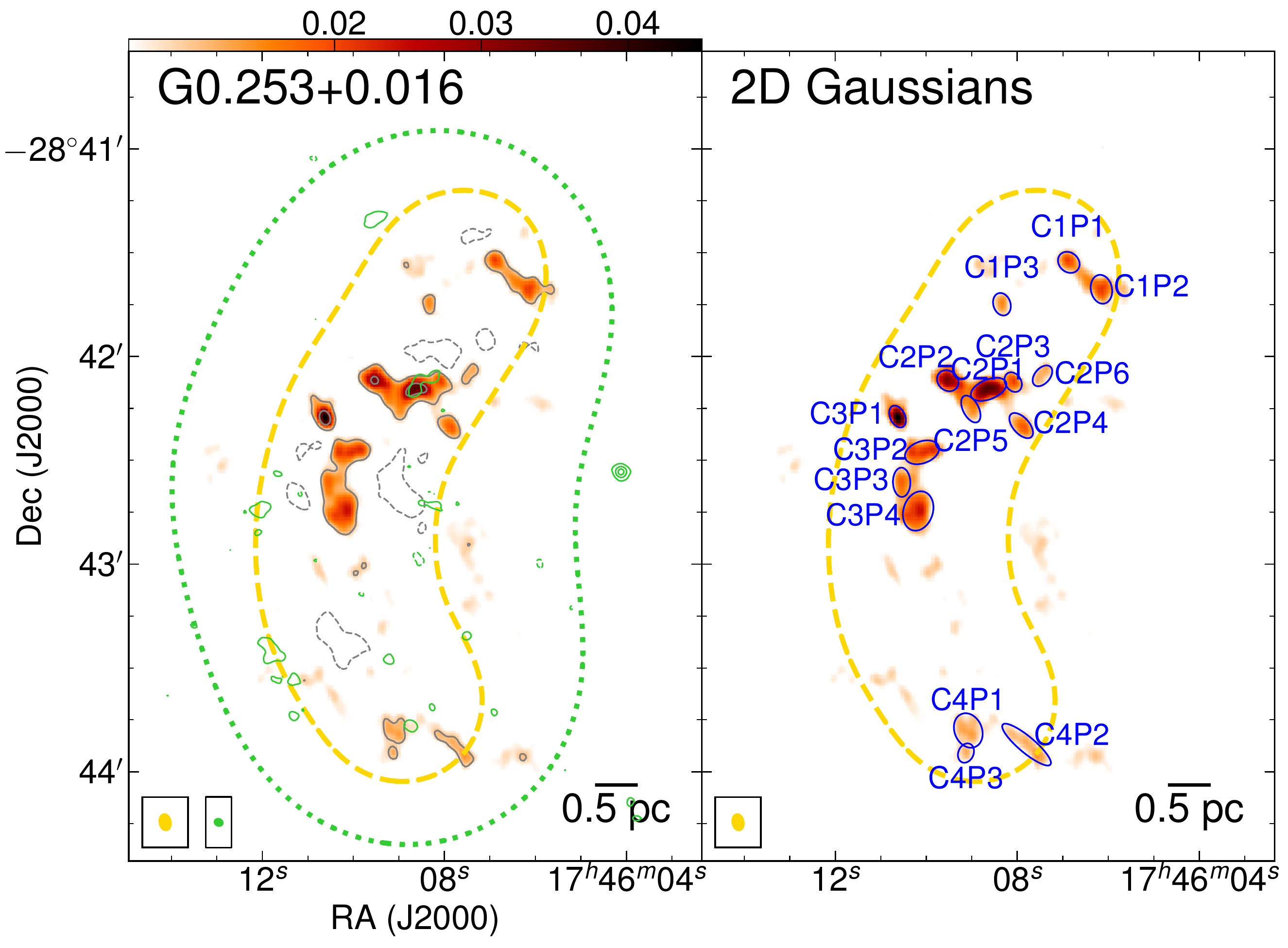} \\
\includegraphics[width=0.9\textwidth]{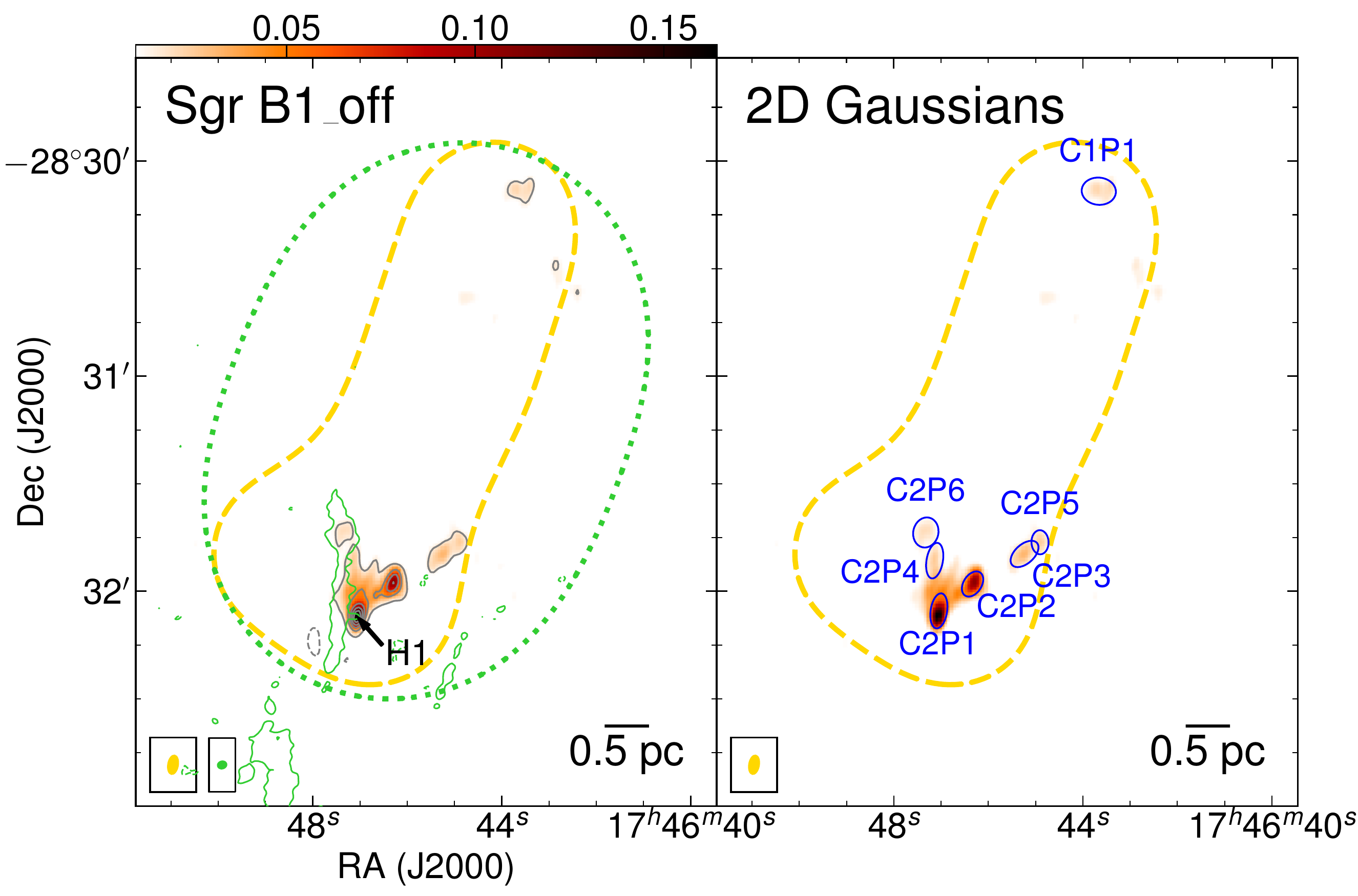}
\end{tabular}
\caption{(Continued)}
\label{fig:smacont}
\end{figure*}

\addtocounter{figure}{-1}
\begin{figure*}[!t]
\begin{tabular}{@{}p{1.0\textwidth}@{}}
\centering
\includegraphics[width=0.8\textwidth]{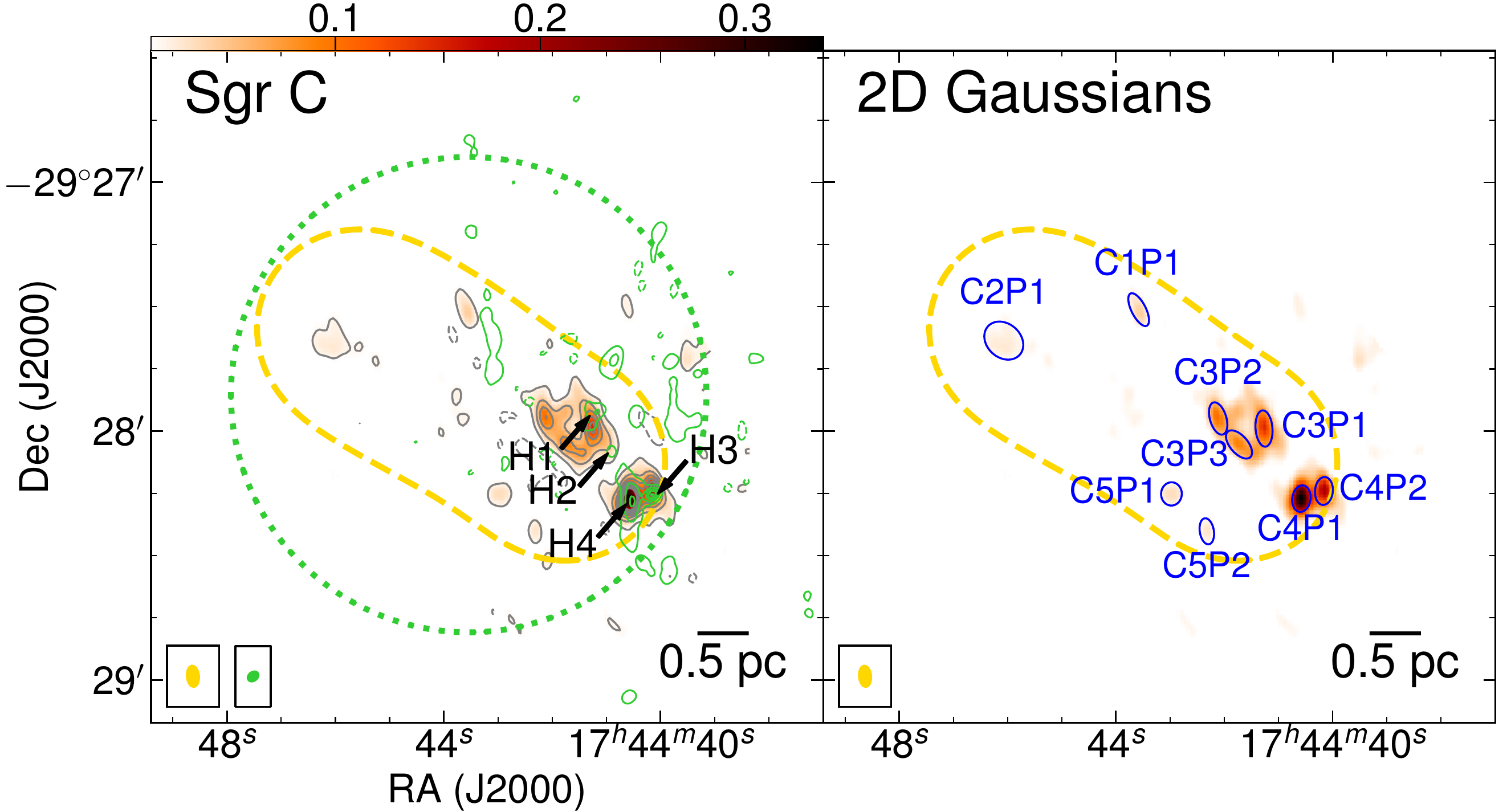} \\
\includegraphics[width=0.8\textwidth]{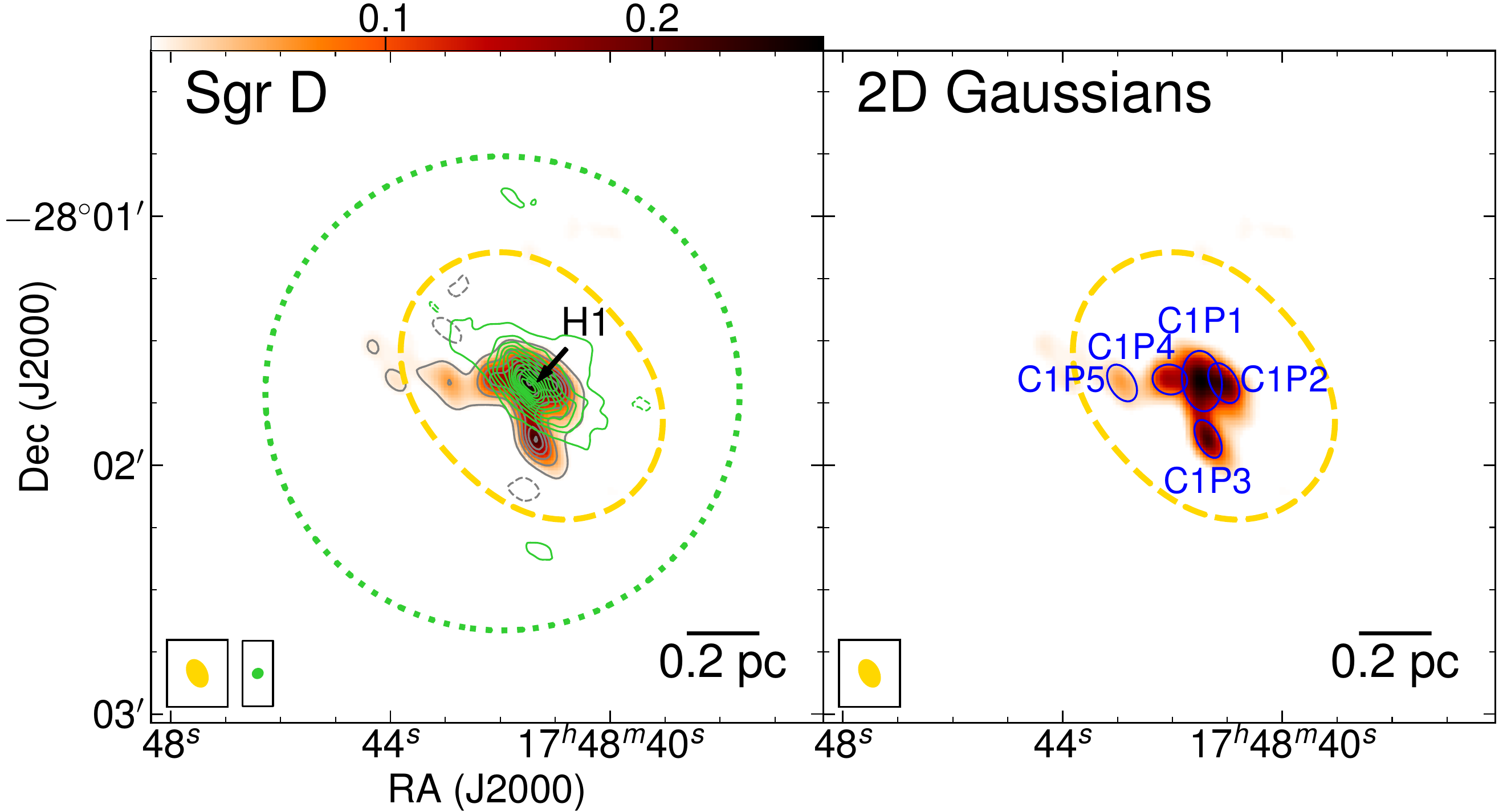}
\end{tabular}
\caption{(Continued)}
\label{fig:smacont}
\end{figure*}

\begin{deluxetable*}{rccccccccc}
\tabletypesize{\scriptsize}
\tablecaption{Properties of cores. \label{tab:cores}}
\tablewidth{0pt}
\tablehead{
\colhead{\multirow{2}{*}{Core ID}} & R.A.~\& Decl. & Deconvl.~size \& PA & $r_c$ & Flux\tablenotemark{a} & $M_\text{core}$ & $n$(H$_2$) & $\sigma_\text{tot}$\tablenotemark{b} & $\alpha_\text{vir}$ & SF Indicators\tablenotemark{c} \\
 & (J2000) & (\arcsec$\times$\arcsec, \arcdeg) & (pc) &  (mJy) & (\msol{}) & (10$^5$~\cc) & (\kms) & & 
}
\startdata
20~\kms{} C1P1 & 17:45:37.58, $-$29:03:48.83 & 7.2$\times$3.2, 48.8   & 0.09   & 324 & 483 & 20.2 & 1.27 & 0.37 & W2, W3 \\
C1P2 & 17:45:38.18, $-$29:03:40.31 & 11.1$\times$2.8, 28.4 & 0.11    &  96 & 143  & 3.8 & 1.29 & 1.49 & W1 \\
C1P3 & 17:45:39.17, $-$29:03:41.03 & 12.2$\times$3.3, 5.2   & 0.12   &   56 & 84   & 1.5 & 1.83 & 5.77 & \\
C2P1 & 17:45:38.23, $-$29:04:26.60 & 10.1$\times$4.1, 36.0 & 0.13   & 95   & 141  & 2.4 & 2.51 & 6.55 &  \\
C2P2 & 17:45:38.62, $-$29:04:18.69 & 9.1$\times$5.0, 22.7   & 0.13   & 60   & 89    & 1.3 & 1.44 & 3.59 &  \\
C2P3 & 17:45:39.04, $-$29:04:13.24 & 7.0$\times$4.0, 34.7   & 0.10   &  44  & 66    & 2.0 & 1.13 & 2.37 &  \\
C3P1 & 17:45:37.81, $-$29:05:02.41 & 6.4$\times$3.1, 178.0 & 0.09   & 104 & 155  & 8.0 & 1.93 & 2.43 & W5 \\
C3P2 & 17:45:37.62, $-$29:05:16.65 & 9.0$\times$0.5, 3.1     & 0.04   & 32   &   47  & 22.4 & 1.39* & 2.00 & W8 \\
C3P3 & 17:45:38.28, $-$29:04:58.59 & 9.0$\times$5.1, 90.8   & 0.13   & 56   &   84  & 1.2 & 1.47 & 3.99 &  \\
C4P1 & 17:45:37.64, $-$29:05:43.65 & 7.8$\times$5.4, 68.0   & 0.13   & 468(459) & 684 & 11.4 & 1.56 & 0.53 & W11--W13, H1  \\ 
C4P2 & 17:45:38.23, $-$29:05:32.72 & 14.0$\times$3.5, 30.5 & 0.14   & 196(195) & 291 & 3.9 & 1.43* & 1.12 &  \\
C4P3 & 17:45:35.36, $-$29:05:55.53 & 4.1$\times$1.7, 99.0   & 0.05   & 52   & 78  & 19.2 & 1.27 & 1.25 & W16 \\
C4P4 & 17:45:36.25, $-$29:05:49.03 & 5.0$\times$2.7, 56.0   & 0.07   & 38   & 57  & 5.2   & 2.15 & 6.83 & W15 \\
C4P5 & 17:45:36.74, $-$29:05:45.93 & $<$5.2$\times$3.1, 33.2 &$<$0.08& 18   & 26  & $>$1.8 & 1.08 & $<$4.12 & W14 \\
C4P6 & 17:45:37.16, $-$29:05:55.13 & 3.4$\times$2.6, 6.2     & 0.06   & 19   & 28  & 4.9 & 1.20 & 3.42 & W17 \\
C5P1 & 17:45:36.71, $-$29:06:17.50 & 7.1$\times$3.9, 73.4   & 0.10   & 95   & 142  & 4.4 & 1.21 & 1.25 &  \\
C5P2 & 17:45:36.43, $-$29:06:19.55 & 6.6$\times$4.4, 13.1   & 0.11   & 76   & 113  & 3.3 & 1.06 & 1.22 &  \\
50~\kms{} C1P1 & 17:45:52.08, $-$28:58:55.91 & 4.3$\times$3.1, 0.5 & 0.07  & 72 & 107  & 10.0 & 5.17 & 20.9 & W2 \\
C1P2 & 17:45:52.57, $-$28:58:57.61 & 4.2$\times$2.2, 81.8   & 0.06  & 46   & 69  & 11.2 & 5.21 & 27.2 &  \\
\gzp{} C1P1&17:46:06.87, $-$28:41:32.79&5.4$\times$3.7, 72.0 & 0.09 & 31  & 45    & 2.3 & 2.35 & 12.4 &  \\
C1P2 & 17:46:06.15, $-$28:41:40.57 & 7.0$\times$5.2, 12.1       & 0.12 & 57  & 85    & 1.8 & 2.35 & 8.95 &  \\
C1P3 & 17:46:08.34, $-$28:41:44.90 & 4.7$\times$3.7, 27.0       & 0.08 & 16  & 25    & 1.6 & 1.39 & 7.50 &  \\
C2P1 & 17:46:08.63, $-$28:42:09.50 & 10.4$\times$3.2, 111.1   & 0.11 &78(76)& 113  & 2.7 & 2.61 & 7.94 &  \\
C2P2 & 17:46:09.53, $-$28:42:07.00 & 5.3$\times$3.7, 74.2       & 0.09 & 42  & 62    & 3.3 & 1.24 & 2.51 &  \\
C2P3 & 17:46:08.09, $-$28:42:07.40 & 4.2$\times$2.6, 55.2       & 0.06 & 22  & 32    & 4.1 & 1.48 & 5.18 &  \\
C2P4 & 17:46:07.92, $-$28:42:19.96 & 7.7$\times$2.8, 48.0       & 0.09 & 38  & 57    & 2.6 & 2.62 & 12.9 &  \\
C2P5 & 17:46:09.02, $-$28:42:15.18 & 7.1$\times$2.9, 32.0       & 0.09 & 20  & 31    & 1.5 & 1.86 & 11.7 &  \\
C2P6 & 17:46:07.45, $-$28:42:05.59 & $<$7.0$\times$4.4, 139.0& $<$0.11& 20  & 29    & $>$0.8 & \nodata & \nodata &  \\
C3P1 & 17:46:10.63, $-$28:42:17.35 & 4.9$\times$2.5, 36.0       & 0.07 & 46   & 69  & 7.4 & 3.24* & 12.1 & W2 \\
C3P2 & 17:46:10.09, $-$28:42:27.69 & 9.7$\times$4.2, 108.5     & 0.13 & 51   & 76  & 1.3 & 2.21 & 9.32 &  \\
C3P3 & 17:46:10.54, $-$28:42:36.53 & 6.8$\times$2.7, 4.4         & 0.08 & 23   & 34   & 2.0 & 1.60 & 7.35 &  \\
C3P4 & 17:46:10.18, $-$28:42:44.65 & 10.8$\times$7.6, 157.6   & 0.18 & 79   & 118 & 0.7 & 3.84* & 25.9 &  \\
C4P1 & 17:46:09.08, $-$28:43:48.09 & 9.2$\times$7.1, 27.0       & 0.16 & 46   & 69   & 0.6 & 1.96 & 10.2 &  \\
C4P2 & 17:46:07.80, $-$28:43:52.17 & 17.5$\times$2.4, 51.1     & 0.13 & 64   & 95   & 1.6 & 1.98 & 6.15 &  \\
C4P3 & 17:46:09.13, $-$28:43:54.60 & 4.2$\times$1.5, 132.0     & 0.05 & 15   & 22   & 6.4 & 2.43 & 15.1 &  \\
\sgb{} C1P1& 17:46:43.66, $-$28:30:08.43&9.2$\times$5.5, 84.0& 0.14   & 71           & 106  & 1.3 & 3.19* & 15.6 &  \\
C2P1 & 17:46:47.05, $-$28:32:05.50 & 8.4$\times$3.7, 171.6     & 0.11   & 222(221) & 329  & 8.7 & 0.84 & 0.27 & W5, H1 \\
C2P2 & 17:46:46.33, $-$28:31:58.00 & 6.1$\times$3.6, 130.0     & 0.09   & 144         & 215  & 9.5 & 1.84 & 1.68 & W3 \\
C2P3 & 17:46:45.23, $-$28:31:49.62 & 8.3$\times$3.4, 121.8     & 0.10   & 82           & 123  & 3.7 & 1.80 & 3.20 &  \\
C2P4 & 17:46:47.14, $-$28:31:51.49 & 8.6$\times$3.6, 170.4     & 0.11   & 40(39)     & 58    & 1.5 & 2.47 & 13.3 &  \\
C2P5 & 17:46:44.91, $-$28:31:46.36 & 4.6$\times$3.5, 31.7       & 0.08   & 50           & 74    & 5.2 & \nodata & \nodata &  \\
C2P6 & 17:46:47.32, $-$28:31:43.63 & 6.7$\times$6.4, 151.0     & 0.13   & 44(41)     & 61   &  1.0 & 1.48 & 5.39 &  \\
Sgr~C C1P1 &17:44:43.58, $-$29:27:30.71 &7.3$\times$0.4, 34.0&0.03& 54 & 80  & 72.8 & 1.22* & 0.73 & W2  \\
C2P1 & 17:44:46.07, $-$29:27:38.18 & 9.5$\times$6.9, 61.0       & 0.16 & 80 & 119  & 1.0 & 0.97 & 1.46 & W3 \\
C3P1 & 17:44:41.27, $-$29:27:59.38 & 7.1$\times$2.8, 4.3         & 0.09 & 206(205) & 306 & 15.7 & 1.60 & 0.86 & W8, W9, H1, H2 \\
C3P2 & 17:44:42.11, $-$29:27:56.96 & 6.3$\times$2.5, 22.8       & 0.08 & 122 & 183  & 13.3 & 1.85* & 1.70 & W7, W10  \\
C3P3 & 17:44:41.73, $-$29:28:03.22 & 7.0$\times$1.4, 51.8       & 0.06 & 118 & 177  & 26.2 & 1.39 & 0.78 & W11 \\
C4P1 & 17:44:40.58, $-$29:28:16.28 & 4.2$\times$2.8, 144.0     & 0.07 & 462(457) & 681  & 77.0 & 1.80 & 0.37 & W13, H4 \\
C4P2 & 17:44:40.16, $-$29:28:14.43 & 4.5$\times$2.9, 160.0     & 0.07 & 335(330) & 492  & 47.6 & 1.70 & 0.48 & W12, H3 \\
C5P1 & 17:44:42.98, $-$29:28:15.14 & 4.1$\times$2.2, 99.0       & 0.06 & 42   & 63    & 10.5 & 2.19 & 5.23 & W14 \\
C5P2 & 17:44:42.32, $-$29:28:24.15 & 3.7$\times$1.7, 16.0       & 0.05 & 30   & 44    & 12.7 & 2.31 & 6.96 &  \\
Sgr~D C1P1 & 17:48:41.46, $-$28:01:39.74 & 13.1$\times$8.2, 3.3 & 0.06 & 616(152)  & 15 & 2.4  & 0.89 & 3.74 & W3, W4, H1 \\
C1P2 & 17:48:41.06, $-$28:01:40.22 & 7.9$\times$4.5, 30.7       & 0.03   & 276(144)     & 14 & 12.0  & 1.69 & 8.23 & W5, H1 \\
C1P3 & 17:48:41.35, $-$28:01:53.67 & 7.2$\times$2.8, 28.2       & 0.03   & 216              & 21 & 42.0  & 2.54 & 9.30 & W7 \\
C1P4 & 17:48:42.05, $-$28:01:39.36 & 6.8$\times$3.1, 100.6     & 0.03   & 168(95)       & 9  & 17.3  & 1.84 & 11.4 & W4, H1 \\
C1P5 & 17:48:42.92, $-$28:01:40.20 & 6.9$\times$4.0, 34.9       & 0.03   & 78               & 8 & 9.5  & 1.43* & 9.53 & W2
\enddata
\tablenotetext{a}{The 1.3~mm continuum fluxes have been corrected for primary-beam response. Note that we take the fluxes inside the FWHM of the 2D Gaussians, which are half of those from the full size of the Gaussian profiles. Fluxes in parentheses are free-free emission subtracted, based on which cores masses and gas densities are derived.}
\tablenotetext{b}{The total line widths marked with asterisks are derived from the SMA \methanol{} line. Otherwise they are derived from the SMA \nthp{} line \citep{kauffmann2017a}.}
\tablenotetext{c}{W and H refer to \water{} masers and \hii{} regions, respectively, with details in Sections~\ref{subsec:results_vlacont} \& \ref{subsec:results_masers}.}
\tablecomments{Uncertainties of the core properties are discussed in \autoref{subsec:results_errors}.}
\end{deluxetable*}

\subsection{SMA Dust Emission}\label{subsec:results_cores}

The SMA 1.3~mm continuum emission maps of the six clouds are shown in \autoref{fig:smacont}. We identified compact structures with peak values above the 5$\sigma$ level and areas larger than the synthesized beams, and within FWHM of the SMA primary beams. Then we fit 2D Gaussians using the interactive tool in CASAviewer to obtain their positions, deconvolved FWHM sizes, and fluxes. To have uniform noise levels so that we can apply the same fitting criteria across the maps, we performed the fit in the images without primary beam corrections. We took the deconvolved FWHM of the 2D Gaussians as the sizes of the compact structures. The fluxes inside the deconvolved FWHM of the 2D Gaussians are half of the measured fluxes of the whole Gaussian profiles, which we took as the fluxes of the compact structures after applying the primary beam correction. In \autoref{subsec:results_cores}, we derived the mean densities inside the deconvolved FWHM sizes using these fluxes.

The dendrogram algorithm is a widely used method for source identification in radio astronomy \citep{rosolowsky2008dendro}. We compared our result with the outcome of dendrogram, shown in \autoref{sec:appd_a}, and found that they are generally consistent. However, dendrogram is not able to separate closely packed structures (e.g., the two emission peaks in the southwestern end of Sgr~C). It also misses several compact structures that are slightly smaller than the synthesized beam size but are spatially coincident with \water{} masers and therefore are likely protostellar cores in nature (e.g., in the southern part of \ctw{}). In light of this, we chose to rely on manual identification and added these structures for consideration.

In the end, we identified 58 structures, marked by ellipses in \autoref{fig:smacont}. They are named by the indices of `clumps' they belong to, plus the indices of peaks inside the clumps in decreasing order of peak intensities. Here the clumps do not have physical meanings but are for name tagging.

We stress that the identification of compact structures is unlikely to be complete. Some features, especially those in crowded environments (e.g., the C4 clump in \ctw{}, the C2 clump in Sgr~C), may have been missed. Nevertheless, we intended to study characteristic physical properties of dense gas in these clouds, and structures identified using this approach make up a good sample for our purpose.

At the wavelength of 1.3~mm, the continuum emission in molecular clouds is often attributed to thermal dust emission associated with dense gas \citep[e.g.,][]{beuther2002mambo}, but can also be free-free emission from embedded \hii{} regions \citep[e.g.,][]{motte2003}. To examine potential contribution from free-free emission, we compared the 1.3~mm continuum with the radio continuum data in \autoref{subsec:results_vlacont}. Two compact structures, C2P1 and C2P2 in \cfi{}, are associated with radio continuum emission of similar or even higher fluxes than the 1.3~mm continuum emission. As discussed in \autoref{subsec:results_vlacont}, the radio continuum in \cfi{} arises from several known \hii{} regions. The 1.3~mm continuum emission of the two compact structures therefore is likely dominated by free-free emission from the \hii{} regions. These two structures are excluded from \autoref{tab:cores}. A few compact structures in \ctw{}, \sgb{}, Sgr~C, and Sgr~D are also found to be associated with compact radio continuum emission, which is much weaker than the 1.3~mm continuum emission.

We obtained dust emission fluxes of the compact structures after excluding the contribution from free-free emission in the 1.3~mm continuum emission. We used a flat spectral index from centimeter to 1.3~mm, assuming slightly optically thick thermal free-free emission. If there is any optically thick free-free emission from hyper-compact \hii{} regions, the spectral index between the frequencies of the VLA and SMA observations may be positive (rising), and the free-free contribution in the 1.3~mm continuum emission will be greater. However, hyper-compact \hii{} regions are rare \citep[the only known cases in the CMZ are six hyper-compact \hii{} regions in Sgr B2;][]{depree2015}, so we did not consider optically thick free-free emission in our assumption. Then we subtracted the radio continuum fluxes inside the FWHM of the compact structures from the corresponding 1.3~mm continuum fluxes and obtained the dust emission fluxes, which are listed in parentheses in \autoref{tab:cores}. Following the nomenclature of \citet{zhang2009}, these structures with typical radii of 0.1~pc are defined as cores. Excluding the two compact structures in \cfi{} that are dominated by free-free emission, we identified 56 cores in the six clouds, as listed in \autoref{tab:cores}.

We derived core masses following
\begin{equation}\label{equ:coremass}
M_\text{core}=R\frac{S_\nu d^2}{B_\nu(T_\text{dust})\kappa_\nu},
\end{equation}
where $R$ is the gas-to-dust mass ratio, $S_\nu$ is the dust emission flux, $d$ is the distance, $B_\nu(T_\text{dust})$ is the Planck function at the dust temperature $T_\text{dust}$, and $\kappa_\nu$ is the dust opacity. We assumed $R$=100, and $\kappa_\nu$=0.899~cm$^2$\,g$^{-1}$ \citep[MRN model with thin ice mantles, after 10$^5$ years of coagulation at 10$^6$~\cc{};][]{ossenkopf1994}. We assumed $T_\text{dust}=20$~K for the cores, except for those in Sgr~D where $T_\text{dust}$ is taken to be 25~K, which are estimated from multi-band SED fitting of \textit{Herschel} data \citep{kauffmann2017a}. The masses of the cores are listed in \autoref{tab:cores}. With a dust emission rms of 3~\mjypbm{}, the 5$\sigma$ mass sensitivity is 22~\msol{} per beam for the CMZ clouds.

Assuming a spherical geometry with a radius $r_a$ that is equivalent to half of the geometric mean of the deconvolved angular sizes of the cores, densities of molecular gas in the cores are derived with
\begin{equation}\label{equ:density}
\begin{split}
n(\text{H}_2)&=\frac{3M_\text{core}}{4\pi r_a^3d^3}\frac{1}{2.8m_\text{H}} \\
&=R\frac{S_\nu}{B_\nu(T_\text{dust})\kappa_\nu}\frac{3}{4\pi r_a^3d}\frac{1}{2.8m_\text{H}},
\end{split}
\end{equation}
where 2.8 is the molecular weight per H$_2$ molecule \citep{kauffmann2008} and $m_\text{H}$ is the mass of a hydrogen atom.

We caution that these cores are defined in terms of their spatial scales, but they are more massive than dense cores in nearby molecular clouds at the same spatial scale \citep[e.g.,][]{alves2007}, and their densities are an order of magnitude higher ($\gtrsim$10$^5$~\cc{} vs.\ $\sim$10$^4$~\cc{}). They may each form a cluster of stars instead of a single star or a multiple star system as assumed for those dense cores in nearby clouds. With higher angular resolutions, they may be further resolved into objects that map to individual protostars \cite[e.g.,][]{ginsburg2018}.

\sgb{} is included in the SMA sample of \citet{walker2018} with a similar observation setup. They identified two cores e1 and e2, corresponding to C2P1 and C2P2 in this cloud in \autoref{fig:smacont} and \autoref{tab:cores}. The masses we derived are 40\% to 50\% smaller than their results, because we only considered fluxes inside the deconvolved FWHM sizes therefore the measured fluxes are smaller. They also estimated densities of these cores to be 10$^5$--10$^6$~\cc{}, much higher than those of dense cores in nearby clouds.

\subsection{Virial States of the Cores}\label{subsec:results_virial}

We studied virial states of the cores in these clouds. The virial parameter is defined as \citep{bertoldi1992}
\begin{equation}\label{equ:virial}
\alpha_\text{vir}=\frac{5\sigma_\text{tot}^2r_ad}{\text{G}M_\text{core}},
\end{equation}
in which $r_a$ is the angular radius of the core as defined above, and $\sigma_\text{tot}$ is the total one-dimensional line width including both thermal and non-thermal components. The properties can be found in \autoref{tab:cores}.

The total line width $\sigma_\text{tot}$ was measured with the \nthp{} 3--2 line \citep{kauffmann2017a}, which has a critical density of $\gtrsim$10$^6$~\cc{} at a temperature of $\gtrsim$50~K \citep{shirley2015}. In the starless core candidates where gas temperatures are low (see \autoref{subsubsec:disc_sf_criteria}), \nthp{} may be chemically biased toward denser regions where CO is frozen out onto dust grains, therefore it may preferentially trace smaller line widths from smaller spatial scales \citep{caselli2002}. When \nthp{} is not detected, the \methanol{} line in our SMA 1.3~mm line data (not combined with single-dish data) is used, which has been shown to best spatially correlated with the dust emission among the 1.3~mm molecular lines \citep{lu2017}. However, \methanol{} is likely influenced by shocks, so we only used it as a second choice. Two cores, C2P6 in \gzp{} and C2P5 in \sgb{}, are not detected in \nthp{} or \methanol{} lines, therefore their virial parameters cannot be determined. We fit a single Gaussian to the mean spectrum of each core, shown in \autoref{sec:appd_c}, to obtain the intrinsic line width $\sigma_v$ that is deconvolved from the channel width. When both the \nthp{} and \methanol{} lines are detected, the line widths measured from them are usually consistent within a factor of 1.5. 

Our VLA observations cover five VLA \amm{} lines, but we did not use them to measure line widths of the cores for three reasons. First, for the lower \amm{} transitions, ($J$, $K$)=(1,~1), (2,~2), and (3,~3), the hyperfine components tend to be blended together and strong absorption features are frequently seen toward the cores. Second, for the higher \amm{} transitions, ($J$, $K$)=(4,~4) and (5,~5), the signal-to-noise ratio is much lower \citep[e.g., Figure~15 of][]{lu2017}. Third, the critical densities of the \amm{} lines are of the order 10$^3$~\cc{} at a temperature of $\gtrsim$50~K \citep{shirley2015}, therefore may not be as good as \nthp{} for tracing of gas in the cores. Nonetheless, we attempted to fit the \amm{} spectra when they are optically thin and absorption is not significant (e.g., toward C1P1 in \ctw{}), and the measured line widths agree with those based on \nthp{} within a factor of 1.5.

We assume a gas temperature $T_\text{gas}$ of 100~K. This is the typical temperature for cores at 0.2~pc scales based on non-LTE modeling of \amm{} (2,~2) and (4,~4) lines \citep{lu2017}, and it generally agrees with those in previous observations \citep{ao2013,ginsburg2016,krieger2017}. Then we derive the line width $\sigma_\text{tot}$:
\begin{equation}\label{equ:sigmav}
\sigma_\text{tot}=\sqrt{\sigma_v^2-\frac{k_BT_\text{gas}}{\mu_mm_p}+\frac{k_BT_\text{gas}}{\mu_p m_p}},
\end{equation}
in which the mean molecule weight $\mu_p$ is 2.33, assuming 90\% H and 10\% He, and $\mu_m$ is 29 or 32 (i.e., the molecule weight of \nthp{} or \methanol{}, depending on which line is used to measure the linewidth).

The derived virial parameters $\alpha_\text{vir}$ are listed in \autoref{tab:cores}. Out of the 54 cores whose virial parameters can be determined, 17 have $\alpha_\text{vir} \le 2$. These cores are likely gravitationally bound and unstable to collapse.

\begin{deluxetable*}{cccc}[!t]
\tabletypesize{\scriptsize}
\tablecaption{Summary of uncertainties in derived core properties. \label{tab:uncertainty}}
\tablewidth{0pt}
\tablehead{
\colhead{Core properties} & Related equations & Considered quantities and random errors\tablenotemark{a} & Uncertainties\tablenotemark{b}
}
\startdata
Mass ($M_\text{core}$)                    & \autoref{equ:coremass} & $\kappa_\nu$ (28\%), $S_\nu$ (15\%), $d$ (1.2\%), $R$ (50\%)  & 59\% \\
Density ($n$(\text{H}$_2$))              & \autoref{equ:density}     & $\kappa_\nu$ (28\%), $S_\nu$ (15\%), $d$ (1.2\%), $R$ (50\%), $r_a$ (10\%) & 66\% \\
Virial parameter ($\alpha_\text{vir}$)& \autoref{equ:virial}         & $\kappa_\nu$ (28\%), $S_\nu$ (15\%), $d$ (1.2\%), $R$ (50\%), $r_a$ (10\%), $\sigma_\text{tot}$ (50\%)& 120\%
\enddata
\tablenotetext{a}{Details of the quantities can be found in Sections~\ref{subsec:results_cores} \& \ref{subsec:results_virial}. $\kappa_\nu$ -- dust opaticy. $S_\nu$ -- dust emission flux. $d$ -- distance. $R$ -- gas-to-dust ratio. $r_a$ -- angular size. $\sigma_\text{tot}$ -- line width.}
\tablenotetext{b}{The dust temperature $T_\text{dust}$ has a large systematic error for the protostellar core candidates, and its effect on uncertainties of the derived properties is not considered here. We consider the effect of $T_\text{dust}$ separately in \autoref{subsec:results_errors}.}
\end{deluxetable*}

\subsection{Uncertainties of Core Properties}\label{subsec:results_errors}

We reported uncertainties in the derived masses, densities, line widths, and virial parameters of the cores. The uncertainties are summarized in \autoref{tab:uncertainty}.

The derived core masses depend on the dust opacity, the gas-to-dust mass ratio, dust temperatures, dust emission fluxes, and distances.  We followed \citet{sanhueza2017} to adopt uncertainties of 28\% and 15\% for the dust opacity and measured dust emission fluxes at 1.3~mm. The uncertainty in the distance to Sgr~A* is small \citep[0.4\%;][]{gravity2018}. However, given that the clouds may be on an orbit of radius$\sim$100~pc around Sgr~A* \citep{molinari2011,kruijssen2015}, we adopt an uncertainty of $\pm$100~pc (1.2\%) for the distance.

The gas-to-dust mass ratio has a large uncertainty. The value of 100 adopted for \autoref{equ:coremass} is characteristic for nearby clouds, although values up to 150 have been suggested \citep{draine2011}. On the other hand, the value for Galactic Center regions may be as low as $\sim$50 \citep{giannetti2017}. Therefore, the uncertainty in the gas-to-dust ratio is adopted to be 50\%.

The dust temperature may have a large systematic error for the cores that are internally heated by protostars. It could reach 50~K around hot molecular cores at the radius of 0.1~pc \citep{longmore2011}, in which case the derived core masses using \autoref{equ:coremass} would decrease by a factor of 3. This only affects cores with significant internal heating (potentially those with star formation indicators in \autoref{tab:cores}), and may not be an issue for cores without signatures of star formation. Further discussion about the impact of the dust temperature is in \autoref{subsubsec:disc_sf_criteria}.

We propagated uncertainties (random errors) in the dust opacity, the gas-to-dust ratio, dust emission fluxes, and the distance, but excluded the systematic error in the dust temperature, and obtained an uncertainty of 59\% for the masses. For cores with significant internal heating therefore potentially higher dust temperatures (e.g., assuming $T_\text{dust}=50$~K), the derived masses could systematically decrease by a factor of 3.

The derived densities of the cores depend on the angular sizes and all the quantities that determine the masses. The measured angular sizes usually have uncertainties of 10\%. We propagated these random errors but excluded the systematic error in the dust temperature, and obtained an uncertainty of 66\% for the densities. Similar to the masses, for cores with significant internal heating with an assumed $T_\text{dust}=50$~K, the densities could systematically decrease by a factor of 3.

The fitting errors of the line widths, as shown in \autoref{sec:appd_c}, are usually $\sim$4\%--50\% depending on the signal-to-noise ratios. However, there are several other uncertainties in the line widths. First, the line widths measured using \nthp{} may be overestimated, when the lines are optically thick and the hyperfine structure of \nthp{} is considered \citep{caselli2002}. Second, absorption features are seen in several \nthp{} spectra, probably due to missing flux of interferometers (see \autoref{sec:appd_c}), which may lead to underestimated line widths. A third issue is the choice of the component to be fit when there are multiple velocity components, especially in the case of \gzp{} (\autoref{app_fig:core_spec_3}) where several components of similar brightnesses are seen in the cores, making it ambiguous which component should be considered. In general, we adopted an uncertainty of 50\% for all the line widths.

Finally, the random errors of the masses, the line widths, and the angular sizes all propagate into that of the derived virial parameters. We estimated a large uncertainty of 120\% (or a factor of 2.2) for the virial parameters without considering the systematic error in the dust temperature, and an even larger uncertainty (a factor of $>$4) for cores with significant internal heating whose masses may be systematically overestimated by a factor of 3. In addition to the uncertainties in the derived virial parameters, there are several factors that may affect the critical virial parameter. First, the magnetic field at 1 pc scales in \gzp{} is suggested to be $\sim$5~mG with large uncertainties \citep{pillai2015}, and it is unclear whether at 0.1 pc the magnetic field is similar. If so, the support against gravitational collapse from the magnetic field would be significant---for example, assuming $B=5$~mG, the critical virial parameter would be as low as $<$1, and most of the cores would be gravitationally unbound. Second, we have ignored rotation of the cores in the plane of the sky, which may be able to support them against collapse and make the critical virial parameter smaller.

\begin{figure*}[!t]
\begin{tabular}{@{}p{0.5\textwidth}@{}p{0.5\textwidth}@{}}
\centering
\includegraphics[width=0.5\textwidth]{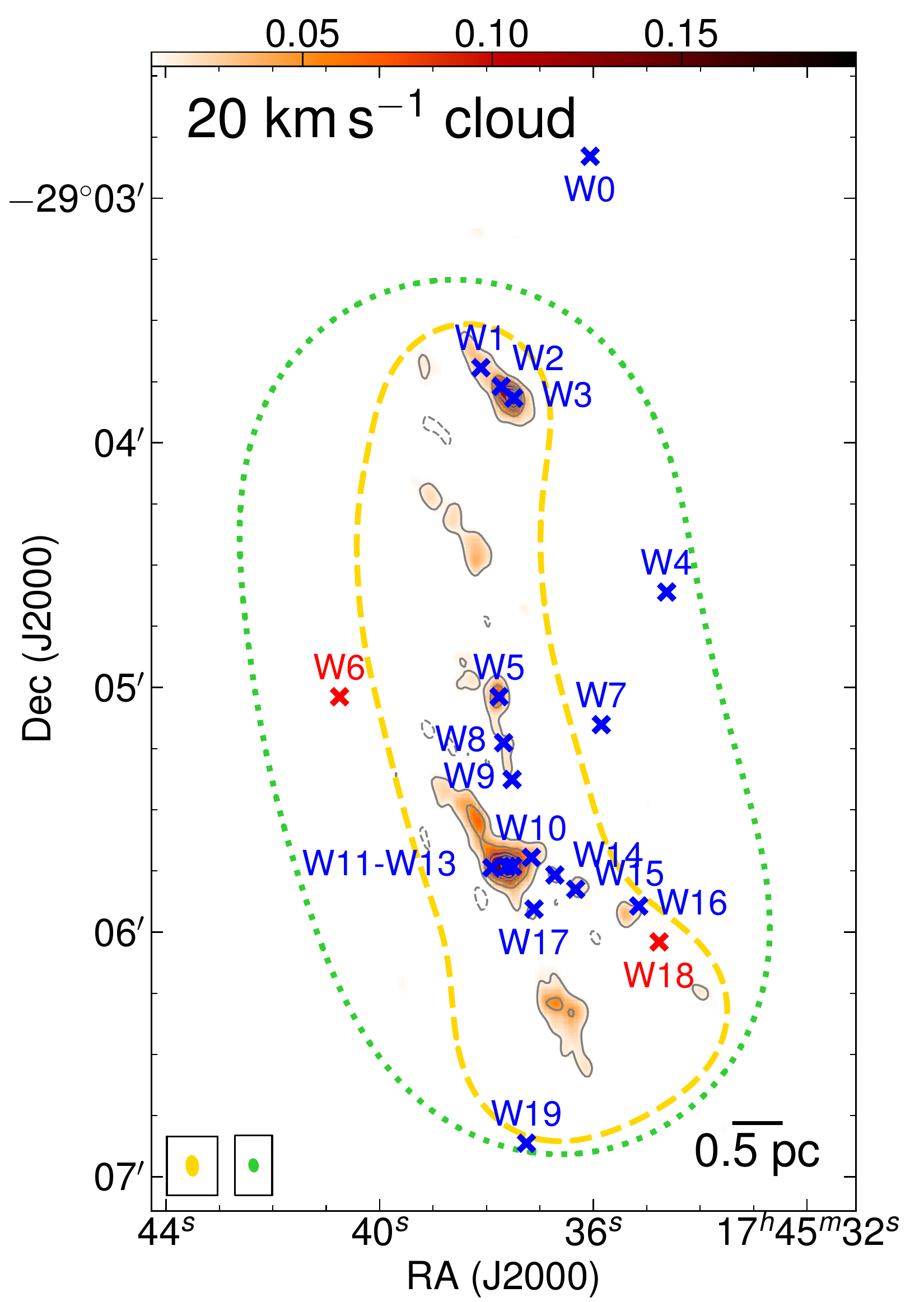} & \includegraphics[width=0.5\textwidth]{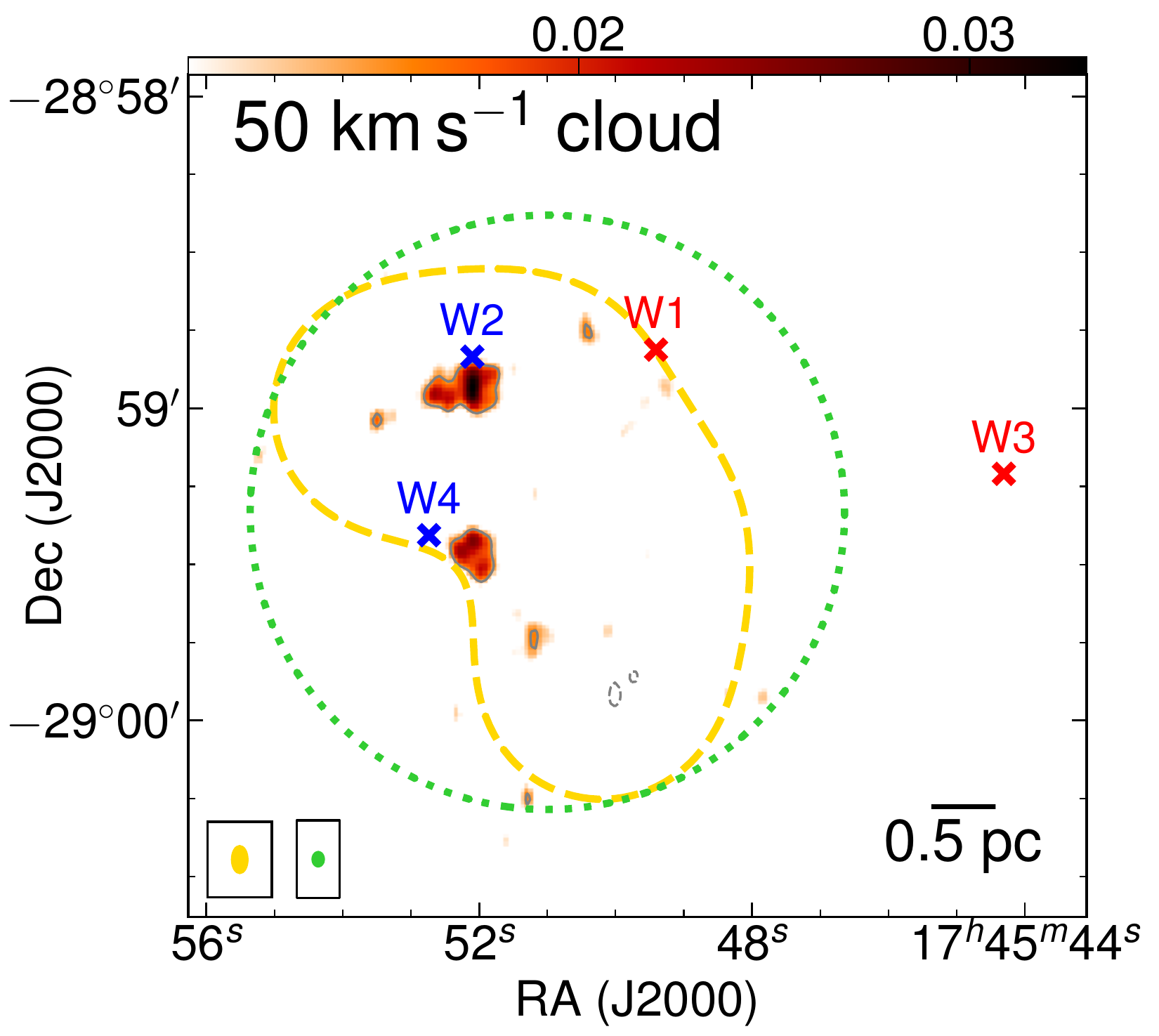}
\end{tabular}
\caption{VLA \water{} masers in the six clouds. The background images and contours show the SMA 1.3~mm continuum emission, and the dashed and dotted loops show the mosaic field of the SMA and VLA, which are identical to those in \autoref{fig:smacont}. \water{} masers are marked by crosses, among which red ones are those with AGB star counterparts.}
\label{fig:masers}
\end{figure*}

\addtocounter{figure}{-1}
\begin{figure*}[!t]
\begin{tabular}{@{}p{0.5\textwidth}@{}p{0.5\textwidth}@{}}
\centering
\includegraphics[width=0.5\textwidth]{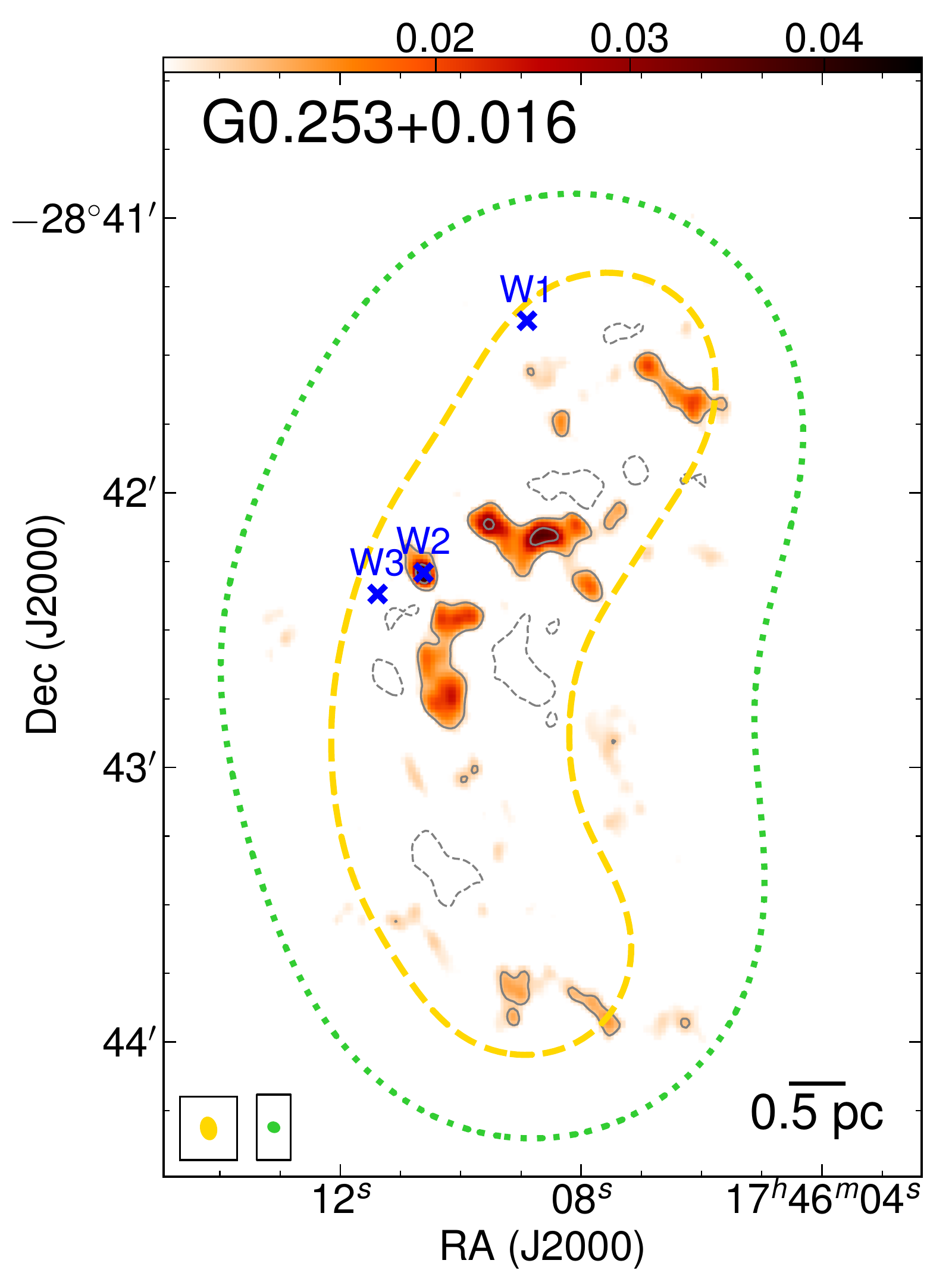} & \includegraphics[width=0.5\textwidth]{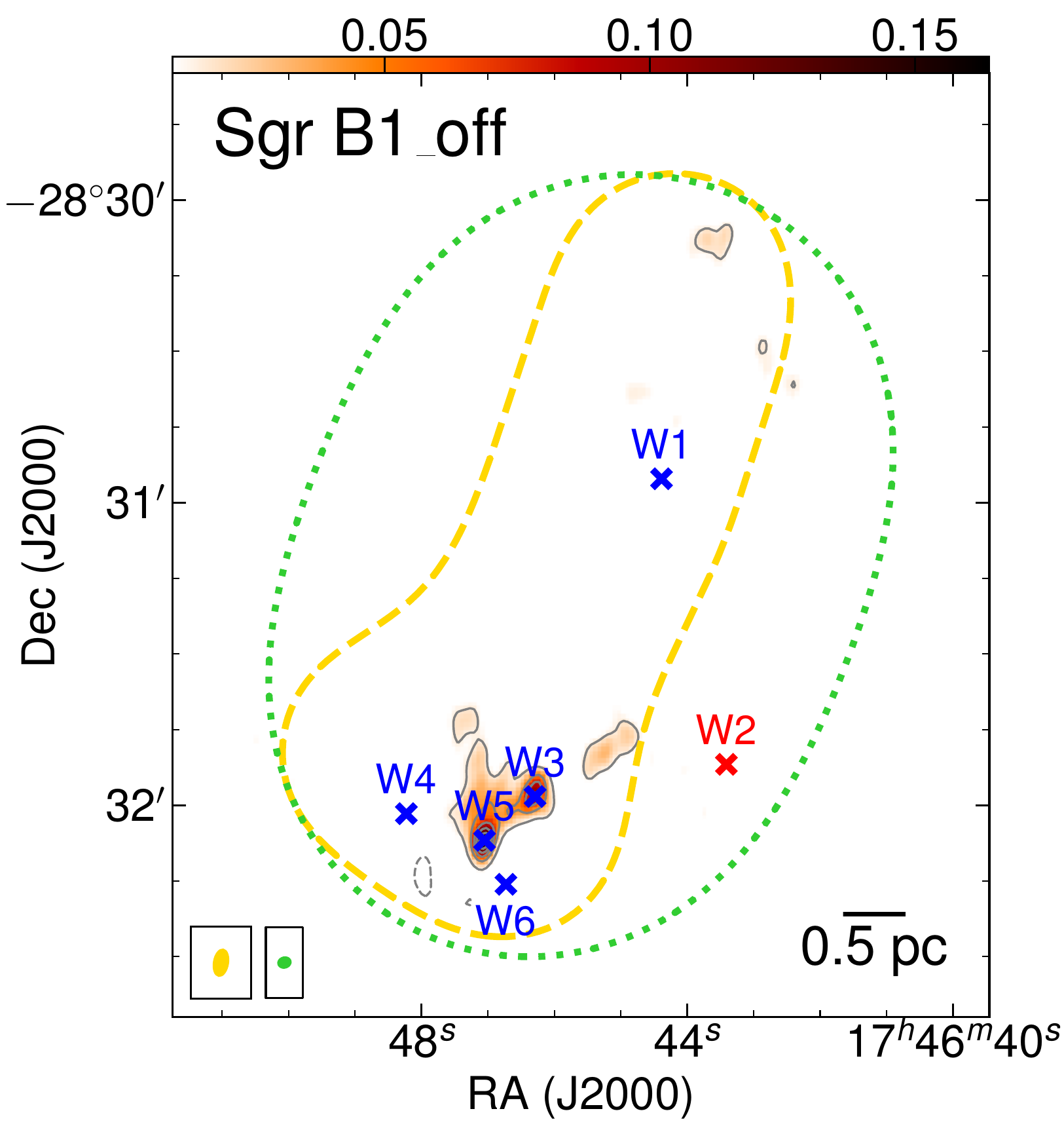}
\end{tabular}
\caption{(Continued)}
\label{fig:masers}
\end{figure*}

\addtocounter{figure}{-1}
\begin{figure*}[!t]
\begin{tabular}{@{}p{0.5\textwidth}@{}p{0.5\textwidth}@{}}
\centering
\includegraphics[width=0.5\textwidth]{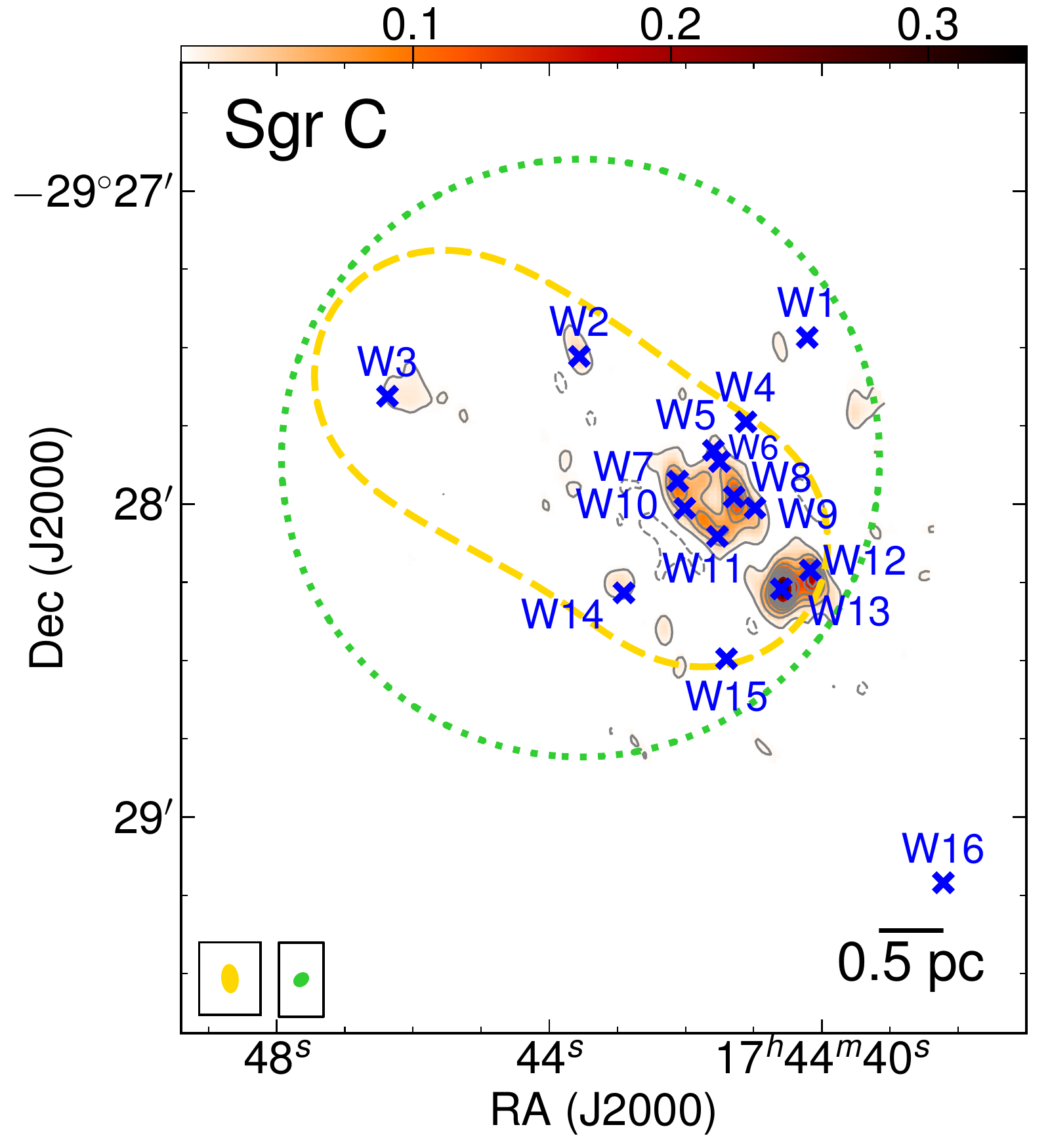} & \includegraphics[width=0.5\textwidth]{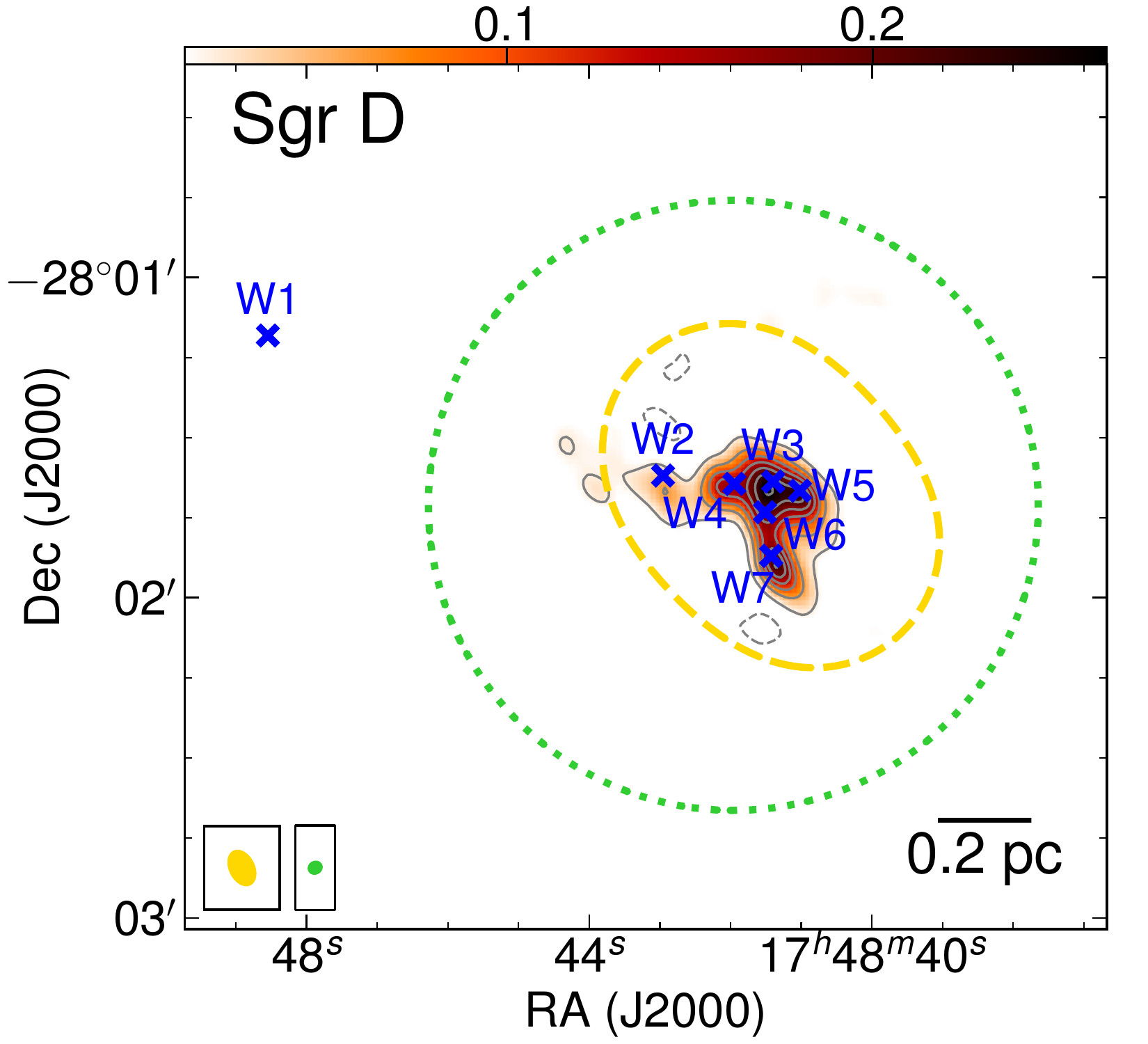}
\end{tabular}
\caption{(Continued)}
\label{fig:masers}
\end{figure*}

\begin{deluxetable*}{rcccccc}
\tabletypesize{\scriptsize}
\tablecaption{Properties of \hii{} regions. \label{tab:hii}}
\tablewidth{0pt}
\tablehead{
\colhead{\multirow{2}{*}{\hii{} region ID}} & R.A. \& Decl.\tablenotemark{a} & $F_\nu$\tablenotemark{b} & $\log_\text{10}N_\text{c}$ & Spectral type\tablenotemark{c} & ZAMS mass\tablenotemark{c} & References \& Alternative identifiers \\
 & (J2000) & (mJy) & (s$^{-1}$) & &  (\msol{}) & 
 }
\startdata
20~\kms{} H1 & 17:45:37.59, $-$29:05:43.60 & 1.5     & 46.00 & B0.5 & 12 & \citealt{lu2017} \\
                 H2 & 17:45:38.00, $-$29:05:45.37 & 148.3 & 48.00 & O9   & 20 & \citealt{ho1985}, Sgr~A-G \\
50~\kms{} H1 & 17:45:52.23, $-$28:59:27.78 & 481.9 & 48.51 & O7.5   & 26 & \citealt{mills2011}, Sgr~A-A \\
                 H2 & 17:45:52.23, $-$28:59:37.97 & 129.1 & 47.94 & O9   & 20 & \citealt{mills2011}, Sgr~A-B \\
                 H3 & 17:45:52.65, $-$28:59:59.00 & 165.7 & 48.05 & O8.5& 21 & \citealt{mills2011}, Sgr~A-C \\
                 H4 & 17:45:51.59, $-$29:00:22.40 & 111.0 & 47.88 & O9    & 19  & \citealt{mills2011}, Sgr~A-D \\
\sgb{} H1 & 17:46:47.10, $-$28:32:07.09 & 0.5   & 45.53 & B1-B0.5 & 11 & \nodata \\
Sgr C H1 & 17:44:41.20, $-$29:27:55.39 & 2.3   & 46.19 & B0.5 & 13 & \nodata \\
          H2 & 17:44:40.92, $-$29:28:05.00 & 0.7   & 45.68 & B1-B0.5 & 11 & \nodata \\
          H3 & 17:44:40.19, $-$29:28:15.20 & 5.0   & 46.53 & B0.5 & 14 & \citealt{forster2000}, 359.44$-$0.10 A\\
          H4 & 17:44:40.51, $-$29:28:16.99 & 14.9 & 47.00 & B0    & 15 & \citealt{forster2000} \\
Sgr D H1 & 17:48:41.46, $-$28:01:40.90 & 1025.8   & 47.77 & O9 & 19 & \citealt{liszt1992}
\enddata
\tablenotetext{a}{Listed coordinates are those of the emission peaks in each \hii{} region.}
\tablenotetext{b}{Fluxes have been corrected for primary-beam response.}
\tablenotetext{c}{Spectral types and ZAMS masses of the (UC) \hii{} regions are estimated following \citet{panagia1973} and \citet{davies2011}.}
\end{deluxetable*}

\subsection{VLA Radio Continuum Emission}\label{subsec:results_vlacont}
Radio continuum emission at 23~GHz obtained by the VLA is displayed as green contours in \autoref{fig:smacont}. There are several known \hii{} regions: one in \ctw{} \citep{ho1985}, four in \cfi{} \citep{goss1985,mills2011}, and one in Sgr~D \citep{liszt1992}. Our observations confirmed radio continuum emission from them. We also detected several fainter compact sources that are associated with dust emission, which may be embedded UC~\hii{} regions. Among them, the nature of the radio continuum in \gzp{} has been discussed in \citet{rodriguez2013} and \citet{mills2015}; one UC \hii{} region in Sgr~C has been studied in \citet{forster2000} and \citet{kendrew2013}; the UC \hii{} region in \ctw{} has been reported in \citet{lu2017}. In addition, we found one compact radio continuum source in \sgb{} that is associated with a core, and several in Sgr~C that have likely dust emission counterparts. Their nature will be discussed in \autoref{subsubsec:disc_sf_hii}. The detections are named with the letter `H' plus a number by decreasing declinations in each cloud, and are marked in \autoref{fig:smacont}. In general, their morphologies are not ellipse-like, so we did not fit 2D Gaussians to them, but measured their fluxes above the 3$\sigma$ level contour and listed the results in \autoref{tab:hii}.

Filamentary radio continuum emission of $>$1~pc is seen in \ctw{} and \sgb{}. Such structure in the CMZ has been suggested to have non-thermal origins \citep{ho1985,lu2003,zhao2016}. As these are not related to recent star formation, we do not consider them further in this paper. Several compact radio continuum emission peaks without dust emission counterparts are also found (e.g., in Sgr~C). They may be associated with more evolved \hii{} regions at late evolutionary phases, therefore are not considered either.

\subsection{VLA \water{} Masers}\label{subsec:results_masers}
In our previous work \citep{lu2015b}, we reported the detection of 18 \water{} masers in \ctw{}. Here we searched for \water{} masers in all the six clouds in our sample. All point sources with peak intensities above the 8$\sigma$ level were identified, where 1$\sigma$ is $\sim$5~\mjypbm{} in 0.2~\kms{} velocity bin (see \autoref{tab:imaging}), but we excluded those found in dynamic-range limited channels, where the rms is significantly higher than the theoretical sensitivity, as these are likely sidelobes of strong sources\footnote{This could leave out faint masers in such channels, but we note that dynamic-range limited channels are only found within $\pm$2 channels (i.e., within a range of 1~\kms{}) of the brightest channels after self-calibration. This is a small fraction of the velocity ranges of the clouds, which are usually $>$10~\kms{}, although it may become a problem if star formation is highly clustered around the brightest masers in space and velocity.}. Multiple velocity components along the same line of sight were counted as a single maser. In addition, several strong masers were found outside the FWHM of the VLA primary beams. For example, in \ctw{}, we found two masers close to or outside of the VLA primary beam boundaries in addition to the 18 masers reported in \citet{lu2015b}. Such masers are also seen in \cfi{}, Sgr~C, and Sgr~D. We took them into account if their peak intensities are above the 10$\sigma$ level. A total of 56 \water{} masers were identified in the six clouds.

We fit 2D Gaussians to the integrated intensity maps of the masers to determine their positions and integrated fluxes. The results are listed in \autoref{tab:masers}, while the positions are marked in \autoref{fig:masers} and the complete spectra shown in \autoref{sec:appd_b}. The masers are named by decreasing declinations in each cloud. Note that in \ctw{}, we started with `W0' that is outside of the VLA primary beams, in order to be consistent with the catalog in \citet{lu2015b}, while in the other clouds we started with `W1'.

The coordinates and fluxes of the masers in \ctw{} are slightly different from those reported in \citet{lu2015b}, but are still within the pointing or flux calibration uncertainties (position differences $<$0.3\arcsec{}, flux differences $<$10\%). This is likely because we applied the self-calibration solutions which slightly changed the phase and amplitude of the data.

While a good spatial correlation between \water{} masers and dust emission is generally found in \autoref{fig:masers}, it is interesting to note that several \water{} masers in \gzp{} (e.g., W1 and W3) and \sgb{} (e.g., W1) do not seem to be associated with any dust emission. They may be unrelated to star formation and will be further discussed in \autoref{subsubsec:disc_sf_h2omasers}.

\section{Discussion}\label{sec:disc}

\subsection{Signatures of Embedded Star Formation}\label{subsec:disc_sf}

We discuss signatures of star formation associated with the cores, and compare densities and virial states of the protostellar and starless core candidates.

\begin{deluxetable*}{rcrrrrcc}
\tabletypesize{\scriptsize}
\tablecaption{Properties of H$_2$O masers. \label{tab:masers}}
\tablewidth{0pt}
\tablehead{
\colhead{\multirow{2}{*}{Maser ID}} & R.A. \& Decl. & $v_\text{peak}$\tablenotemark{a} & $F_\text{peak}$\tablenotemark{a,b} & $F_\text{integrated}$\tablenotemark{b} & $L_\text{H$_2$O}$ & \multirow{2}{*}{Cores/clumps} & \multirow{2}{*}{Other masers} \\
 & (J2000) & (\kms{}) & (mJy per channel) & (mJy$\cdot$\kms{}) & (10$^{-7}$~\lsol{}) & & 
 }
\startdata
20~\kms{} W0   & 17:45:36.06, $-$29:02:49.73 & $-$29.7 & 493   & 1090 &16.6 & \nodata \\
W1   & 17:45:38.10, $-$29:03:41.58 & 14.8      & 713   & 491   & 7.5     & C1P2 \\
W2   & 17:45:37.73, $-$29:03:46.18 & 27.8      & 240   & 946   & 14.4   & C1P1 \\
W3   & 17:45:37.49, $-$29:03:49.02 & 28.0      & 1568 & 4320 & 65.7   & C1P1 \\
W4   & 17:45:34.63, $-$29:04:36.62 & 2.4        & 686   & 852   & 12.9   & \nodata \\
W5   & 17:45:37.76, $-$29:05:02.28 & 18.6    & 15377 & 7153 & 108.7 & C3P1 \\
W6   & 17:45:40.75, $-$29:05:02.29 & 12.7      & 317   & 418   & 6.4    & AGB star & \\
W7   & 17:45:35.85, $-$29:05:09.13 & 51.0      & 274   & 242   & 3.7    & \nodata \\
W8   & 17:45:37.68, $-$29:05:13.68 & 46.2      & 38     & 49     & 0.7    & C3P2 \\
W9   & 17:45:37.52, $-$29:05:22.75 & $-$40.0 & 50     & 100   & 1.5    & C3P2 \\
W10 & 17:45:37.16, $-$29:05:42.07 & 10.8      & 124   & 337   & 5.1    & C4 \\
W11 & 17:45:37.62, $-$29:05:44.24 & 4.4        & 919   & 3738 & 56.8  & C4P1 \\
W12 & 17:45:37.53, $-$29:05:44.11 & 9.3        & 192   & 411   & 6.2     & C4P1 \\
W13 & 17:45:37.92, $-$29:05:45.05 & 26.4      & 49     & 153   & 2.3    & C4P1 \\
W14 & 17:45:36.72, $-$29:05:46.23 & $-$24.6 & 204   & 327   & 5.0    & C4P5 \\
W15 & 17:45:36.33, $-$29:05:49.82 & 13.1      & 1454 & 2762 & 42.0  & C4P4 \\
W16 & 17:45:35.15, $-$29:05:53.92 & $-$4.4   & 64     & 94     & 1.4    & C4P3 \\
W17 & 17:45:37.10, $-$29:05:54.75 & $-$3.8   & 222   & 360   & 5.5    & C4P6 \\
W18 & 17:45:34.78, $-$29:06:02.90 & 20.7      & 56     & 95     & 1.4    & AGB star \\
W19 & 17:45:37.25, $-$29:06:52.03 & $-$46.3 & 186   & 349   & 5.3    & \nodata \\
50~\kms{} W1 & 17:45:49.41, $-$28:58:48.72 & $-$3.9 & 844 & 1432 & 21.8 & AGB star & \\
                 W2 & 17:45:52.10, $-$28:58:50.06 & 37.6    & 94   & 263   & 4.0   & C1P1 \\
                 W3 & 17:45:44.31, $-$28:59:12.59 & 77.6    & 589 & 2790 & 42.4 & AGB star & \\
                 W4 & 17:45:52.73, $-$28:59:24.40 & 156.0  & 64   & 17     & 0.3   & \nodata \\
\gzp{} W1 & 17:46:08.90, $-$28:41:22.44 & 70.8 & 43   & 36   & 0.5  & \nodata \\
          W2 & 17:46:10.62, $-$28:42:17.44 & 39.0 & 262 & 541 & 8.2   & C3P1 \\
          W3 & 17:46:11.38, $-$28:42:22.13 & 28.4 & 267 & 372 & 5.6   & \nodata \\
\sgb{} W1 & 17:46:44.39, $-$28:30:55.28 & 111.4 & 50   & 73   & 1.1    & \nodata \\
          W2 & 17:46:43.41, $-$28:31:51.90 & 59.8   & 70   & 188 & 2.8   & AGB star \\
          W3 & 17:46:46.29, $-$28:31:58.28 & 27.9   & 57   & 21   & 0.3   & C2P2 \\
          W4 & 17:46:48.23, $-$28:32:01.68 & 31.7   & 869 & 945 & 14.4 & \nodata \\
          W5 & 17:46:47.05, $-$28:32:06.97 & 30.5   & 360 & 364 & 5.5   & C2P1 & class~\textsc{ii} \methanol{} \\
          W6 & 17:46:46.73, $-$28:32:15.69 & $-$42.2 & 59 & 24  & 0.4   & \nodata \\
Sgr C W1 & 17:44:40.21, $-$29:27:28.09 & $-$52.8 & 302 & 469   & 7.1 & \nodata \\
        W2   & 17:44:43.56, $-$29:27:31.71 & $-$47.7 & 551 & 587   & 8.9 & C1P1 \\
        W3   & 17:44:46.37, $-$29:27:39.35 & 2.2        & 31   & 40     & 0.6 & C2P1 \\
        W4   & 17:44:41.11, $-$29:27:44.26 & 3.9        & 48   & 301   & 4.6 & \nodata \\
        W5   & 17:44:41.59, $-$29:27:49.73 & $-$53.8 & 44   & 43     & 0.7 & C3 \\
        W6   & 17:44:41.50, $-$29:27:51.88 & $-$55.3 & 94   & 96     & 1.5 & C3 \\
        W7   & 17:44:42.11, $-$29:27:55.62 & $-$53.0 &5341& 9037 & 137.4 & C3P2 \\
        W8   & 17:44:41.29, $-$29:27:58.65 & $-$56.0 & 798 & 2495 & 37.9 & C3P1 \\
        W9   & 17:44:40.98, $-$29:28:00.78 & $-$50.9 & 88   & 105   & 1.6  & C3P1 \\
        W10 & 17:44:42.01, $-$29:28:00.84 & $-$53.0 & 180 & 373   & 5.7  & C3P2 \\
        W11 & 17:44:41.53, $-$29:28:06.22 & $-$51.3 &3764& 2760 & 42.0 & C3P3 \\
        W12 & 17:44:40.17, $-$29:28:12.68 & $-$58.7 &24754&42370& 644.0 & C4P2 & class~\textsc{ii} \methanol{} \\
        W13 & 17:44:40.60, $-$29:28:16.28 & $-$57.8 &7263& 12700 & 193.0 & C4P1 & OH (1665~MHz), class~\textsc{ii} \methanol{} \\
        W14 & 17:44:42.90, $-$29:28:17.04 & $-$67.1 & 376 & 596   & 9.0 & C5P1 \\
        W15 & 17:44:41.40, $-$29:28:29.67 & $-$61.6 &8179& 8152 & 123.9 & \nodata \\
        W16 & 17:44:38.22, $-$29:29:12.61 & $-$0.5   &2075& 8360 & 127.1 & \nodata \\
Sgr D W1 & 17:48:48.55, $-$28.01.10.88 & $-$13.3 & 7191 & 11450 & 14.8 & \nodata & class~\textsc{ii} \methanol{} \\
          W2 & 17:48:42.96, $-$28.01.37.12 & $-$22.8 & 90     & 110     & 0.1   & C1P5 \\
          W3 & 17:48:41.39, $-$28.01.38.25 & $-$9.3   & 1885 & 5144   & 6.6 & C1P1 \\
          W4 & 17:48:41.95, $-$28.01.38.69 & $-$21.3 & 27     & 40       & 0.05   & C1P4 \\
          W5 & 17:48:41.02, $-$28.01.39.97 & $-$24.1 & 30     & 86       & 0.1   & C1P2 \\
          W6 & 17:48:41.51, $-$28.01.44.00 & 3.1        & 5550 & 7302   & 9.4 & C1P1 \\
          W7 & 17:48:41.42, $-$28.01.52.17 & $-$24.9 & 301   & 576     & 0.7   & C1P3
\enddata
\tablenotetext{a}{For masers with multiple velocity components along the line of sight, \vlsr{} and flux of the strongest peak is listed, while the complete spectra can be found in \autoref{sec:appd_b}.}
\tablenotetext{b}{Peak fluxes and integrated fluxes have been corrected for primary-beam response.}
\end{deluxetable*}

\subsubsection{\water{} Masers}\label{subsubsec:disc_sf_h2omasers}
\water{} masers have been detected in both low-mass ($\le$2~\msol{}) and high-mass ($\ge$8~\msol{}) star forming regions \citep{furuya2003,szymczak2005,urquhart2011} and are suggested to be associated with protostellar outflows \citep{elitzur1989,codella2004}. However, they may also be detectable toward the atmosphere of AGB stars. We compare our maser detections with the AGB star catalogs of \citet{lindqvist1992}, \citet{sevenster1997}, \citet{sjouwerman1998,sjouwerman2002}, and \citet{messineo2002}, which are based on detections of OH/SiO masers, and with the catalog of \citet{robitaille2008}, which is based on infrared color criteria. Five of the \water{} masers have AGB star counterparts and are marked as red crosses in \autoref{fig:masers}: W6 and W18 in \ctw{}, W1 and W3 in \cfi{}, and W2 in \sgb{}. We thus excluded them in the following analysis. It is also possible that the AGB star catalogs are incomplete, therefore there may be more contamination from uncataloged AGB stars.

Another possibility is that the masers are created by pc-scale shocks, similar to the case of wide-spread class~\textsc{i} \methanol{} masers found in the CMZ \citep{yusefzadeh2013}. However, as we have argued in \citet{lu2015b}, this is unlikely for most \water{} masers we detected, given their strong spatial correlation with the cores and their largely scattered velocities. For the eight \water{} masers not associated with detectable dust emission in \cfi{} (W4), \gzp{} (W1, W3), \sgb{} (W1, W4, W6), and Sgr~C (W4, W15), however, this is a viable scenario. Alternatively, these masers may be associated with low-mass protostellar cores that are missed by our observations (e.g., below the 5$\sigma$ mass sensitivity of 22~\msol{}) or uncataloged AGB stars.

There are also six \water{} masers detected outside of the SMA mosaic fields and not associated with known AGB stars or other types of masers, including W0, W4, W7, and W19 in \ctw{}, and W1 and W16 in Sgr~C (but excluding W1 in Sgr~D that is associated with a class~\textsc{ii} \methanol{} maser, see \autoref{subsubsec:disc_sf_other}), therefore their association with dust emission is unknown and their nature cannot be determined.

Thus, we conclude that most (37 out of 56, a percentage of 66\%) of the detected \water{} masers are likely associated with star formation activities. It is unclear whether they trace low-mass or high-mass star formation. If we adopt the empirical correlation between the luminosities of \water{} masers and young stellar objects \citep[e.g.,][]{urquhart2011}, then the more luminous \water{} masers ($\gtrsim$10$^{-6}$~\lsol{}) would be associated with high-mass young stellar objects. As listed in \autoref{tab:masers}, there are 19 such masers in our observations, and we note that some of them are associated with UC \hii{} regions or class~\textsc{ii} \methanol{} masers, which signify high-mass star formation (see the next two sections). However, the scatter in the correlation of \citet{urquhart2011} is large, and due to the time variability of \water{} masers, their luminosities can change by several orders of magnitude over several years \citep{felli2007}. We cannot rule out the possibility that some of the fainter masers are associated with high-mass star formation, or that some of the luminous \water{} masers trace low-mass star formation.

\subsubsection{\hii{} Regions}\label{subsubsec:disc_sf_hii}
\hii{} regions are created by high-mass protostars of O or early-B types \citep{churchwell2002}. As shown in \autoref{subsec:results_vlacont}, we confirm the existence of \hii{} regions in \ctw{}, \cfi{}, and Sgr~D using the VLA radio continuum emission. In addition, several potential UC \hii{} regions of $<$0.1~pc scales are identified in \ctw{}, \sgb{}, and Sgr~C, and marked in \autoref{fig:smacont}. We do not know their spectral indices, therefore are unable to verify whether the radio continuum emission represents a thermal free-free component. However, the close spatial correlations with compact dust emission suggest that they are more likely to be UC \hii{} regions embedded in cores. Note that in \gzp{} we detect radio continuum emission towards the core C2P1, but this emission has been suggested to be unrelated to star formation \citep[][labeled as C3 in their Figure~2]{mills2011}. C2P1 is gravitationally unbound according to our virial analysis in \autoref{subsec:results_virial} and is unlikely to form stars. Therefore, this emission is not identified as an UC \hii{} region.

The ionizing photon fluxes of \hii{} regions $N_\text{c}$ are estimated from their radio continuum emission, assuming optically thin free-free emission and an electron temperature of 10$^4$~K, following \citet{mezger1974}. Then assuming that each of the \hii{} regions is powered by a single star, we determine spectral types of the ionizing sources by comparing to the fluxes of ZAMS stars in \citet{panagia1973} and \citet{davies2011}, and estimate their stellar masses. The results are listed in \autoref{tab:hii}. 

\subsubsection{Other Types of Masers from Literature}\label{subsubsec:disc_sf_other}
Early evolutionary phases of star formation in these clouds are also revealed by OH masers and \methanol{} masers. OH masers have also been detected toward AGB stars, as discussed in \autoref{subsubsec:disc_sf_h2omasers}. Meanwhile, radiatively excited class~\textsc{ii} \methanol{} masers have been suggested to uniquely trace high-mass star formation \citep{menten1991a,ellingsen2006,breen2013}.

We compare our results with the OH masers from catalogs in \citet{karlsson2003} and \citet{cotton2016}. Among the four ground-state OH maser lines, the sole detection of the one at 1720~MHz, without accompanying main OH lines (at 1665/1667~MHz) or other maser species, usually traces supernova remnants \citep{wardle2002}. Similarly, the sole detection of the 1612~MHz OH maser is usually indicative of AGB stars, as discussed in \autoref{subsubsec:disc_sf_h2omasers}. After excluding the supernova remnant or AGB star candidates, only one OH maser at 1665~MHz from  \citet{cotton2016} is found to be spatially coincident with the \water{} maser W13 and the core C4P1 in Sgr~C.

The class~\textsc{ii} \methanol{} maser catalog is taken from \citet{caswell2010}. Four masers are found, in \sgb{} towards W5/C2P1, in Sgr~C toward W12/C4P2 and W13/C4P1, and in Sgr~D towards W1. Their \water{} maser counterparts are bright ($\gtrsim$10$^{-6}$~\lsol{}), consistent with being associated with high-mass star formation (\autoref{subsubsec:disc_sf_h2omasers}).

Therefore, we conclude that our \water{} maser observations recover all previously detected star formation sites traced by OH and class~\textsc{ii} \methanol{} masers. Nevertheless, the detection of class \textsc{ii} \methanol{} masers helps to confirm high-mass star formation. We listed all these detections in the last column of \autoref{tab:masers}.

\subsubsection{Densities and Virial States of Protostellar and Starless Cores}\label{subsubsec:disc_sf_criteria}

We classify the cores in a straightforward way as `protostellar', which are associated with \water{} masers and/or (UC)~\hii{} regions, and `starless', where none of the star formation indicators is detected. Excluding Sgr~D, we find 21 protostellar core candidates and 28 starless core candidates in the five CMZ clouds, as indicated in \autoref{tab:cores}. With our classification, the starless core sample may be contaminated (e.g., some of the objects may already harbor protostars), while the protostellar core sample is precise (i.e., all the cores in this sample are likely forming stars, minus potential contamination by AGB stars) but is likely incomplete.

\begin{figure}[!t]
\centering
\includegraphics[width=0.48\textwidth]{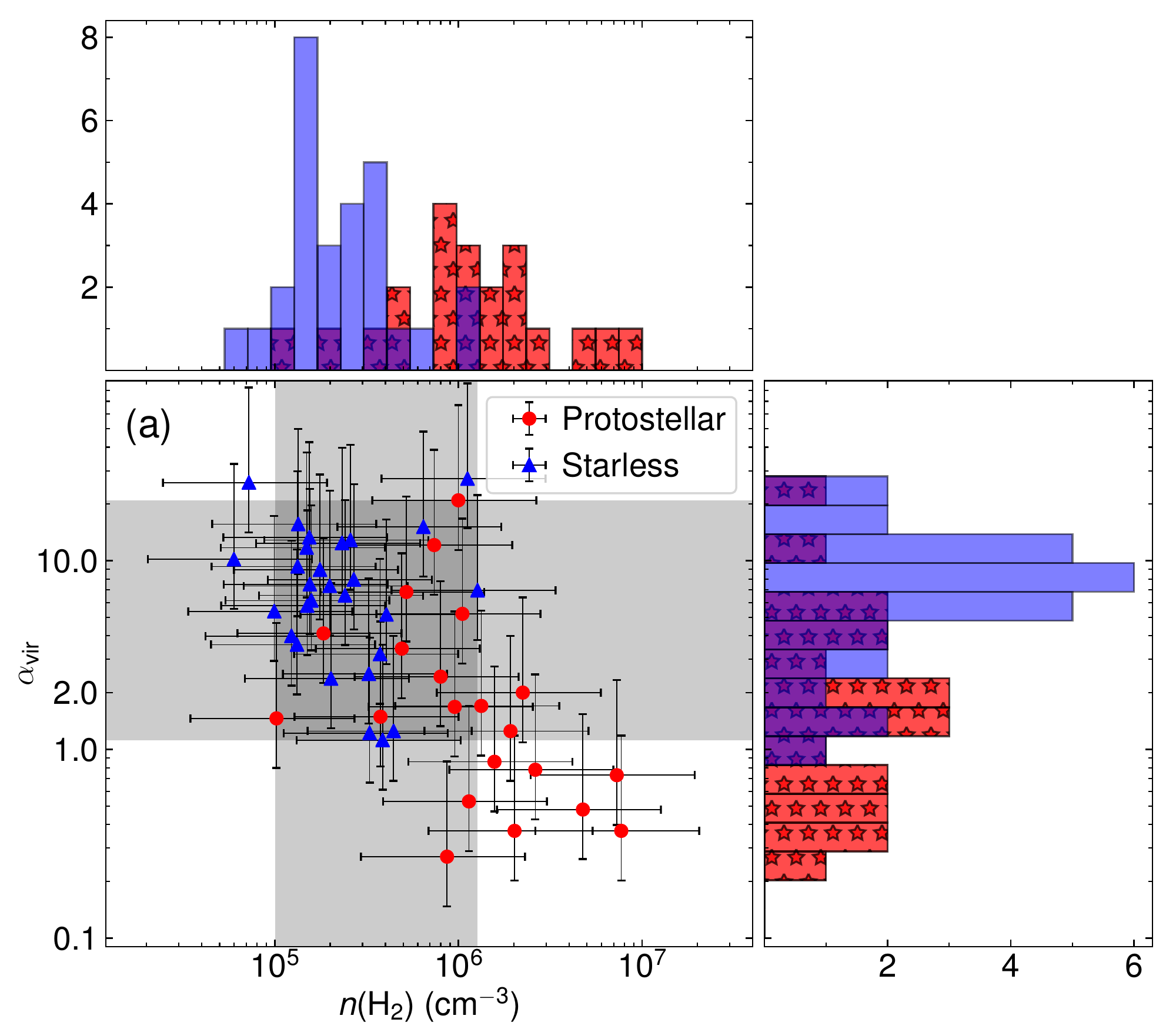} \\
\includegraphics[width=0.48\textwidth]{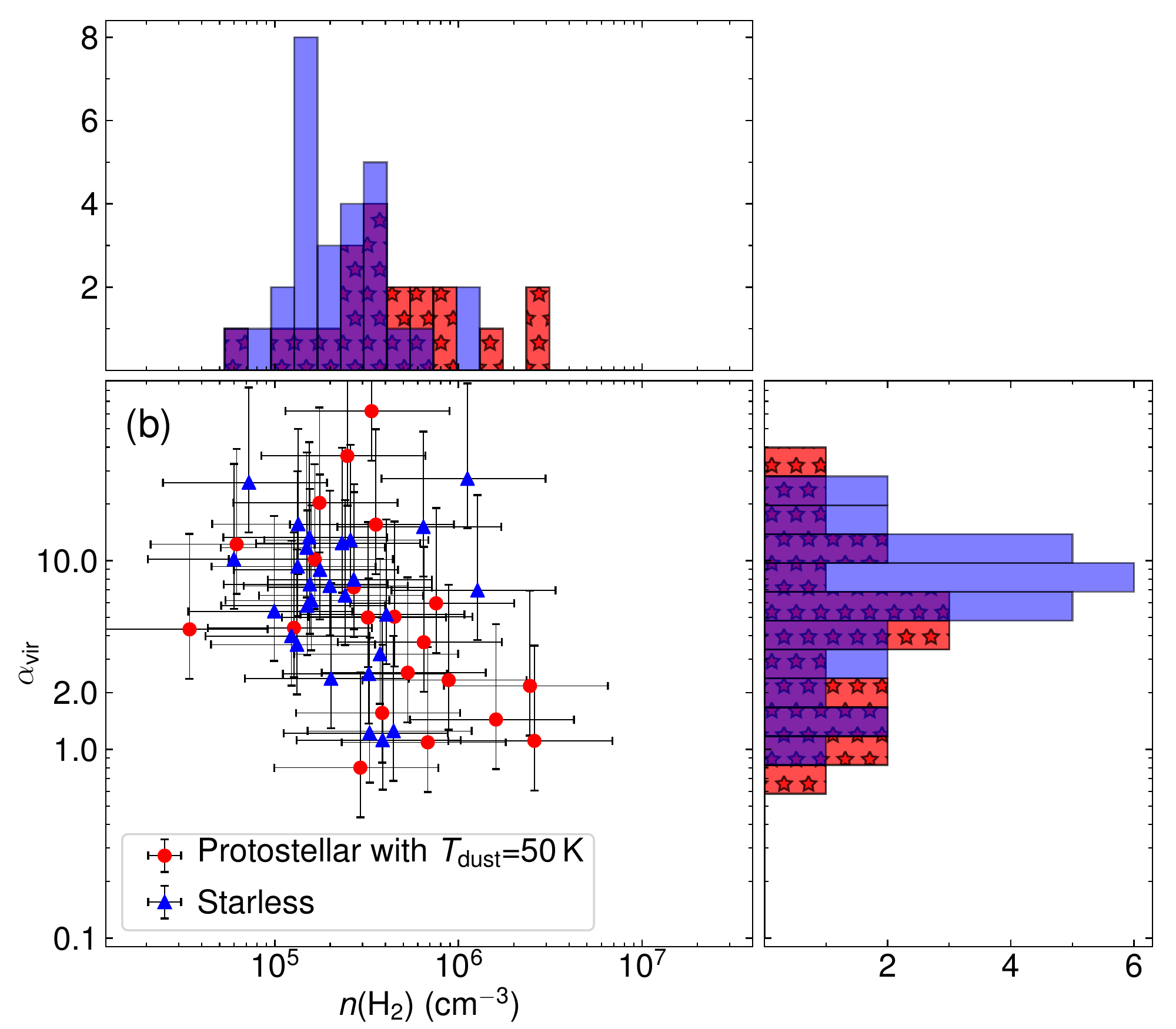}
\caption{Densities and virial parameters of the protostellar and starless core candidates in the five clouds. Horizontal error bars show uncertainties in densities of 66\%, while vertical error bars show uncertainties in virial parameters of 120\% (factor 2.2). (a) The vertical shaded area spans densities of 1.0$\times$10$^5$~\cc{} to 12.7$\times$10$^5$~\cc{}, representing the sufficient and the necessary density criteria for protostellar cores, respectively. Similarly, the horizontal shaded area spans between virial parameters of 20.9 and 1.12, representing the sufficient and the necessary virial parameter criteria for protostellar cores. (b) Same as the first panel but the densities and virial parameters of the protostellar core candidates are derived assuming a dust temperature of 50~K.}
\label{fig:sfcriteria}
\end{figure}

One property that may be able to modulate star formation in these cores is the density $n$(H$_2$). A wide variety of recent papers \citep{kruijssen2014,rathborne2014b,KK2015,federrath2016,krumholz2017,ginsburg2018}
have predicted or measured a density threshold for star formation in the CMZ of 10$^5$--10$^7$~\cc{}, which is much higher than the threshold in the Galactic disk clouds, $\sim$10$^4$~\cc{} \citep{lada2012}. The other property significantly affecting star formation is the virial parameter $\alpha_\text{vir}$ (see \autoref{subsec:results_virial}), which determines the gravitational boundness of the cores. In \autoref{fig:sfcriteria}a, we plot these two properties of the protostellar and starless core candidates.

It is clear from \autoref{fig:sfcriteria}a that the protostellar core candidates tend to have higher densities and smaller virial parameters than the starless core candidates. We run a Kolmogorov-Smirnov (K-S) test to quantify how different the two samples are in terms of densities and virial parameters. Usually when the p-value is much smaller than 0.05, the difference between the two samples is significant, and when the $p$-value is much larger than 0.05, the difference is not significant and we cannot rule out the possibility that the two samples are drawn from the same distribution. The $p$-values from the K-S test of the densities or virial parameters of the two samples are $<$4$\times$10$^{-4}$, suggesting the difference between the two samples is statistically significant.

However, we do not find a clear density or virial parameter threshold between the protostellar and starless core candidates. The lowest density found in protostellar core candidates is 1.0$\times$10$^5$~\cc{}, which is one order of magnitude lower than the highest density found in starless core candidates, 12.7$\times$10$^5$~\cc{}. On the other hand, 7 out of the 21 protostellar core candidates have virial parameters $>$2, while small virial parameters of 1.1--1.3 are found in three starless core candidates. In \autoref{fig:sfcriteria}a, we show these criteria as shaded regions.

If we consider the densities and virial parameters jointly, then the cores having both virial parameters $<$6 (or a physically more meaningful threshold of $<$2) and densities above 4.5$\times$10$^5$~\cc{} are all protostellar candidates. Likewise, the core having both virial parameters $>$2 and densities below 4.5$\times$10$^5$~\cc{} are all starless candidates (except C4P5 in \ctw{}, which is protostellar but spatially unresolved, so its density is a lower limit and its virial parameter is an upper limit). However, this does not suggest a clear criterion for separating the two samples, given the exceptions discussed below.

One protostellar core candidates, C2P1 in Sgr~C, has a density of 1.0$\times$10$^5$~\cc{} that is 10 times lower than the median density of the protostellar core candidates and than most of the starless core candidates, even though its virial parameter is $<$2 suggesting that it is gravitationally bound and unstable to collapse. Its small density indicates that it is not as compact as the other protostellar cores, which can happen if it is at an earlier evolutionary phase than the others and collapse has just started, or if it only harbors lower mass protostars.

As stated above, there are seven protostellar core candidates with virial parameters $>$2. If outflows already exist in these cores, the line widths may be broadened therefore the virial parameters may be overestimated. Another possibility is that they further fragment into multiple substructures, each of which is gravitationally bound but as a whole they are not.

The starless core candidates, which may be contaminated with star forming cores, tend to have lower densities and larger virial parameters than the protostellar core candidates in \autoref{fig:sfcriteria}a. In particular, all the cores in \gzp{} show large virial parameters and may be unbound. In fact, it is a question whether or not these substructures should be called `cores', because if they are indeed unbound, they will be transient objects and likely disperse in a dynamical time scale.

One potential bias of the analysis here is that dust temperatures in the protostellar core candidates may be higher than the assumed 20~K because of internal heating, in which case the masses and the densities would be overestimated (see \autoref{subsec:results_errors}). Assuming a higher dust temperature of 50~K for the protostellar core candidates, the densities will be 3 times smaller while the virial parameters will be 3 times larger. As shown in \autoref{fig:sfcriteria}b, in this case the difference between the protostellar and starless core candidates is not significant, with $p$-values of $>$0.05 in the K-S test of densities or virial parameters between the two samples.

Another potential bias is that the low-/intermediate-mass star formation in the CMZ clouds is largely unknown, which may be taking place at lower densities. Our SMA dust continuum observations are not sensitive to low-mass protostellar cores below 22~\msol{} (corresponding to a density of 0.7$\times$10$^5$~\cc{} assuming a typical radius of 0.1~pc). Although we detect several \water{} masers of lower luminosities (several times 10$^{-8}$~\lsol{}) that are usually found toward low-mass young stellar objects \citep{furuya2003}, they are insufficient to account for the expected low-mass star formation. Assuming a stellar initial mass function (IMF) from \citet{kroupa2001}, about 100 low-mass ($\leq$2~\msol{}) stars will form in company with each high-mass ($\geq$8~\msol{}) star. Therefore, a large fraction of the low-mass star formation activity is not revealed by our \water{} maser observations. Future interferometer observations with high angular resolution and sensitivity that is able to resolve low-mass protostellar cores will help to address the issue of low-mass star formation in the CMZ.

\begin{deluxetable*}{ccccccc}
\tabletypesize{\scriptsize}
\tablecaption{SFRs of the five clouds in our sample plus Sgr~B2 from the literature.\label{tab:sfr}}
\tablewidth{0pt}
\tablehead{
\colhead{\multirow{2}{*}{Cloud}} & Mass\tablenotemark{a} & Bound Mass  & Bound Mass Fraction & Masses of embedded high-mass protostars\tablenotemark{b} & $M_\text{cluster}$ & SFR \\
 & (10$^4$~\msol{}) & (10$^2$~\msol{}) & (\%) & (\msol{}) & (\msol{}) & (10$^{-3}$ \msol{}\,yr$^{-1}$) 
 }
\startdata
20~\kms{} & 32.0 & 22.5 & 0.7 & 12(H1), 8(W1), 9(W2), 13(W3), 14(W5), 12(W15)  & 603$\pm$193 & 2.0$\pm$0.6 \\
50~\kms{} & 6.1   & 1.1   & 0.2 & \nodata  & $<$91           & $<$0.3 \\
\gzp{}        & 8.8   & 0.7   & 0.08 & 8(W2)    & 91$\pm$82   & 0.3$\pm$0.3 \\
\sgb{}        & 13.7 & 5.4   & 0.4 & 11(H1)   & 91$\pm$82   & 0.3$\pm$0.3 \\
Sgr~C       & 2.4   & 21.0 & 9    & 13(H1), 11(H2), 14(H3), 15(H4), 8(W2), 15(W7), 12(W11), 8(W14) & 803$\pm$223 & 2.7$\pm$0.7 \\
Sgr~B2     & 140  & 450  & 3.2 &  271 high-mass protostars \citep{ginsburg2018} & (2.6$\pm$0.1)$\times$10$^4$  & 86$\pm$3
\enddata
\tablenotetext{a}{The cloud masses in \citet{kauffmann2017a}, which adopted a distance of 8.34~kpc, have been scaled to the distance of 8.1~kpc.}
\tablenotetext{b}{Indicators of embedded high-mass protostars are noted in parentheses. For Sgr~B2 we directly quote the number from \citet{ginsburg2018}. The stellar masses associated with UC \hii{} regions are taken from \autoref{tab:hii}. For \water{} masers, we first use the correlation between \water{} maser luminosities and bolometric luminosities in \citet{urquhart2011} to estimate luminosities of the young stellar objects, then estimate the stellar masses assuming the luminosity comes from a single protostar following the mass-luminosity relation in \citet{davies2011}. These masses do not enter the calculation of SFRs. They only demonstrate the range of masses (all $\ge$8~\msol{}, therefore in the high-mass regime).}
\end{deluxetable*}

\subsection{SFRs of the Clouds}\label{subsec:disc_sfr}

We attempt to estimate SFRs of the CMZ clouds based on the \water{} masers and UC \hii{} regions, which characterize star formation in deeply embedded phases. In this analysis, we exclude the more evolved \hii{} regions that are not embedded in cores (e.g., H2 in \ctw{} and H1--H4 in \cfi{}). We expect the resulting SFRs are more closely related to the observed gas than the SFRs estimated in longer time scales based on \hii{} regions or infrared luminosities, although evolutionary cycling between gas and stars still causes the ratio between both to evolve with time \citep{KL2014}.

\subsubsection{Derivation of the SFRs}\label{subsubsec:disc_sfr_derive}

First, we define the characteristic time scale. The typical lifetime of the UC \hii{} region phase is $\sim$0.3~Myr \citep{davies2011}. The lifetime of the evolutionary phase traced by the \water{} masers is estimated to be $\sim$0.3~Myr \citep{breen2010a}. In general, we adopt a time scale of 0.3~Myr for the overall star formation activities traced by \water{} masers and UC \hii{} regions.

Second, we estimate the stellar mass that will be formed based on the observed star formation tracers. We assume a canonical multiple-power-law IMF following \citet[][Equation~(2)]{kroupa2001}, with stellar masses between 0.01~\msol{} and 150~\msol{}.

We estimate how massive clusters should be given the observed numbers of high-mass protostars. The numbers of high-mass protostars are estimated by counting UC \hii{} regions and luminous \water{} masers ($\gtrsim$10$^{-6}$~\lsol{}) associated with cores. When both UC~\hii{} regions and \water{} masers are detected, we count them as one. This approach alleviates the problem of Poisson noise associated with the detection of \water{} masers. However, it still suffers from several uncertainties. As discussed in \autoref{subsubsec:disc_sf_h2omasers}, using luminous \water{} masers as indicators of high-mass star formation is highly uncertain, and the multiplicity of protostars is also an issue.

In \autoref{sec:appd_d}, we obtain a relation between cluster masses $M_\text{cluster}$ and numbers of high-mass protostars (see \autoref{app_fig:mcluster_N}b) by running Monte-Carlo simulations. The fractional uncertainty in $M_\text{cluster}$ decreases when more high-mass protostars are detected (e.g., 86\% for one detection and 25\% for 10 detections).

Stellar masses of the five clouds estimated using this relation are listed in \autoref{tab:sfr}. Then divided by the characteristic time scale of 3$\times$10$^5$~yr, we obtain the SFRs of the clouds as listed in \autoref{tab:sfr} and plotted as blue dots in \autoref{fig:sflaw}. Note that the uncertainties in the SFRs of \gzp{} and \sgb{} are as large as the derived SFRs themselves, therefore the SFRs of these two clouds should be treated as having an upper limit of 0.6$\times$10$^{-3}$~\msolpyr{}.

To validate our approach, we apply it to the whole CMZ, using the \water{} maser survey of \citet{walsh2014}. This survey is a follow up of the HOPS survey \citep{walsh2011} that covers Galactic longitudes between 290\arcdeg{} and 30\arcdeg{} and Galactic latitudes between $-$0\fdg{5} and 0\fdg{5}, and achieves a point source sensitivity of $<$0.2~Jy per 0.42~\kms{} channel for most of the data. We only consider \water{} masers with peak fluxes $\geq$0.6~Jy, corresponding to luminosities of $\gtrsim$10$^{-6}$~\lsol{} at the distance of 8.1~kpc, and only count masers within $|l|<1$\arcdeg{}. We find 112 such masers from the catalog of \citet{walsh2014}, among which 49 are in Sgr~B2. This number should be a lower limit given the issues in detection rate and multiplicity, although the sample may be contaminated by AGB stars. Assuming each of them is associated with a high-mass protostar, we use \autoref{equ:mcluster_N}, which agrees well with our simulations in \autoref{sec:appd_d}. The total stellar mass to be formed is estimated to be 1.1$\times$10$^4$~\msol{}, then dividing by the time scale of 0.3~Myr, we obtain a SFR of 0.04~\msolpyr{}. This is lower by a factor of 1.5--3 than those estimated from infrared luminosities or free-free emission over the same area \citep[0.06--0.12~\msolpyr{};][]{longmore2013a,barnes2017}, which is reasonable given the limitations of our mass measurements.

\subsubsection{Comparing with SFRs in Previous Studies}\label{subsubsec:disc_sfr_compare}

We compare the derived SFRs of the clouds with results in previous studies. \citet{kauffmann2017a} has estimated SFRs of these clouds based on (both compact and UC) \hii{} regions and class~\textsc{ii} \methanol{} masers, which characterize star formation in a time scale of 1.1~Myr. Their results are marked as crosses in \autoref{fig:sflaw}a, and typical uncertainty in their estimate of SFRs is a factor of 2. The most significant difference is that we find $>$10 times lower SFR for \cfi{} ($<$0.3$\times$10$^{-3}$~\msolpyr{} vs.\ 3.2$\times$10$^{-3}$~\msolpyr{}). \citet{kauffmann2017a} took the four \hii{} regions in this cloud into account. \citet{kauffmann2017b} noted the disconnection between the active star formation traced by the four \hii{} regions and a lack of massive clumps in this cloud. Our result suggests inactive star formation in \cfi{} in the last 0.3~Myr (one weak \water{} maser, no signatures of high-mass star formation), which is consistent with the observed dearth of cores.

The SFR of Sgr~C we derive is a factor of 3.4 higher than \citet{kauffmann2017a}: (2.7$\pm$0.7)$\times$10$^{-3}$~\msolpyr{} vs.\ 0.8$\times$10$^{-3}$~\msolpyr{}. Given the large uncertainties in our estimate, this difference is not considered to be significant.

For the other three clouds, including \ctw{}, \gzp{}, and \sgb{}, the SFRs in this work and in \citet{kauffmann2017a} generally agree within a factor of 3, and are $\sim$10 times lower than expected by the linear correlation in \citet{lada2010}.

In addition, \citet{barnes2017} estimated embedded stellar population of \gzp{} and \sgb{} (named as Brick and clouds e \& f, respectively, in their Tables 4 \& 5) using infrared luminosities, and found $>$10 times higher stellar masses, which are upper limits, as the infrared luminosities have non-negligible contributions from other sources (e.g., external radiation, the diffuse infrared field at the Galactic center).

\subsubsection{The SFR of Sgr~B2}\label{subsubsec:disc_sfr_sgrb2}

We estimate the SFR of Sgr~B2 using data from the literature. The star formation at early evolutionary phases in Sgr~B2 was recently studied by \citet{ginsburg2018}, who detected 271 compact continuum emission sources at 3~mm using ALMA, which are argued to be a mix of hyper-compact \hii{} regions and (high-mass) young stellar objects. Assuming that these 271 compact sources represent similar evolutionary phases as our \water{} maser and UC~\hii{} sample, we estimate a total stellar mass of (2.6$\pm$0.1)$\times$10$^4$~\msol{} using \autoref{equ:mcluster_N}, and obtain a SFR of 0.086$\pm$0.003~\msolpyr{} in a time scale of 0.3~Myr.

Our result is 40\% larger than the result of 0.062~\msolpyr{} reported in \citet{ginsburg2018}. The difference comes from both the stellar masses and the assumed time scales. \citet{ginsburg2018} obtained a stellar mass of 3.3$\times$10$^4$~\msol{} when only considering sources not associated with \hii{} regions, which is 30\% larger than our result, mostly because of different stellar masses attributed to each source. This indicates an additional uncertainty of 30\% for the stellar masses in Sgr~B2 from source classification. \citet{ginsburg2018} also used a time scale of 0.74~Myr that is based on the dynamical model of \citet{kruijssen2015}, which is longer than our assumption of 0.3~Myr. Our result is also a factor of 2.4 larger than the result of 0.036~\msolpyr{} in \citet{kauffmann2017a}, which is based on the detection of 49 compact \hii{} regions in a time scale of 1.1~Myr.

Overall, we do not find significantly different SFRs for Sgr~B2 from different approaches, and the discrepancy mostly comes from different assumed time scales. We summarize our estimate in \autoref{tab:sfr}.

\subsubsection{Comparing with the Orbital Model of the CMZ}\label{subsubsec:disc_sfr_orbit}

We compare our results with the orbital model of \citet{kruijssen2015}. This model suggests that all major clouds in the CMZ are subject to the gravitational potential around the Galactic Center and move in several gas streams (see the green curve in \autoref{fig:overview}). It also suggests that star formation in clouds could be triggered by tidal compression during a close passage to the bottom of the gravitational potential well near Sgr~A*.
 
In the model of \citet{kruijssen2015}, \gzp{}, \sgb{}, and Sgr~B2 are moving along one gas stream and have passed the pericenter to Sgr~A*. Sgr~C, \ctw{}, and \cfi{} are in the other gas stream, with Sgr~C in the upstream, \ctw{} approaching the pericenter, and \cfi{} having passed the pericenter. As discussed previously and shown in \autoref{tab:sfr}, we find signatures of increasing SFRs from \gzp{} to \sgb{} to Sgr~B2, which agree with the proposed monotonic increase of the star formation activity along the direction of motion after passing by Sgr~A* in this gas stream \citep{longmore2013b,kruijssen2015}. However, we do not find a similar trend for Sgr~C, \ctw{}, and \cfi{}. The derived SFRs of Sgr~C and \ctw{} are similar given the uncertainties, and are higher than that of \cfi{}. This may suggest that star formation in these clouds is not triggered by tidal compression when passing by the pericenter, but may be owing to self-gravity or impact of other sources (e.g., supernova remnants: \citealt{lu2003,mills2011}; \hii{} regions: \citealt{kendrew2013}).

\begin{figure}[!p]
\centering
\includegraphics[width=0.45\textwidth]{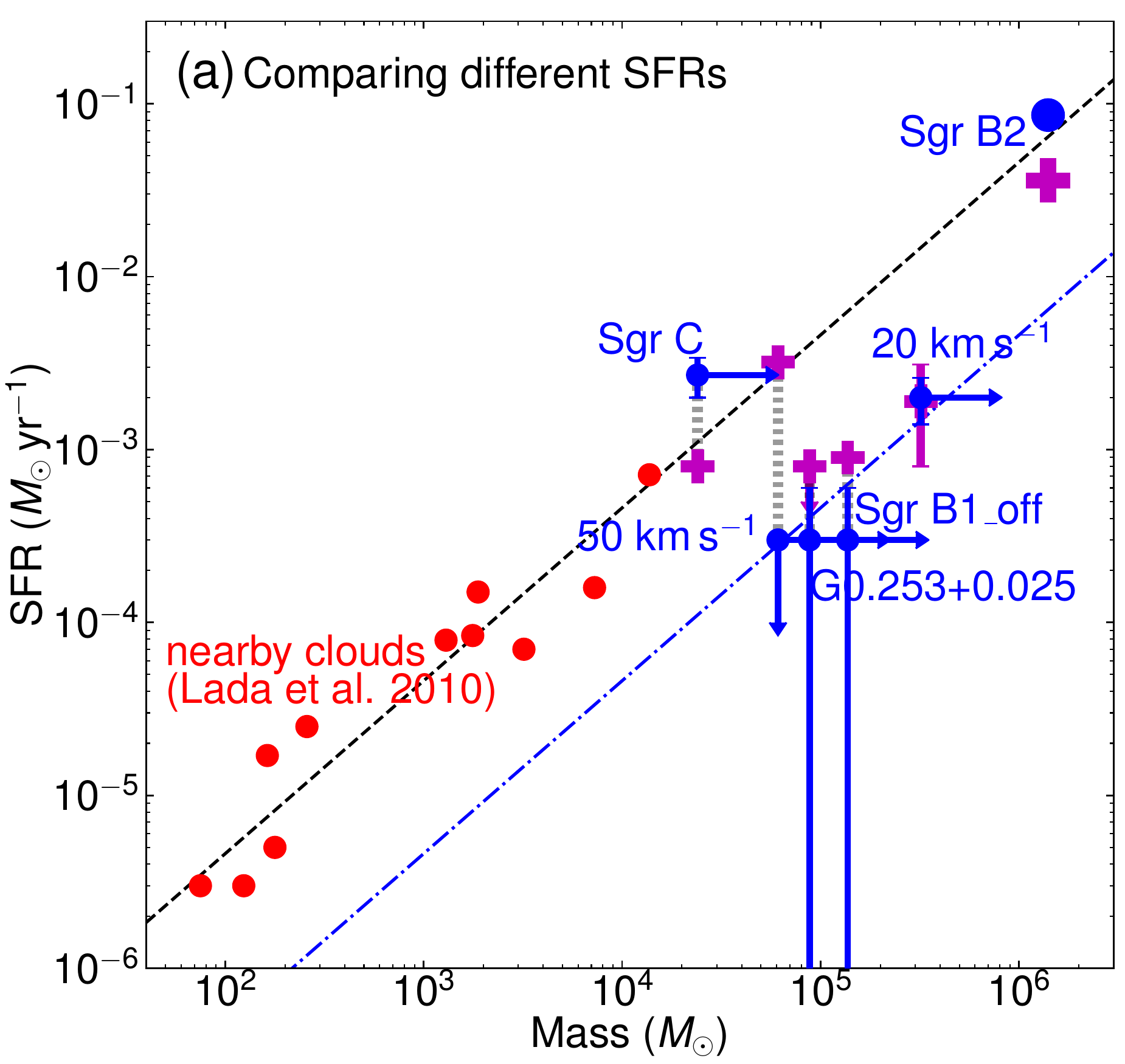} \\ 
\includegraphics[width=0.45\textwidth]{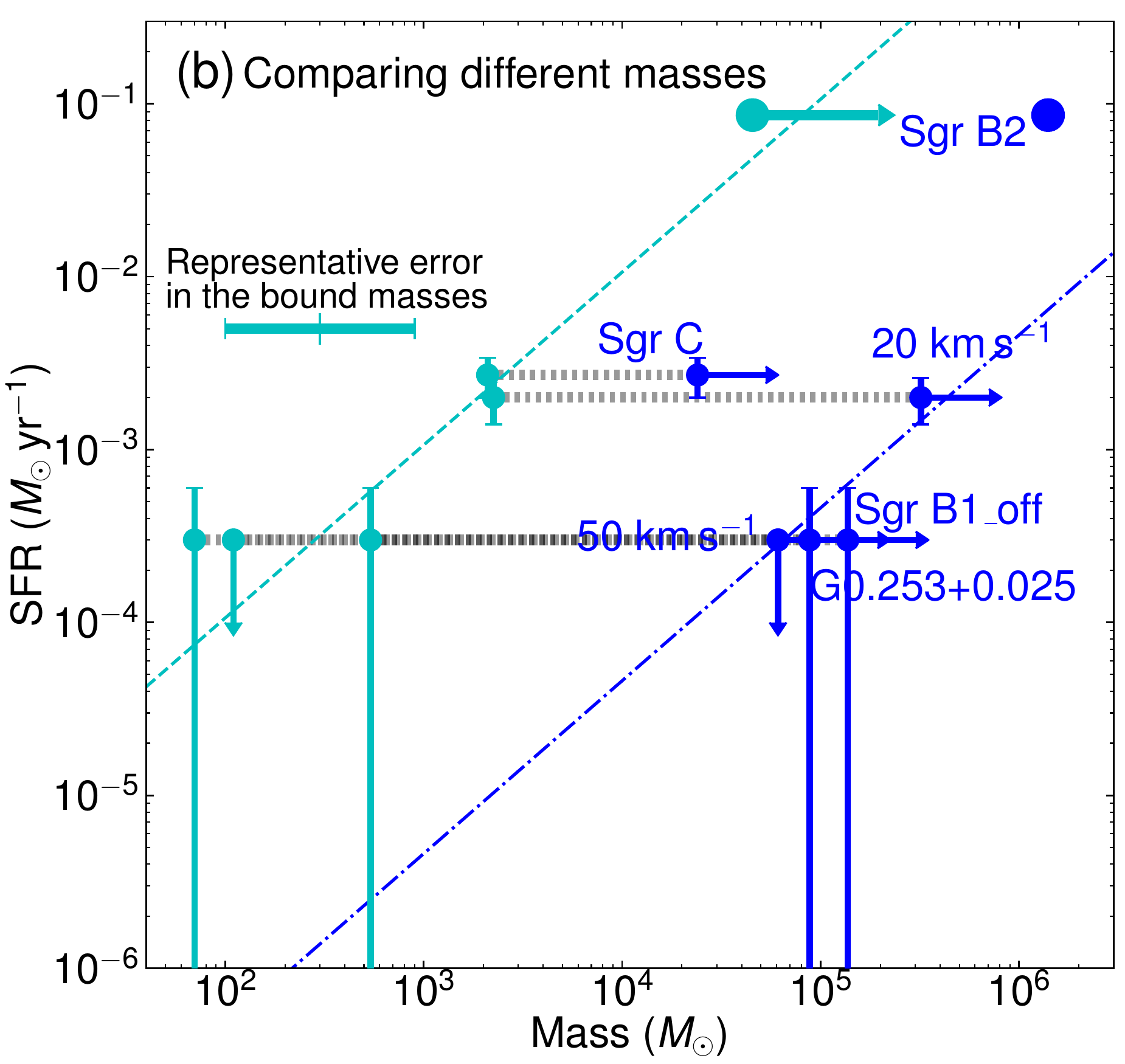}
\caption{SFRs and masses of dense gas in the five CMZ clouds and in a sample of nearby clouds from \citet{lada2010}. Data of Sgr~B2 compiled from the literature are also shown. (a) The blue dots mark the masses and SFRs of the five CMZ clouds in our observations plus Sgr~B2. Errorbars, when presented, represent the errors listed in \autoref{tab:sfr}, and arrows suggest lower or upper limits. Each blue dot is connected to a magenta cross through a vertical line, which shows the SFR derived in \citet{kauffmann2017a}. The black dashed diagonal line marks the linear correlation in \citet{lada2010}, with red dots showing masses and SFRs of the nearby clouds based on which the correlation is derived. The blue dash-dotted diagonal line marks the expected relation when the SFR is 10 times lower. (b) The blue dots, errorbars, and the blue diagonal line are identical to those in (a). In addition, each blue dot is connected to a cyan dot through a horizontal line, which marks the gravitationally bound masses in \autoref{tab:sfr}. The horizontal cyan error bar represents the uncertainty in the gravitationally bound masses. The cyan dashed diagonal line is a linear fit to the cyan dots of the five CMZ clouds in our observations, with a slope of 10$^{-6}$~yr$^{-1}$. The big cyan dot for Sgr~B2 denotes a gravitationally bound mass of 4.5$\times$10$^4$~\msol{}, which is a lower limit.}
\label{fig:sflaw}
\end{figure}

\subsection{Comparing with the Dense Gas Star Formation Relation}\label{subsec:disc_sflaw}
A quantitative comparison between star formation in these five CMZ clouds and that defined by the dense gas star formation relation has been done in \citet{kauffmann2017a}. Here we use the updated SFRs based on the \water{} masers and UC~\hii{} regions to carry out this analysis. As discussed in \autoref{subsec:disc_sfr}, these SFRs characterize embedded star formation at very early evolutionary phases therefore are more closely related to the observed gas.

The cloud masses are taken from \citet{kauffmann2017a}, which are estimated using \textit{Herschel} multi-wavelength data. The mean H$_2$ densities of these clouds are $\gtrsim$10$^4$~\cc{} \citep{kauffmann2017a}, therefore the dense gas fraction as defined in \citet{lada2010} is 100\%---that is, all the gas in these clouds are supposed to be `dense' and will collapse and form stars (but see \citealt{mills2018} for potential multiple density components in \ctw{}, \cfi{}, and \gzp{}, where $\sim$85\% of the gas has a density of $<$10$^4$~\cc{}). We then take the cloud masses to directly compare with the SFRs in the clouds.

The cloud masses and the SFRs of the five CMZ clouds (taken from \autoref{tab:sfr}) are plotted in \autoref{fig:sflaw}a. Given their masses, the SFR in Sgr~C agrees with the linear correlation in \citet{lada2010}, while the SFRs in the other four clouds are $\sim$10 times lower than expected, around a linear relation with a slope of 5$\times$10$^{-9}$~yr$^{-1}$.

The linear relation between SFRs and gas masses can be written as \citep{lada2010}
\begin{equation}\label{equ:sflaw}
\text{SFR} = \frac{\epsilon}{\tau_\text{SF}}\text{Mass},
\end{equation}
in which $\epsilon$ is the SFE (the integrated efficiency of converting gas to stars) in the gas under consideration, and $\tau_\text{SF}$ is the time scale of star formation. Then over a time scale of 0.3~Myr, the SFE of the four clouds (\ctw{}, \cfi{}, \gzp{}, and \sgb{}) is 0.15\%. This is significantly lower than those found in Galactic disk clouds, which are usually a few percent \citep{lada2010,louvet2014}.

In \autoref{subsubsec:disc_sf_criteria} we classify a sample of starless core candidates with large virial parameters, which may not form stars. Especially for \gzp{}, most of the cloud mass seems to be quiescent and irrelevant to high-mass star formation \citep{rathborne2014b}. It might make more sense to compare the masses in gravitationally bound cores, instead of those of the whole clouds, with the SFRs.

We attempt to estimate the gravitationally bound masses of the clouds by summing up masses of the protostellar core candidates and the prestellar core candidates that are gravitationally bound ($\alpha_\text{vir}$$\le$2, see \autoref{tab:cores}). In the following, we consider two systematic errors that significantly affect the estimate of the gravitationally bound masses and show that the uncertainty in the derived masses is a factor of 3.

First, the core identification in \autoref{subsec:results_cores} is very likely incomplete, therefore the derived gravitationally bound masses are likely underestimated. To quantify how much mass may be missed, we estimate upper limits of the core masses in \ctw{} and Sgr~C by taking the total dust emission in the SMA maps into account. Most of the identified cores in these two clouds are protostellar and/or gravitationally bound, therefore we likely miss gravitationally bound gas in the core identification. We do not use the other three clouds for this estimate, because the cores in them are mostly unbound and the majority of the gas is clearly not involved in star formation. Then we derive masses using the total dust emission fluxes in \ctw{} and Sgr~C, which are upper limits to the gravitationally bound masses. These masses are 3 times higher than the derived bound masses.

Second, the derived masses are likely affected by the systematic error in the dust temperature owing to internal heating, since we take all the protostellar core candidates into account. As discussed in \autoref{subsec:results_errors}, the masses of cores with significant internal heating may be overestimated by a factor of 3.

We list the derived gravitationally bound masses in \autoref{tab:sfr}. In particular, Sgr~C shows a high fraction of gravitationally bound mass (9\%), while the other four clouds all have much smaller fractions ($<$1\%). The derived gravitationally bound mass of Sgr~C is similar to that of \ctw{}. This has been noted in \citet{kauffmann2017a} as a shallower mass-size slope in Sgr~C than the other four clouds. It may explain the similar SFRs of Sgr~C and \ctw{} despite the fact that the cloud mass of Sgr~C is only 7.5\% of that of \ctw{}.

Then we plot the gravitationally bound masses against the SFRs in \autoref{fig:sflaw}b. The five clouds can be fit by a linear function, with a slope of 10$^{-6}$~yr$^{-1}$, which indicates a gas consumption time of 1~Myr in these gravitationally bound cores. The linear relation is expected from a constant SFE. Following \autoref{equ:sflaw}, we obtain a SFE of 30\% for the gas in the gravitationally bound/protostellar cores over a time scale of 0.3~Myr. Given the systematic errors in the derived gravitationally bound masses as discussed previously, the SFE may be as low as 10\% and as high as 90\%. This SFE is comparable to the value of 30\%--40\% for dense cores in nearby clouds \citep{alves2007,konyves2015}.

In addition, we include Sgr~B2 in the analysis, while the data are compiled from the literature. The SFR at early evolutionary phases of 0.086$\pm$0.003~\msolpyr{} is based on the work of \citet{ginsburg2018} (see \autoref{subsubsec:disc_sfr_sgrb2}). The cloud mass of Sgr~B2 is taken to be 1.4$\times$10$^6$~\msol{} \citep[][scaled to the distance of 8.1~kpc]{ginsburg2018}. The gravitationally bound mass is difficult to characterize and similar analysis to ours has not yet been published. As an approximate we use the total gas mass of the four protoclusters (Sgr~B2~NE, N, M, and S), 4.5$\times$10$^4$~\msol{} \citep[][scaled to the distance of 8.1~kpc]{schmiedeke2016}, which should be a lower limit given that the gas associated with the distributed protostellar population in Sgr~B2 \citep{ginsburg2018} is not included. About 200 among the 271 compact sources found by \citet{ginsburg2018} are not associated with any of the four protoclusters, therefore the total bound gas mass might be as much as four times larger than the mass in the four protoclusters if we assume the gas mass to be proportional to the number of compact sources. This results in a lower limit for the bound mass fraction of 3.3\% for Sgr~B2 that is potentially several times larger.

Comparing Sgr~B2 with other clouds in \autoref{fig:sflaw}, we note that its SFR agrees with the star formation dense gas relation in \citet{lada2010}, similar to the case of Sgr~C. Its bound gas mass fraction, with a lower limit of 3.3\%, is also similar to Sgr~C but 5--40 times larger than those of the other four clouds. When only considering the mass in the bound gas, Sgr~B2 falls closely around the linear relation by fitting the five clouds in our sample, as shown in \autoref{fig:sflaw}b.

These results may suggest that star formation at the core scale in these CMZ clouds is not different from that in Galactic disk clouds in terms of the core to star-formation efficiency, but at the cloud scale, except for Sgr~B2 and Sgr~C, the star formation is $\sim$10 times less efficient because less than 1\% of gas is confined in gravitationally bound regions. This small fraction of gravitationally bound gas in the clouds may be because of the strong turbulence in these clouds \citep{oka2001,shetty2012,kruijssen2013} as indicated by the large line widths in the cores (\autoref{sec:appd_c}), or because the clouds have only recently condensed, as expected if the CMZ as a whole is undergoing episodic cycles of star formation activity \citep{kruijssen2014,KK2015,krumholz2017}.

\section{CONCLUSIONS}\label{sec:conclusions}
We use the SMA 1.3~mm continuum and the VLA $K$-band continuum and \water{} maser observations to study star formation in five massive molecular clouds in the CMZ and one cloud that is likely outside of the CMZ. The main results are as follows.
\begin{itemize}
\item A total of 56 cores at the 0.2~pc scale are resolved by the SMA continuum emission in the six clouds. Their virial parameters are derived using line widths measured with the SMA \nthp{} or \methanol{} lines.
\item In the five CMZ clouds, signatures of embedded star formation at very early evolutionary phases, as traced by \water{} masers and compact free-free emission from UC~\hii{} regions, are found toward the cores, based on which we classify the cores as protostellar or starless. The protostellar core candidates tend to have higher densities and smaller virial parameters than the starless core candidates (\autoref{fig:sfcriteria}).
\item Based on the detection of bright \water{} masers (with luminosities $\gtrsim$10$^{-6}$~\lsol{}) and UC~\hii{} regions, SFRs within a time scale of 0.3~Myr of the five CMZ clouds are estimated. We also include Sgr~B2 in the analysis after compiling data from previous studies. The observed increasing SFRs from \gzp{} to \sgb{} to Sgr~B2 are expected by the CMZ orbital model of \citet{kruijssen2015}, but the SFRs of the other three clouds do not show a monotonic change that is predicted by this model.
\item Excluding Sgr~B2 and Sgr~C, SFRs of the other four CMZ clouds are $\sim$10 times lower than expected from the dense gas star formation relation extrapolated from nearby clouds in \citet{lada2010} (\autoref{fig:sflaw}a). If the masses in protostellar and/or gravitationally bound cores are used instead, these clouds can be better fit with a linear function, with a SFE of 30\% over a time scale of 0.3~Myr (\autoref{fig:sflaw}b). Among the six CMZ clouds (five in our sample plus Sgr~B2 from the literature), Sgr~B2 and Sgr~C stand out with larger fractions of mass in gravitationally bound regions and higher SFRs per unit cloud mass.
\item We confirm that star formation in four of the CMZ clouds is inactive with respect to the prediction of the dense gas star formation relation, even after taking star formation at very early evolutionary phases into account. This is likely related to their low gravitationally bound gas fractions of $<$1\%, which may be because of high turbulent energy densities in the CMZ and/or because they are dynamically young and have not had the time to populate their high-density component.
\end{itemize}

\acknowledgments
We thank the anonymous referee for helpful comments that improved the manuscript. X.L.\ thanks Andr\'{e}s E.\ Guzm\'{a}n and Shu-ichiro Inutsuka for helpful discussions. The operation staff and postdocs of the SMA are gratefully acknowledged for their help with the SMA observations. This work was supported by JSPS KAKENHI grant No.\ JP18K13589. This work was supported by NSFC grant 11629302. J.M.D.K.\ gratefully acknowledges funding from the German Research Foundation (DFG) in the form of an Emmy Noether Research Group (grant number KR4801/1-1) and from the European Research Council (ERC) under the European Union's Horizon 2020 research and innovation programme via the ERC Starting Grant MUSTANG (grant agreement number 714907). This research made use of Astropy, a community-developed core Python package for Astronomy \citep{astropy2013}, APLpy, an open-source plotting package for Python \citep{aplpy2012}, and astrodendro, a Python package to compute dendrograms of Astronomical data (\url{http://www.dendrograms.org/}). Data analysis was in part carried out on the open use data analysis computer system at the Astronomy Data Center (ADC) of the National Astronomical Observatory of Japan. This research has made use of the NASA/IPAC Infrared Science Archive, which is operated by the Jet Propulsion Laboratory, California Institute of Technology, under contract with the National Aeronautics and Space Administration. This research has made use of NASA's Astrophysics Data System and the SIMBAD database operated at CDS, Strasbourg, France.

\vspace{5mm}
\facilities{SMA, VLA, \textit{Spitzer}, \textit{Herschel}}

\software{MIR, Miriad \citep{sault1995}, CASA \citep{mcmullin2007}, APLpy \citep{aplpy2012}, Astropy \citep{astropy2013}}

\appendix
\section{Identification of substructures using Dendrogram}\label{sec:appd_a}
\begin{figure*}[!h]
\centering
\begin{tabular}{@{}p{0.333\textwidth}@{}p{0.333\textwidth}@{}p{0.333\textwidth}@{}}
\begin{tabular}[c]{@{}c@{}}
\includegraphics[width=0.33\textwidth]{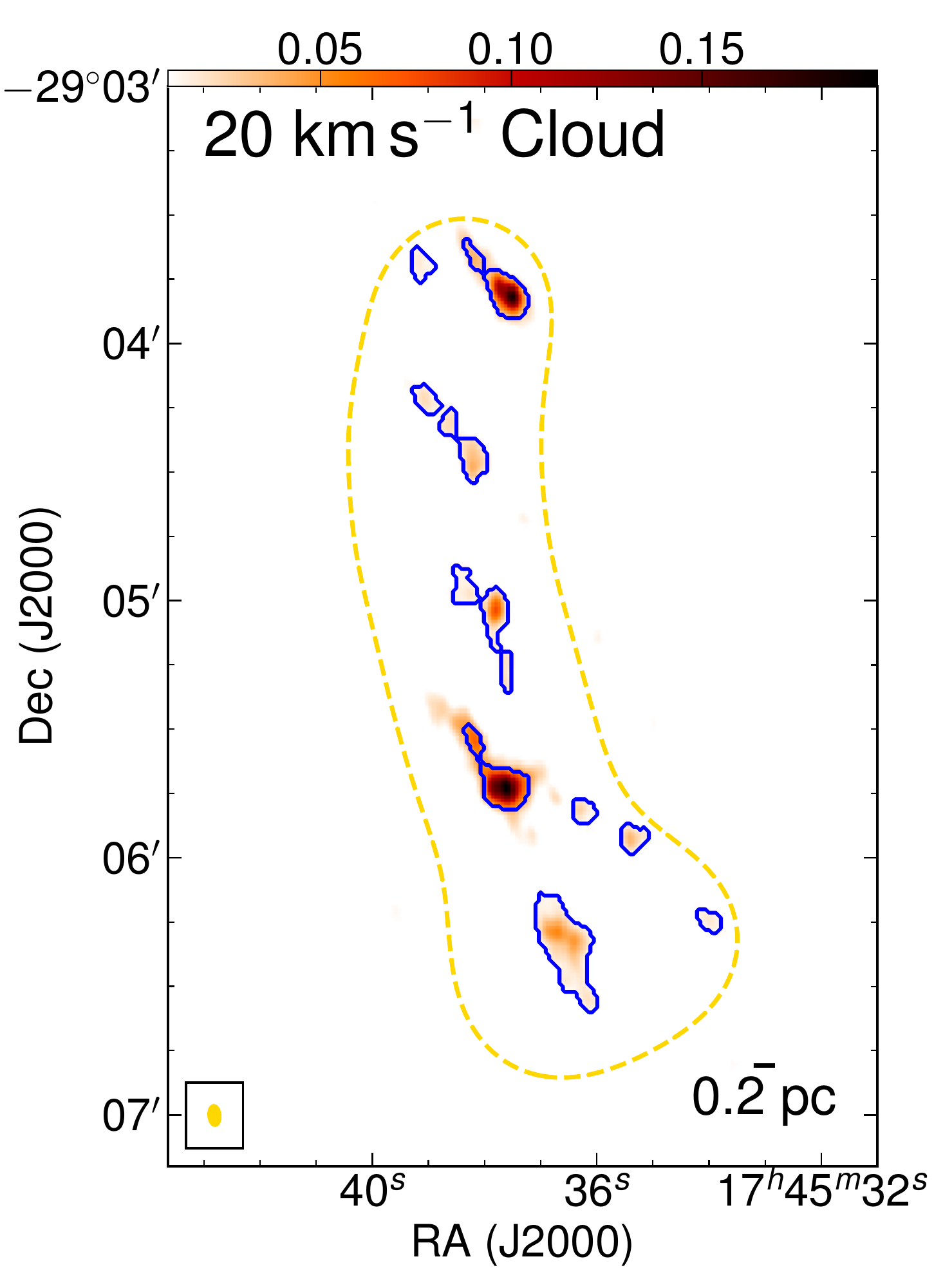} \\
\includegraphics[width=0.32\textwidth]{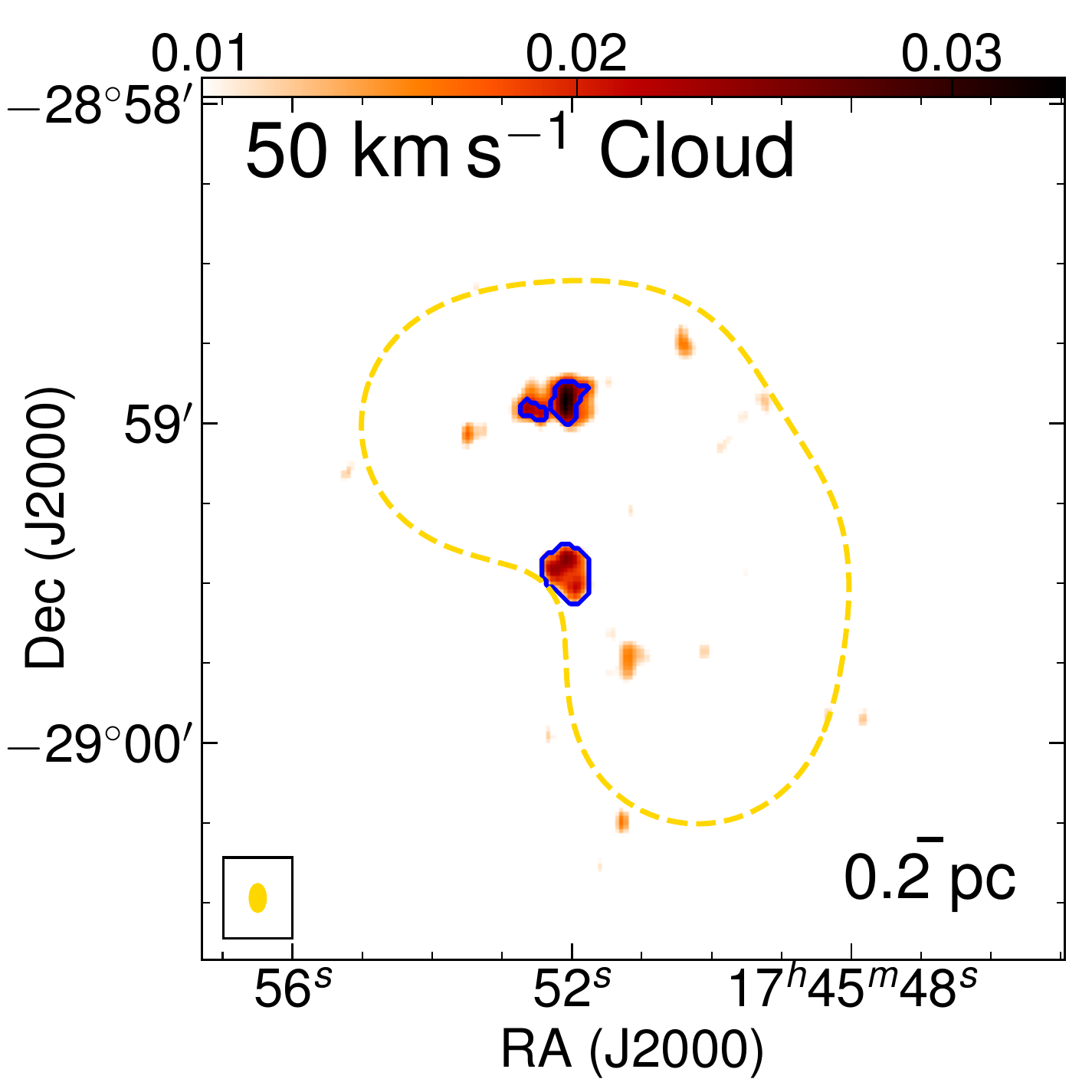}
\end{tabular}
&
\begin{tabular}[c]{@{}c@{}}
\includegraphics[width=0.32\textwidth]{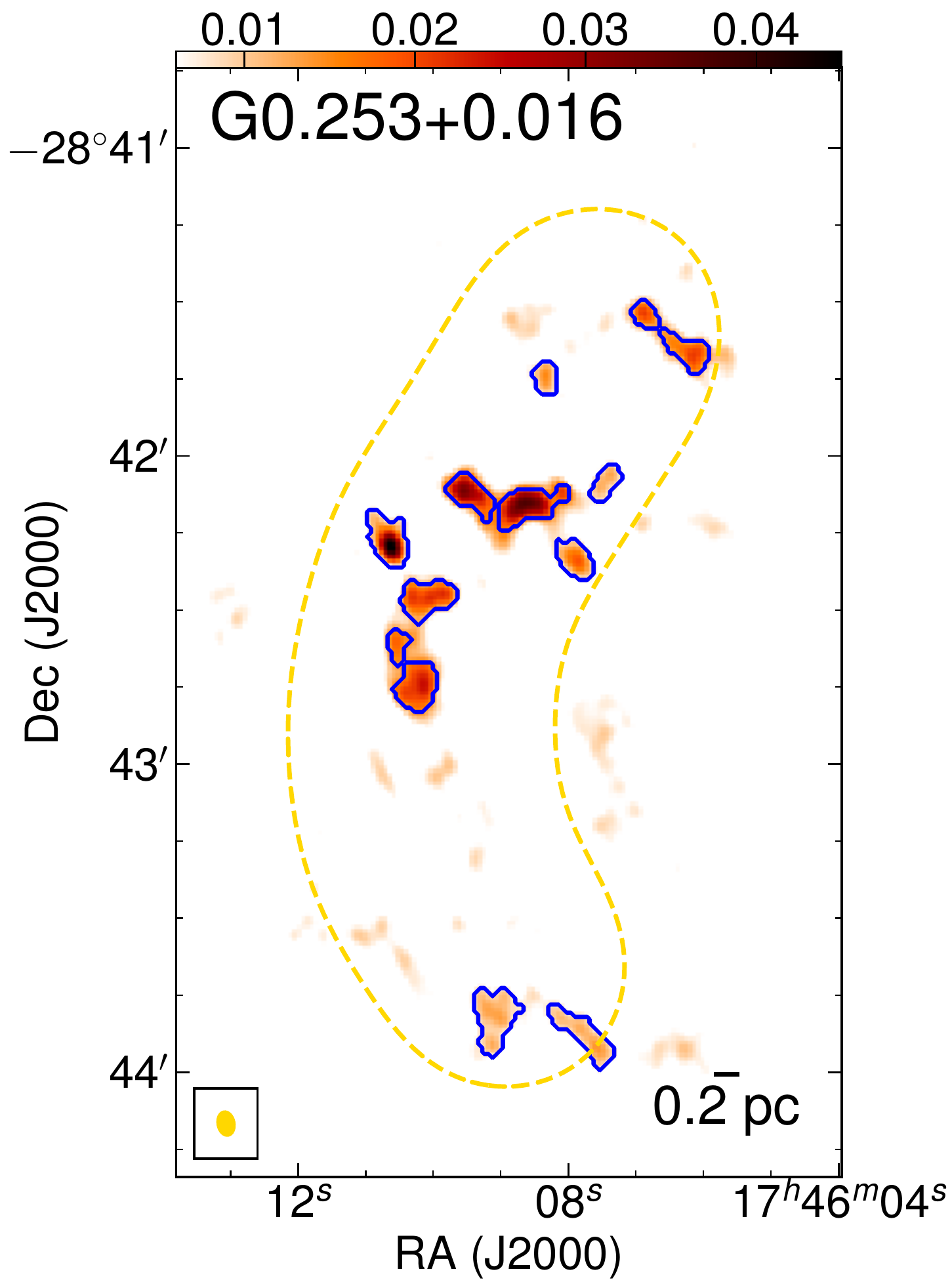} \\ 
\includegraphics[width=0.32\textwidth]{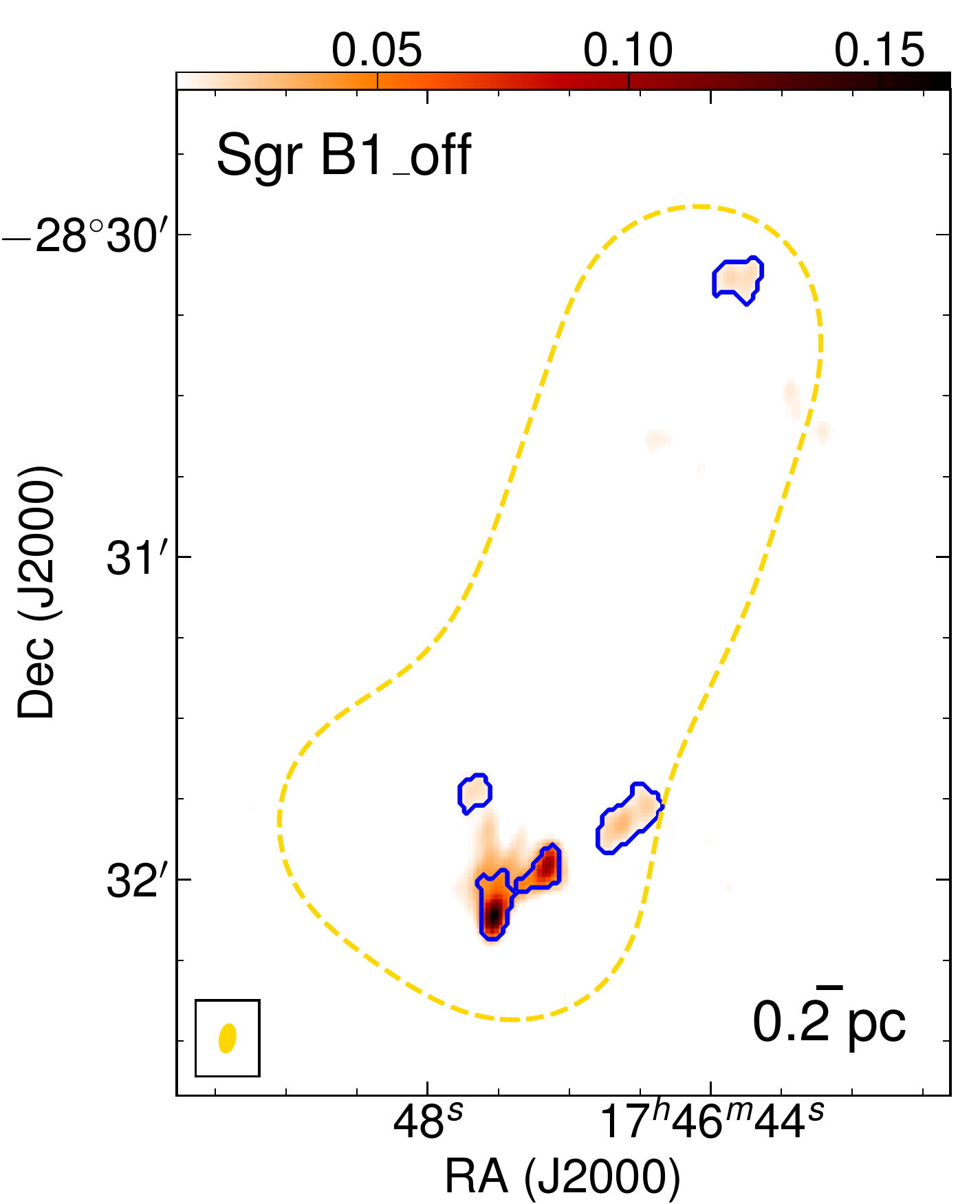}
\end{tabular}
&
\begin{tabular}[c]{@{}c@{}}
\includegraphics[width=0.32\textwidth]{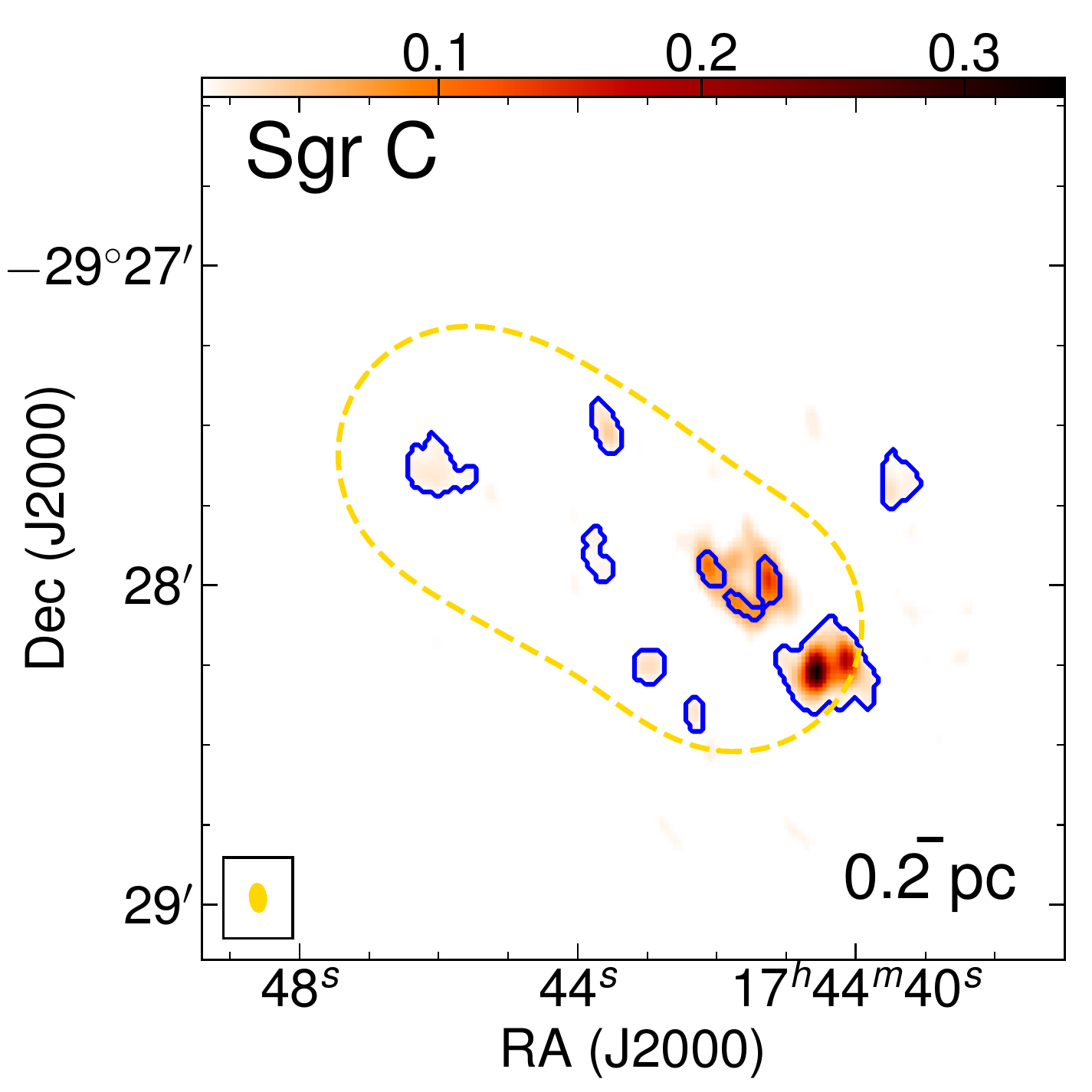} \\
\includegraphics[width=0.32\textwidth]{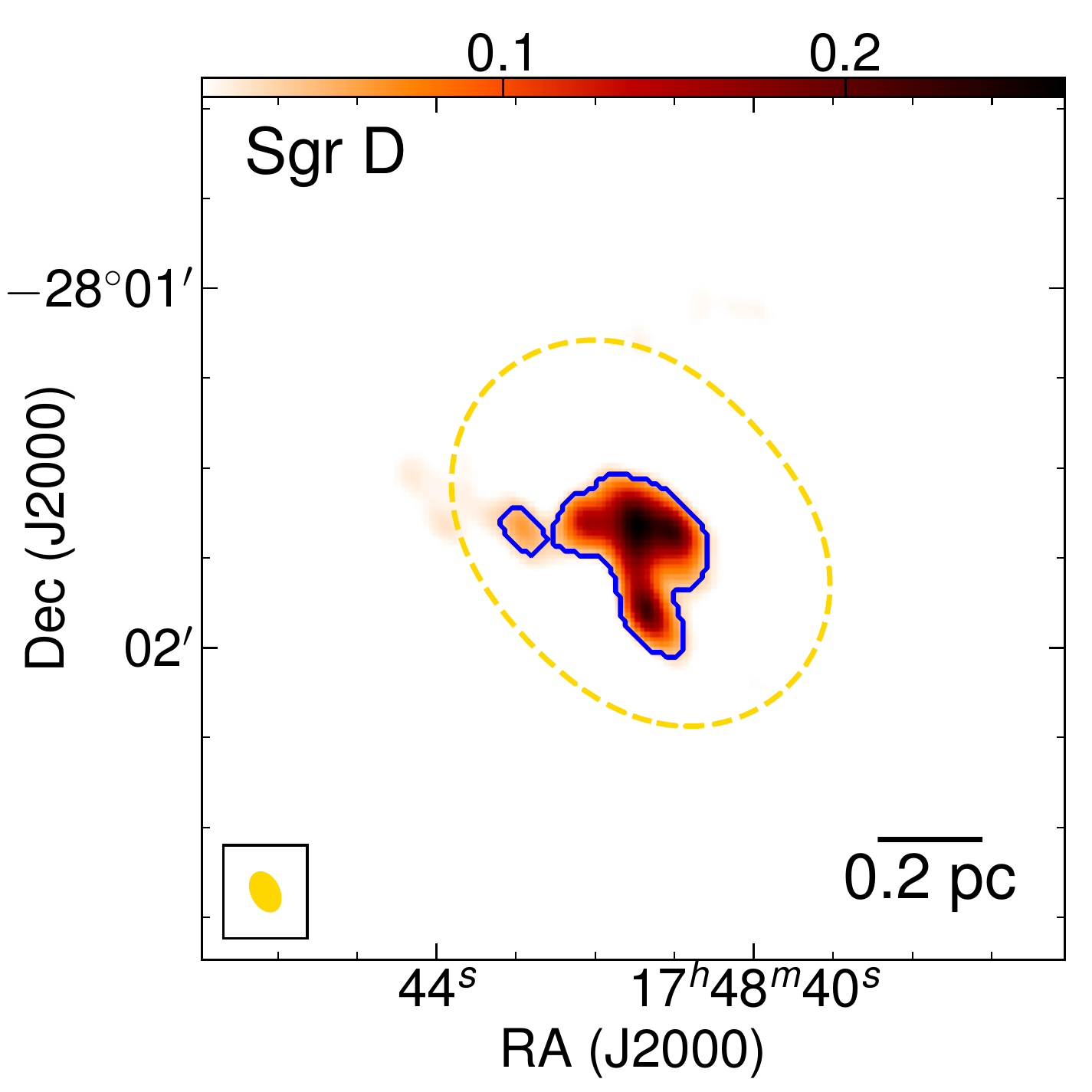}
\end{tabular}
\end{tabular}
\caption{Result of the dendrogram analysis. Background images and dashed loops show the SMA 1.3~mm continuum and the SMA mosaic field, which are identical to those in \autoref{fig:smacont}. Blue contours mark leaves identified using \textit{astrodendro}.}
\label{app_fig:dendro}
\end{figure*}

We identified compact structures (`leaves') in the SMA 1.3~mm continuum emission maps using the dendrogram algorithm \citep{rosolowsky2008dendro} implemented with \textit{astrodendro}\footnote{\url{https://dendrograms.readthedocs.io/en/stable/}}. We adopted a minimum intensity of 4$\sigma$ where the $\sigma$ values can be found in \autoref{tab:imaging}, below which the emission will not be considered, and a minimum significance of 1$\sigma$, which characterizes the significance a local maxima has to reach to be considered as an independent structure. With this setup, the identified structures will have peak intensities above the 5$\sigma$ level. We additionally specified that the number of pixels above the 5$\sigma$ level in one structure must be larger than the pixel number within the FWHM of one synthesized beam. The results are presented in \autoref{app_fig:dendro}, where the identified structures are marked by blue contours.

This procedure occasionally misses obvious structures. For example, in \ctw{} several emission peaks are spatially coincident with \water{} masers therefore should be protostellar core candidates (see \autoref{fig:masers}), but are not identified because their areas are slightly smaller than the beam size. Another example is that in Sgr~C two adjacent bright structures in the southwestern end are identified as one because of their small projected spatial separation, but they should be two independent cores given that they are each associated with a UC~\hii{} region and a \water{} maser (see Figures~\ref{fig:smacont} \& \ref{fig:masers}). Therefore, in \autoref{subsec:results_cores} we used the outcome of \textit{astrodendro} as a reference but manually added or removed structures in consideration of the above situations.

\section{\water{} maser spectra}\label{sec:appd_b}
\begin{figure*}[!h]
\centering
\includegraphics[width=0.5\textwidth]{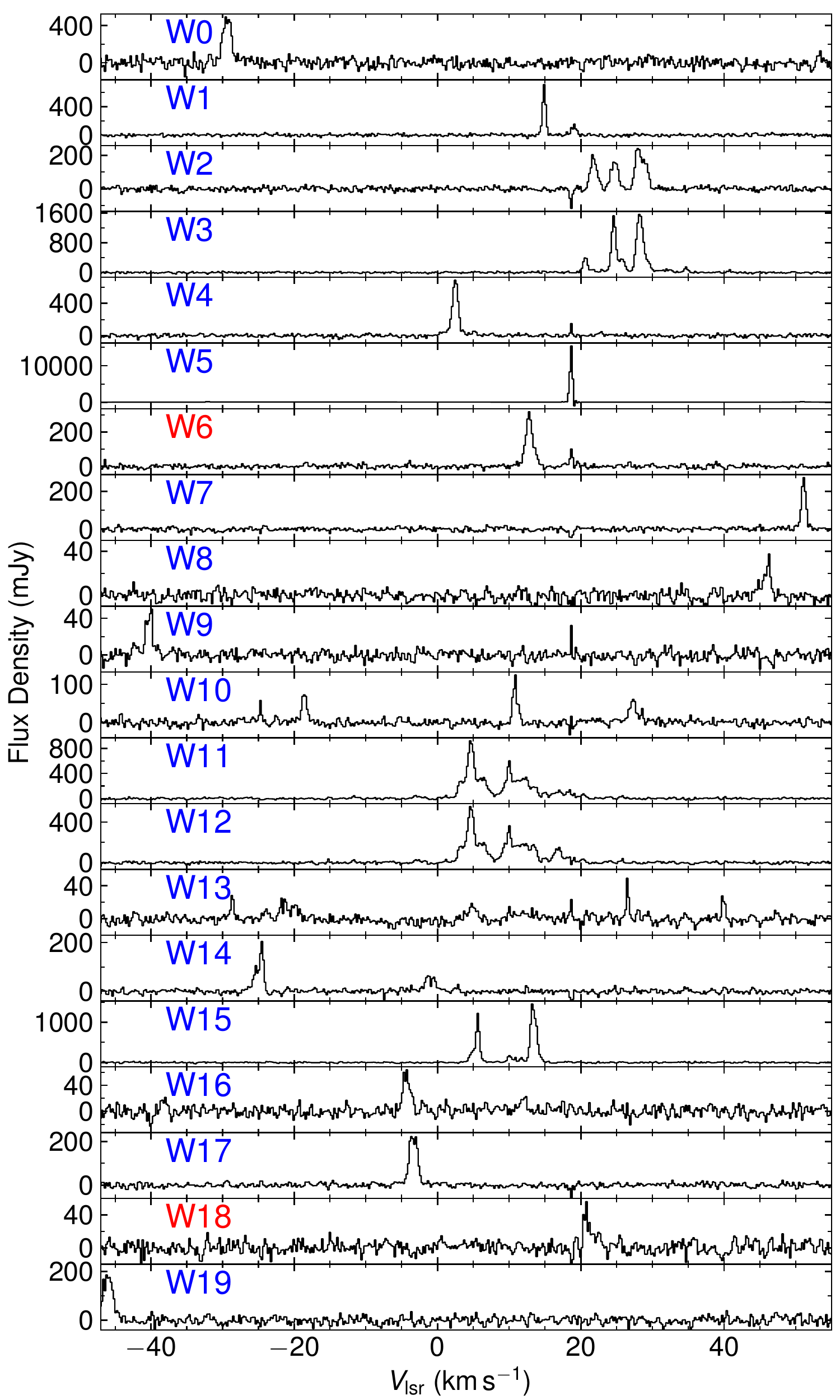} 
\caption{\water{} maser spectra in \ctw{} detected by the VLA.}
\label{app_fig:maser_spec_1}
\end{figure*}

\begin{figure*}[!h]
\centering
\includegraphics[width=0.5\textwidth]{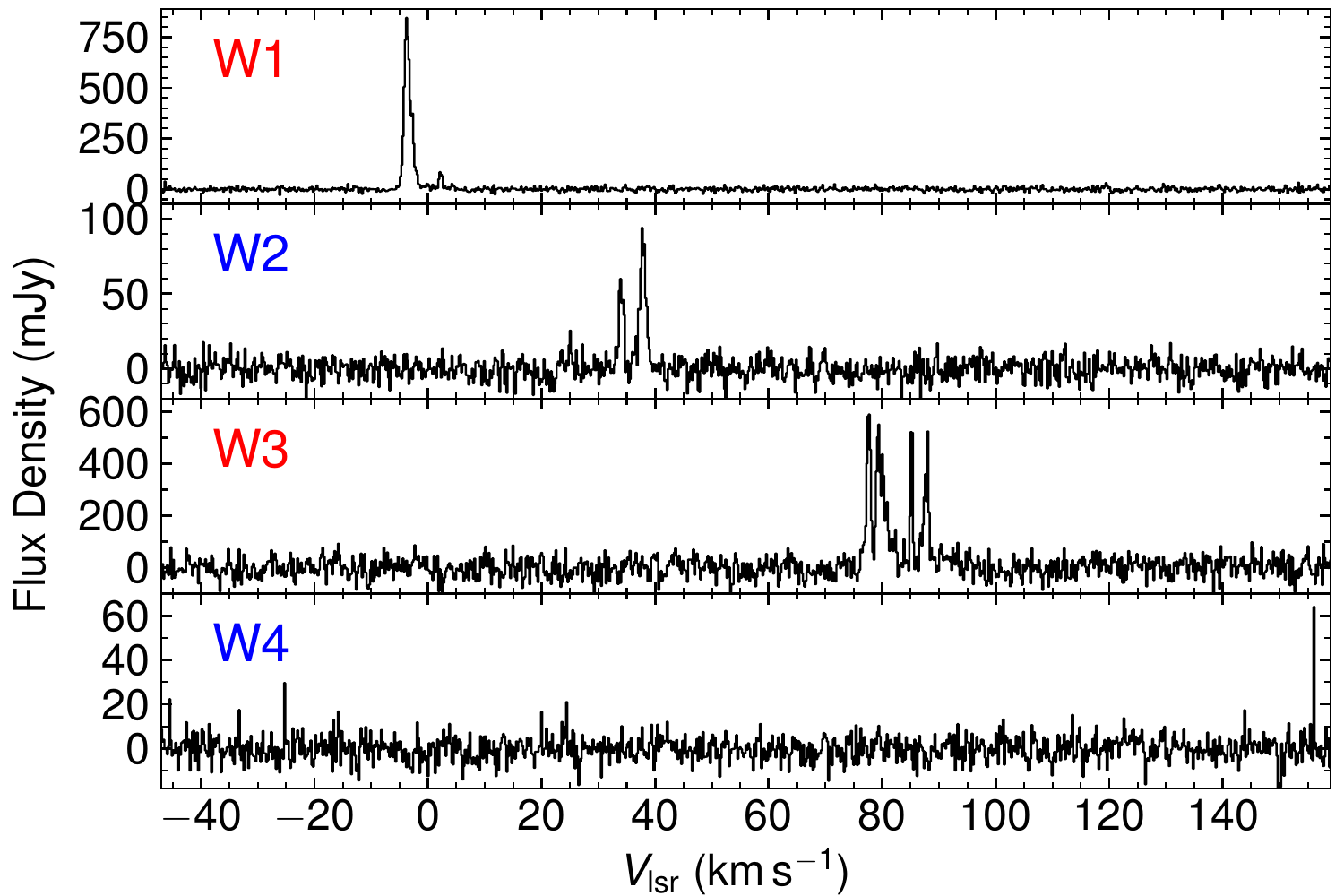} 
\caption{\water{} maser spectra in \cfi{} detected by the VLA.}
\label{app_fig:maser_spec_2}
\end{figure*}

\begin{figure*}[!h]
\centering
\includegraphics[width=0.5\textwidth]{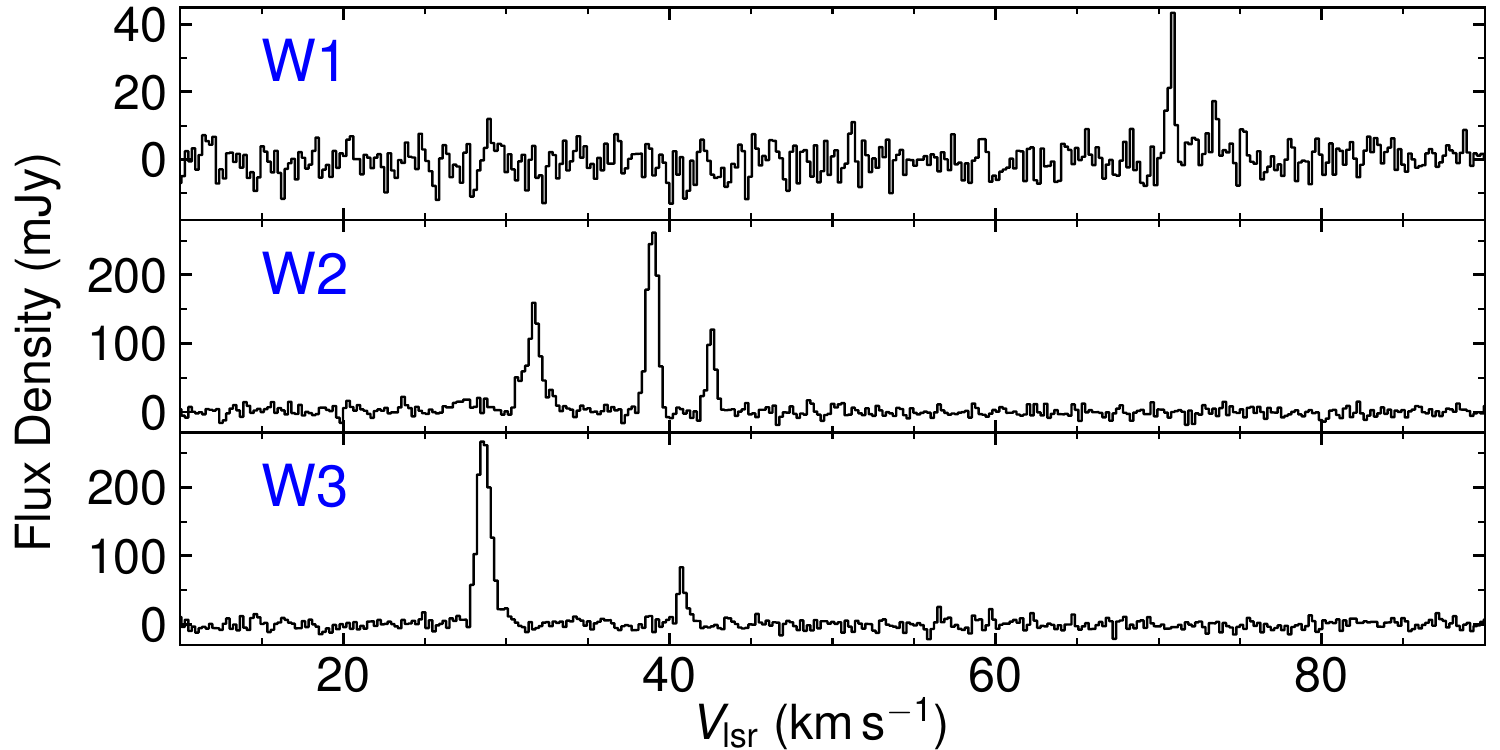} 
\caption{\water{} maser spectra in \gzp{} detected by the VLA.}
\label{app_fig:maser_spec_3}
\end{figure*}

\begin{figure*}[!h]
\centering
\includegraphics[width=0.5\textwidth]{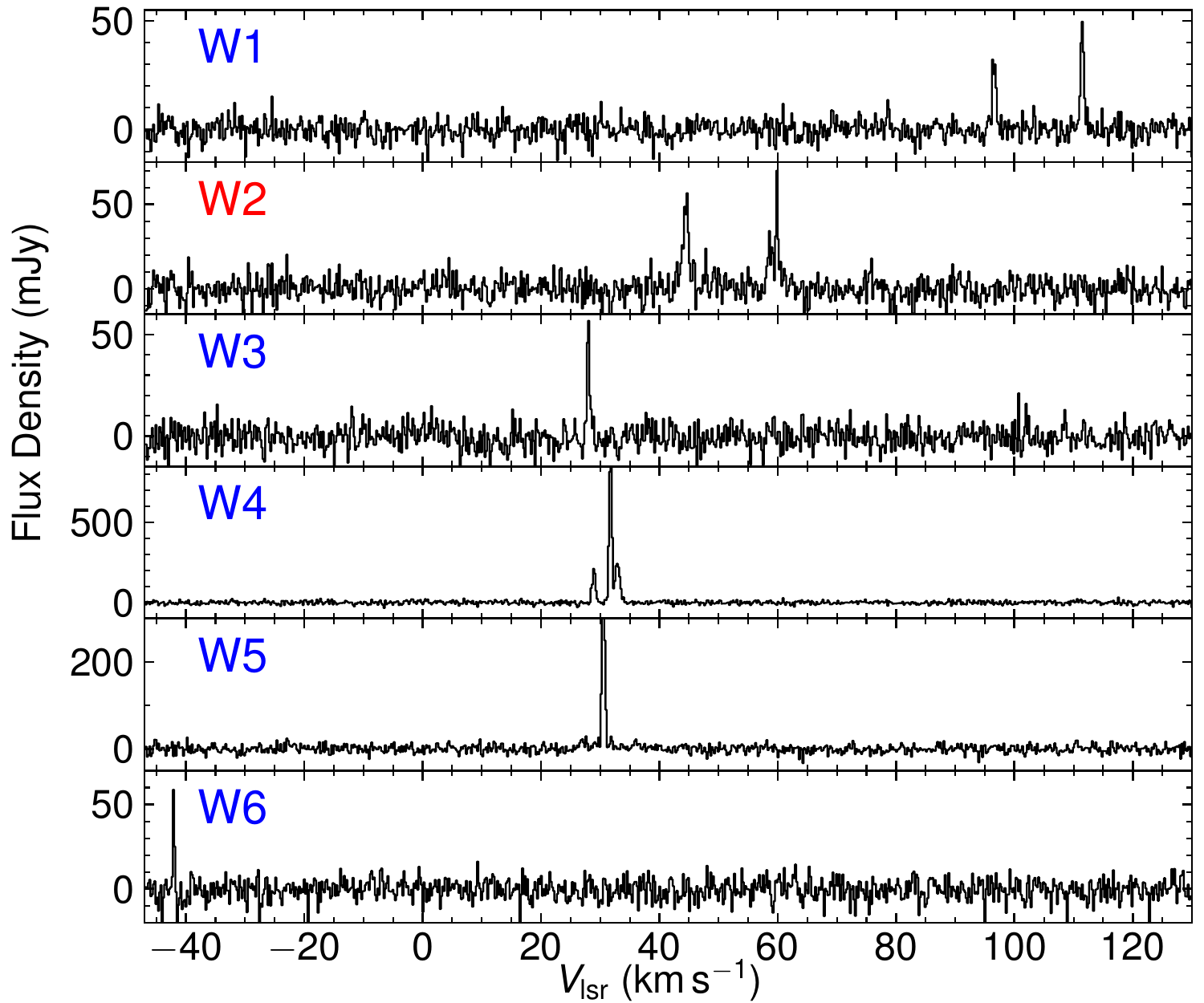} 
\caption{\water{} maser spectra in \sgb{} detected by the VLA.}
\label{app_fig:maser_spec_4}
\end{figure*}

\begin{figure*}[!h]
\centering
\includegraphics[width=0.5\textwidth]{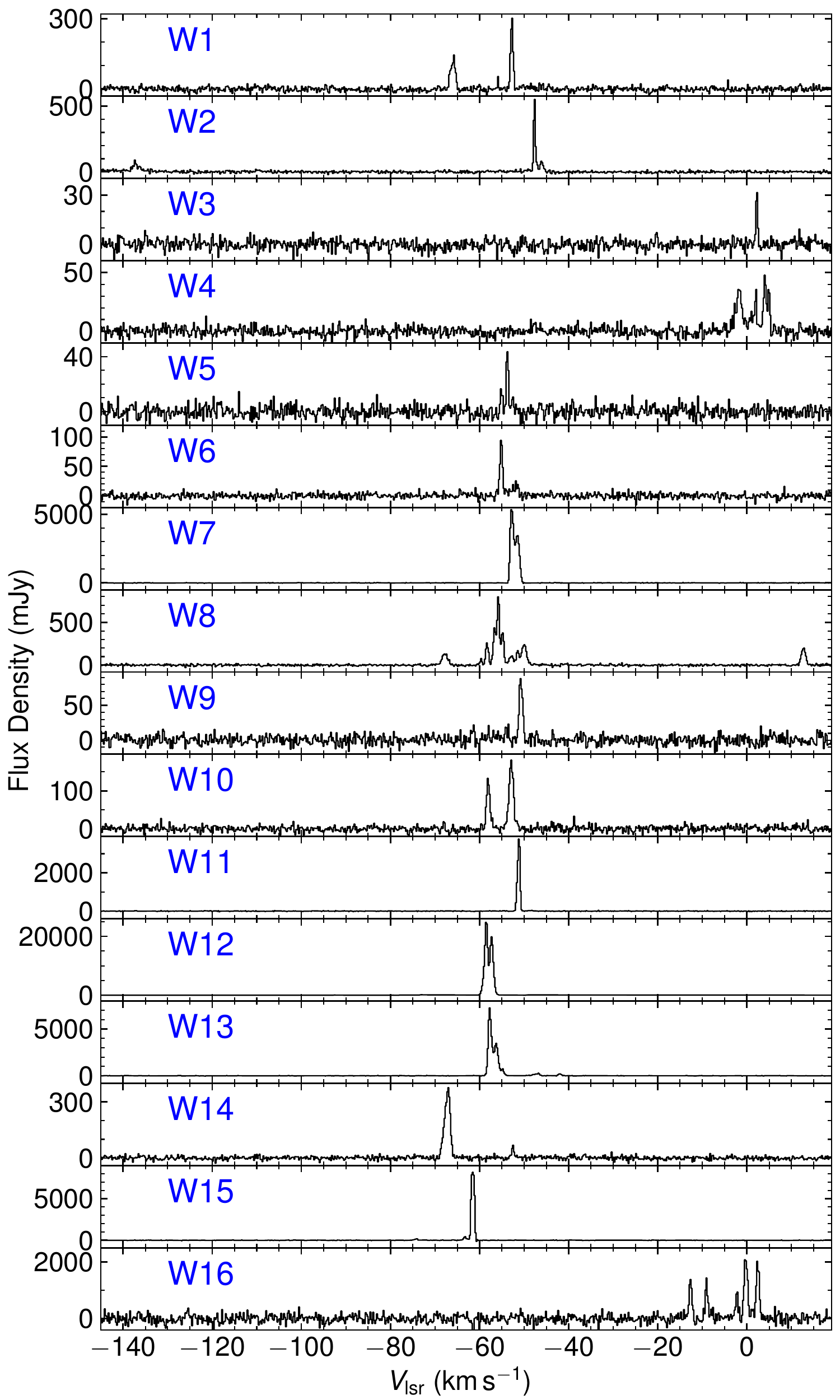} 
\caption{\water{} maser spectra in Sgr~C detected by the VLA.}
\label{app_fig:maser_spec_5}
\end{figure*}

\begin{figure*}[!h]
\centering
\includegraphics[width=0.5\textwidth]{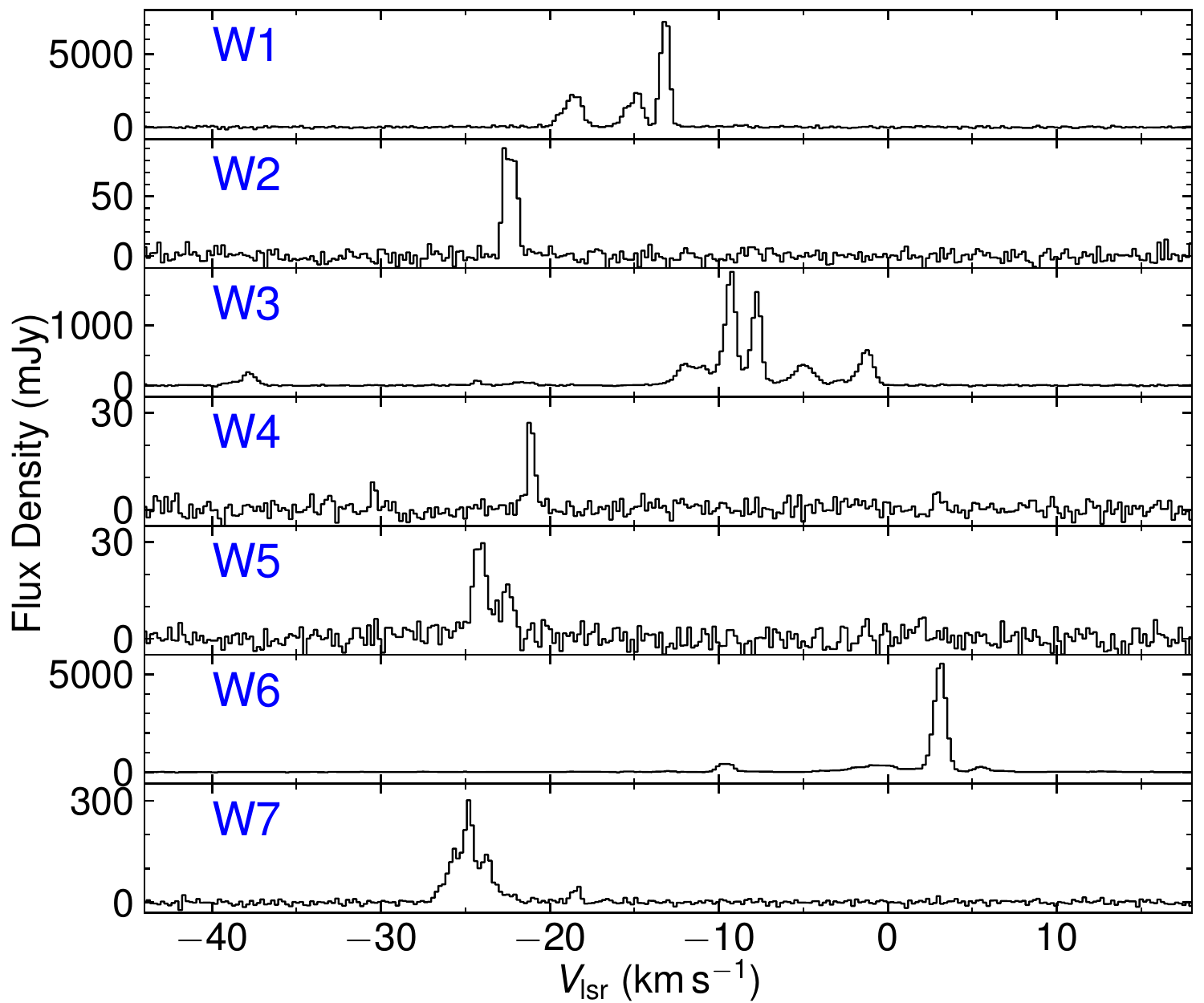} 
\caption{\water{} maser spectra in Sgr~D detected by the VLA.}
\label{app_fig:maser_spec_6}
\end{figure*}

We present the spectra of all the detected \water{} masers in the six clouds in Figures~\ref{app_fig:maser_spec_1}--\ref{app_fig:maser_spec_6}. The x-axis is \vlsr{} in unit of \kms{}, while the y-axis is flux density in unit of mJy. Positions of the masers are marked in the maps in \autoref{fig:masers}. The labels of \water{} masers with OH/IR star counterparts are marked by red.

\section{Line widths in cores}\label{sec:appd_c}
\begin{figure*}[!h]
\centering
\includegraphics[width=1\textwidth]{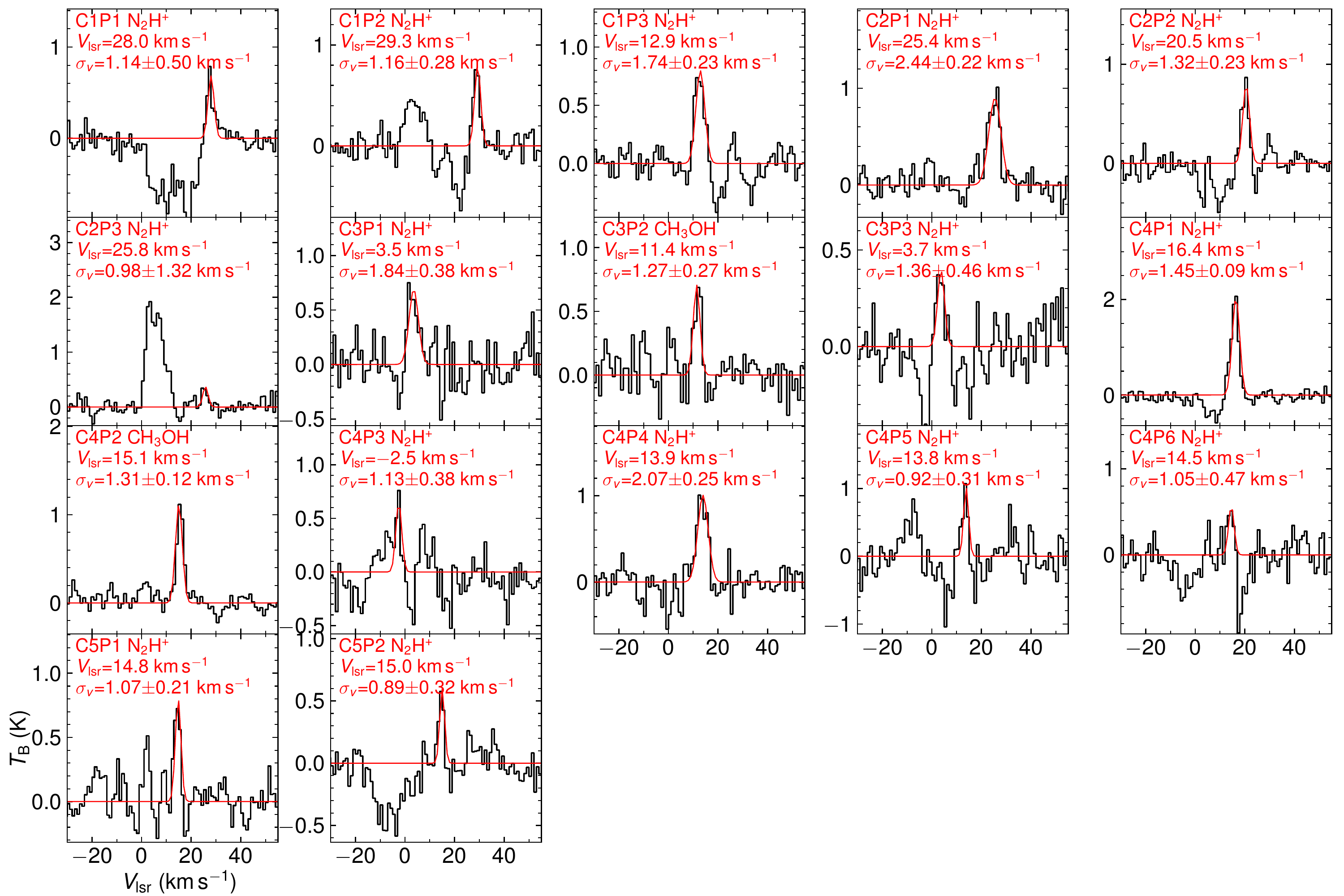} 
\caption{The SMA \nthp{} or \methanol{} spectra and best-fit models of cores in \ctw{}. The best-fit centroid velocities (\vlsr{}) and line widths ($\sigma_v$, deconvolved from channel width) are noted in each panel.}
\label{app_fig:core_spec_1}
\end{figure*}

\begin{figure*}[!h]
\centering
\includegraphics[width=0.8\textwidth]{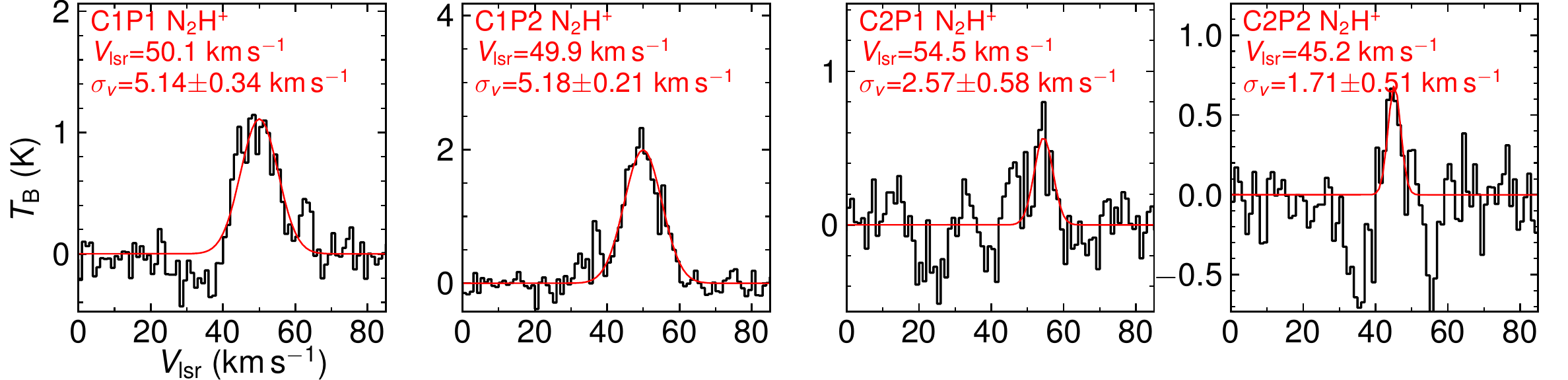} 
\caption{The SMA \nthp{} spectra and best-fit models of cores in \cfi{}. The best-fit line widths (deconvolved from channel width) are noted in each panel.}
\label{app_fig:core_spec_2}
\end{figure*}

\begin{figure*}[!h]
\centering
\includegraphics[width=1\textwidth]{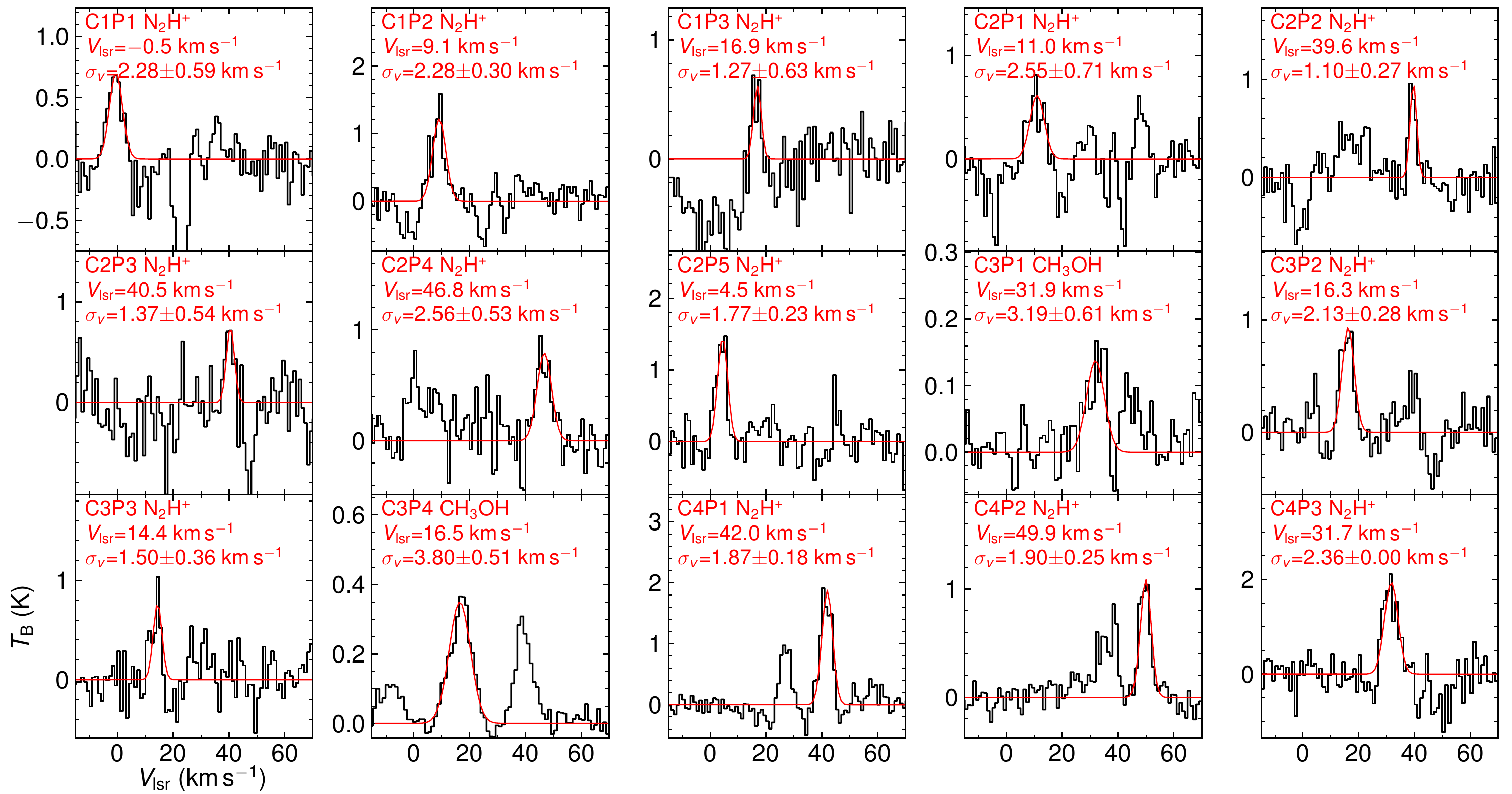} 
\caption{The SMA \nthp{} or \methanol{} spectra and best-fit models of cores in \gzp{}. The best-fit line widths (deconvolved from channel width) are noted in each panel.}
\label{app_fig:core_spec_3}
\end{figure*}

\begin{figure*}[!h]
\centering
\includegraphics[width=1\textwidth]{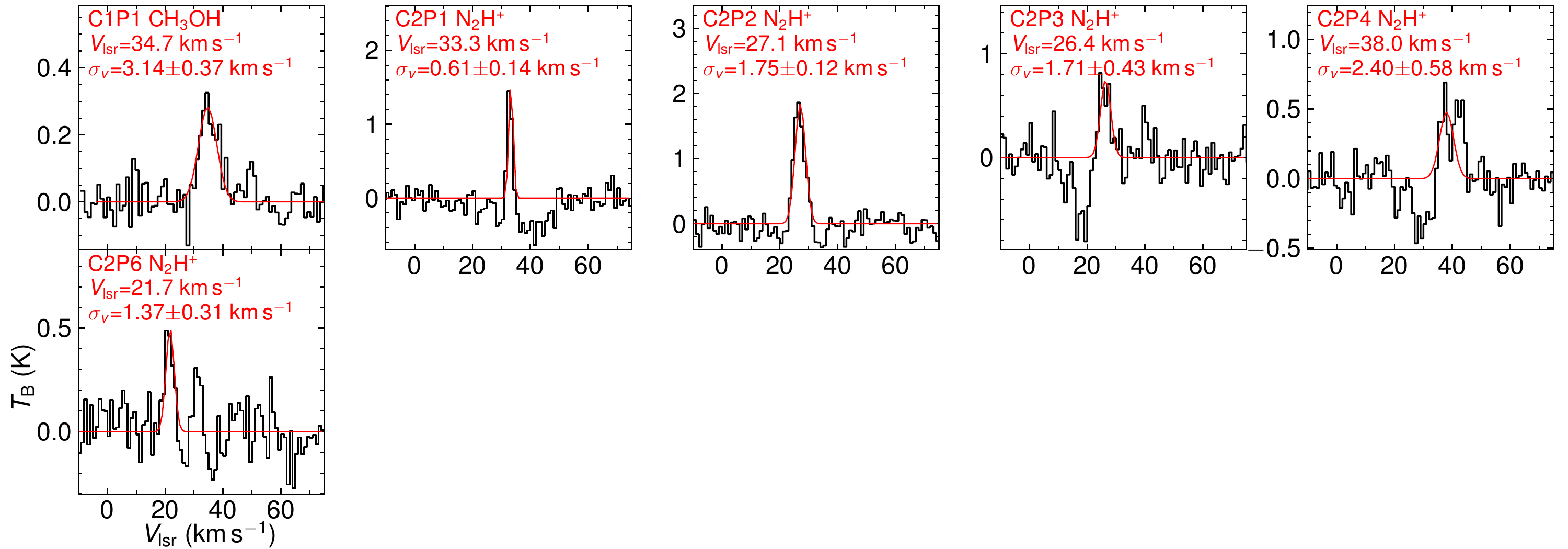} 
\caption{The SMA \nthp{} or \methanol{} spectra and best-fit models of cores in \sgb{}. The best-fit line widths (deconvolved from channel width) are noted in each panel.}
\label{app_fig:core_spec_4}
\end{figure*}

\begin{figure*}[!h]
\centering
\includegraphics[width=1\textwidth]{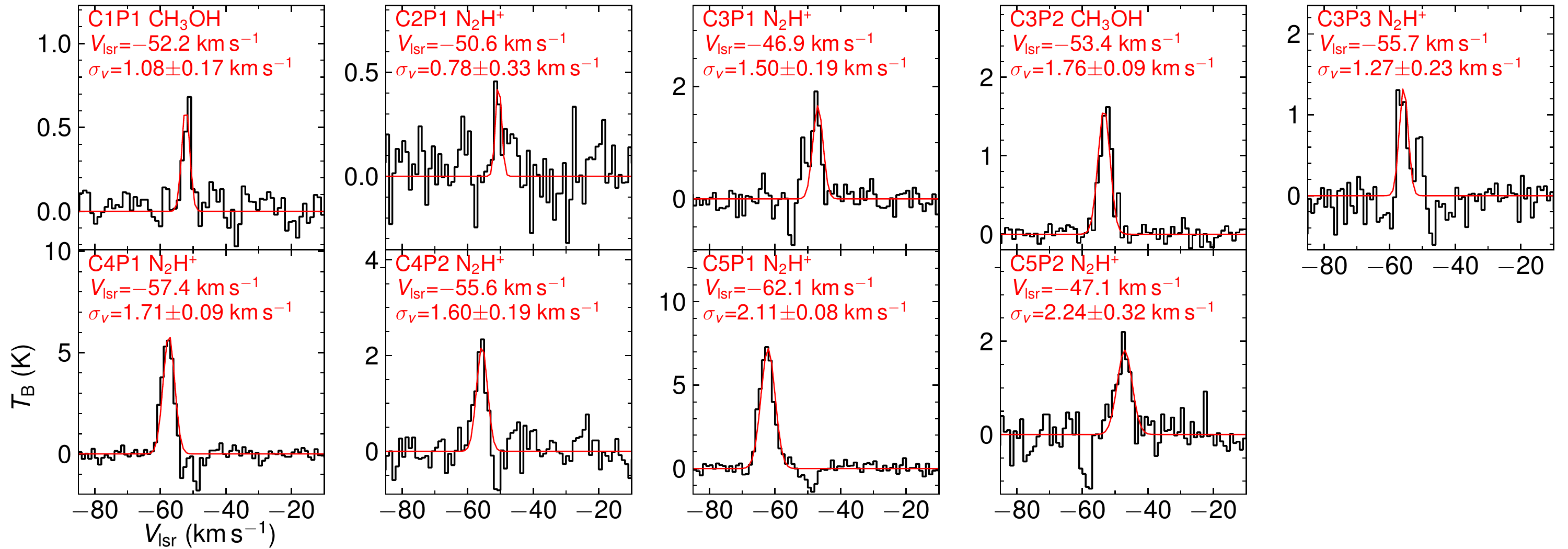} 
\caption{The SMA \nthp{} or \methanol{} spectra and best-fit models of cores in Sgr~C. The best-fit line widths (deconvolved from channel width) are noted in each panel.}
\label{app_fig:core_spec_5}
\end{figure*}

\begin{figure*}[!h]
\centering
\includegraphics[width=1\textwidth]{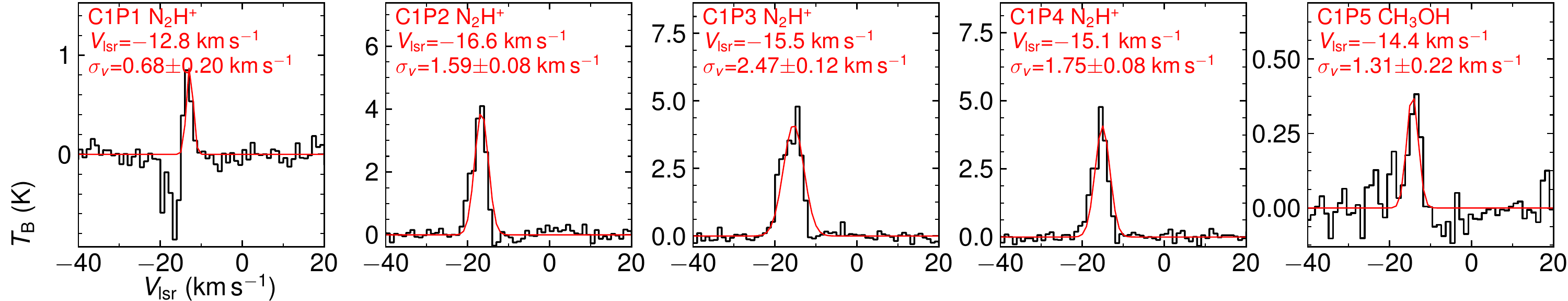} 
\caption{The SMA \nthp{} or \methanol{} spectra and best-fit models of cores in Sgr~D. The best-fit line widths (deconvolved from channel width) are noted in each panel.}
\label{app_fig:core_spec_6}
\end{figure*}

We extract the mean spectra of \nthp{}~3--2 \citep{kauffmann2017a}, or if it is not detected, those of \methanol{}~4$_{22}$--3$_{12}$ in our SMA data, toward the identified cores. Then for each spectrum, a single Gaussian is fit to obtain the line width. The results are shown in Figures~\ref{app_fig:core_spec_1}--\ref{app_fig:core_spec_6}.

\section{Estimate of Cluster Masses through Monte-Carlo Simulations}\label{sec:appd_d}
In order to quantify the uncertainties in cluster masses stemming from the random sampling of stellar masses following the IMF as a probability distribution function, we run Monte-Carlo simulations to generate cluster masses, and find the relation between them and the number of high-mass ($\ge$8~\msol{}) stars in the cluster. The scripts to run these simulations are available at \url{https://github.com/xinglunju/CMZclouds/tree/master/Mcluster}.

Following the Bayes theorem, the posterior probability of cluster masses $M_\text{cluster}$ given that there are $N$ high-mass stars in the cluster is
\begin{equation}
\text{prob}(M_\text{cluster}|N)=\frac{\text{prob}(M_\text{cluster})\text{prob}(N|M_\text{cluster})}{\text{prob}(N)},
\end{equation}
in which $\text{prob}(M_\text{cluster})$ is the prior probability of $M_\text{cluster}$, $\text{prob}(N|M_\text{cluster})$ is the likelihood of having $N$ high-mass stars given a cluster mass of $M_\text{cluster}$, and $\text{prob}(N)$ is the evidence and usually treated as a normalizing factor.

\begin{figure*}[!t]
\centering
\begin{tabular}{@{}p{0.50\textwidth}@{}p{0.50\textwidth}@{}}
\includegraphics[width=0.50\textwidth]{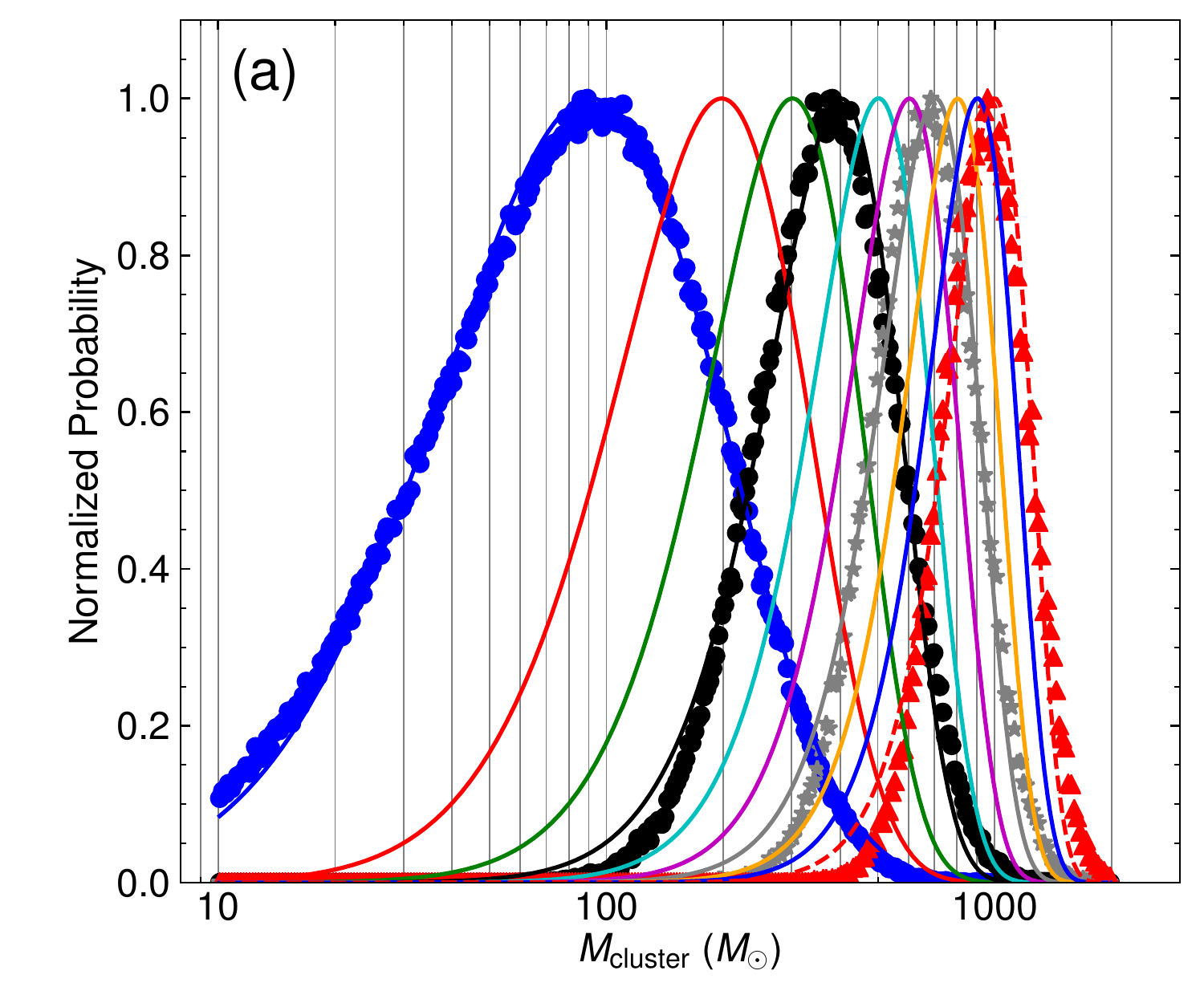} & 
\includegraphics[width=0.50\textwidth]{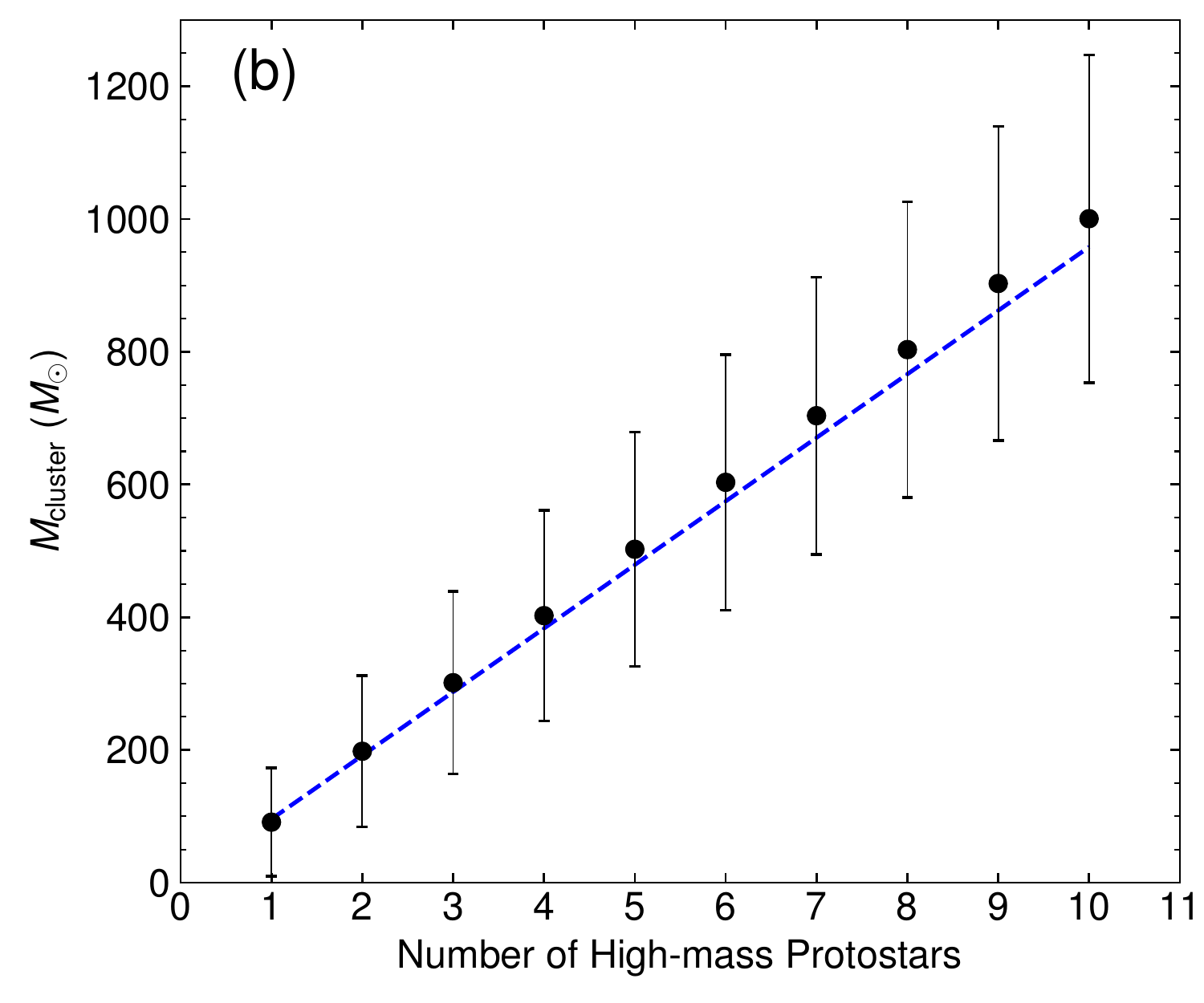}
\end{tabular}
\caption{(a) Normalized probabilities of cluster masses given 1 to 10 high-mass stars produced in the cluster. Data points are results from simulations and solid curves are best-fit Weibull functions. For clarify of presentation, only data points of simulations with $N=1, 4, 7, 10$ are plotted. (b) Expected cluster masses given the detection of a certain number of high-mass protostars. Vertical errorbars show the uncertainties in cluster masses.}
\label{app_fig:mcluster_N}
\end{figure*}

We adopt a flat log-prior probability $\text{prob}(M_\text{cluster})$ in a range of $[10, 2000]$~\msol{}, outside of which $\text{prob}(M_\text{cluster})$ is set to 0. To obtain the likelihood $\text{prob}(N|M_\text{cluster})$, we run Monte-Carlo simulations to generate stellar masses following the IMF \citep[Equations~1 \& 2 of][with stellar masses between 0.01~\msol{} and 150~\msol{}]{kroupa2001} as a probability distribution function until the total mass reaches $M_\text{cluster}$, and count the number of high-mass stars produced in this process. For each $M_\text{cluster}$, we run the simulation 10,000 times, and divide the number of the runs producing $N$ high-mass stars by 10,000 to get the likelihood $\text{prob}(N|M_\text{cluster})$. Finally, the posterior probability $\text{prob}(M_\text{cluster}|N)$ is proportional to $\text{prob}(N|M_\text{cluster})$, and is shown in \autoref{app_fig:mcluster_N}a.

The resulting posterior probability functions $\text{prob}(M_\text{cluster}|N)$ are skewed to the right side compared to a lognormal distribution. To better estimate peak locations, we fit Weibull distribution functions \citep{weibull1951} to $\text{prob}(M_\text{cluster}|N)$, shown in \autoref{app_fig:mcluster_N}a. We use the modes (peak locations) of the Weibull functions (therefore the most likely) as the final estimate of cluster masses, and the rms of the simulated data to approximate the uncertainties in the estimated masses.

In \autoref{app_fig:mcluster_N}b, we plot the cluster masses estimated above given $N$ high-mass stars in the cluster. The result agrees well with the analytical form directly derived from the IMF,
\begin{equation}\label{equ:mcluster_N}
M_\text{cluster}=95.8\times N~\msol,
\end{equation}
which is plotted as a dashed curve in \autoref{app_fig:mcluster_N}b. The simulations are able to give an estimate of uncertainties stemming from the random sampling of stellar masses following the IMF as a probability distribution function. In \autoref{subsec:disc_sfr}, we rely on the data points in \autoref{app_fig:mcluster_N}b to estimate cluster masses.

\end{CJK}
\end{document}